\documentclass[sigconf,nonacm]{acmart}
\setcopyright{none}

\AtBeginDocument{%
  }

\usepackage{amsmath}
\usepackage{bm}

\usepackage[ruled]{algorithm2e} 

\SetKwComment{Comment}{$//$\ }{}

\usepackage{empheq} 
\SetKwComment{Comment}{$//$\ }{}
\usepackage{tikz,pgfplots} 
\pgfplotsset{width=\linewidth, compat=1.9}
\usepackage{subcaption}
\usepackage{multirow}
\usepackage{dblfloatfix}
\newtheorem{theorem}{Theorem}
\usepackage[export]{adjustbox}

\newcommand{\makemath}[1]{{\ensuremath{#1}}}

\newcommand{\zeros}{\makemath{\mathbf{0}}} 
\newcommand{\identity}{\makemath{\mathbf{I}}} 

\newcommand{\mass}{\makemath{\mathbf{M}}} 
\newcommand{\reduction}{\makemath{\mathbf{G}}} 
\newcommand{\contact}{\makemath{\mathbf{C}}} 
\newcommand{\system}{\makemath{\mathbf{A}}} 
\newcommand{\delasus}{\makemath{\mathbf{W}}} 
\newcommand{\jacobian}{\makemath{\mathbf{H}}} 
\newcommand{\nsjacobian}{\makemath{\mathbf{J}}} 

\newcommand{\choleskyfactor}{\makemath{\mathbf{L}}} 
\newcommand{\sparseinverse}{\makemath{\mathbf{S}}} 

\newcommand{\compliance}{\makemath{\mathbf{E}}} 

\newcommand{\bcompliance}{\makemath{\mathbf{E}_b^{}}} 
\newcommand{\ncompliance}{\makemath{\mathbf{E}_n^{}}} 
\newcommand{\fcompliance}{\makemath{\mathbf{E}_f^{}}} 

\newcommand{\weightnsjacobian}{\mathbf{\Omega}} 

\newcommand{\matP}{\makemath{\mathbf{P}}} 
\newcommand{\matQ}{\makemath{\mathbf{Q}}} 

\newcommand{\pos}{\makemath{\mathbf{q}}} 
\newcommand{\vel}{\makemath{\mathbf{v}}} 

\newcommand{\proj}{\makemath{\mathbf{p}}} 
\newcommand{\Lam}{\makemath{\mathbf{\boldsymbol{\Lambda}}}} 
\newcommand{\Pene}{\makemath{\mathbf{\boldsymbol{\Delta}}}} 
\newcommand{\lam}{\makemath{\mathbf{\boldsymbol{\lambda}}}} 
\newcommand{\pene}{\makemath{\mathbf{\boldsymbol{\delta}}}} 
\newcommand{\rhs}{\makemath{\mathbf{b}}} 

\newcommand{\sn}{\makemath{\Tilde{\pos}}} 

\newcommand{\relativevel}{\makemath{\mathbf{u}}} 

\newcommand{\vecd}{\makemath{\mathbf{d}}} %
\newcommand{\vecg}{\makemath{\mathbf{g}}} %
\newcommand{\vech}{\makemath{\mathbf{h}}} %
\newcommand{\vecf}{\makemath{\mathbf{f}}} %

\newcommand{\nsprecond}{\makemath{\mathbf{r}}} 

\newcommand{\veczeros}{\makemath{\mathbf{0}}} 
\newcommand{\vecones}{\makemath{\mathbf{1}}} 

\newcommand{\cbilateral}{\makemath{\vec{b}}} 
\newcommand{\cnormal}{\makemath{\vec{n}}} 
\newcommand{\cfriction}{\makemath{\vec{f}}} 

\newcommand{\cgeneral}{\makemath{\vec{c}}} 

\newcommand{\dt}{\makemath{h}} 
\newcommand{\weight}{\makemath{w}} 

\newcommand{\weightns}{\mathbf{\omega}} 

\newcommand{\lamval}{\makemath{\lambda}} 
\newcommand{\peneval}{\makemath{\delta}} 

\newcommand{\elasticEnergy}{\makemath{\mathbf{\boldsymbol{\psi}}}} 
\newcommand{\localEnergy}{\makemath{\mathbf{\boldsymbol{\zeta}}}} 

\newcommand{\ncpfunc}{\makemath{\mathbf{\boldsymbol{\phi}}}} 

\newcommand{\scfunc}{\makemath{\mathbf{\boldsymbol{\xi}}}} 

\begin{document}

\title{Fast But Accurate: A Real-Time Hyperelastic Simulator with Robust Frictional Contact}

\author{Ziqiu Zeng}
\affiliation{%
  \institution{University of Strasbourg}
  \country{}
}
\affiliation{%
  \institution{CAIR, CAS Hong Kong}
  \country{}
}
\email{zengziqiu1995@gmail.com}

\author{Siyuan Luo}
\authornote{Corresponding author.}
\affiliation{%
  \institution{National University of Singapore}
  \country{}
}
\email{sy.luo@nus.edu.sg}

\author{Fan Shi}
\affiliation{%
  \institution{National University of Singapore}
  \country{}
}
\email{fan.shi@nus.edu.sg}

\author{Zhongkai Zhang}
\affiliation{%
  \institution{CAIR, CAS Hong Kong}
  \country{}
}
\email{zhongkai.zhangzkz@gmail.com}

\begin{abstract}
We present a GPU-friendly framework for real-time implicit simulation of elastic material in the presence of frictional contacts. The integration of hyperelasticity, non-interpenetration contact, and friction in real-time simulations presents formidable nonlinear and non-smooth problems, which are highly challenging to solve. By incorporating nonlinear complementarity conditions within the local-global framework, we achieve rapid convergence in addressing these challenges. While the structure of local-global methods is not fully GPU-friendly, our proposal of a simple yet efficient solver with sparse presentation of the system inverse enables highly parallel computing while maintaining a fast convergence rate. Moreover, our novel splitting strategy for non-smooth indicators not only amplifies overall performance but also refines the complementarity preconditioner, enhancing the accuracy of frictional behavior modeling. Through extensive experimentation, the robustness of our framework in managing real-time contact scenarios, ranging from large-scale systems and extreme deformations to non-smooth contacts and precise friction interactions, has been validated. Compatible with a wide range of hyperelastic models, our approach maintains efficiency across both low and high stiffness materials. Despite its remarkable efficiency, robustness, and generality, our method is elegantly simple, with its core contributions grounded solely on standard matrix operations.
\end{abstract}

\begin{teaserfigure}
  \includegraphics[width=\textwidth]{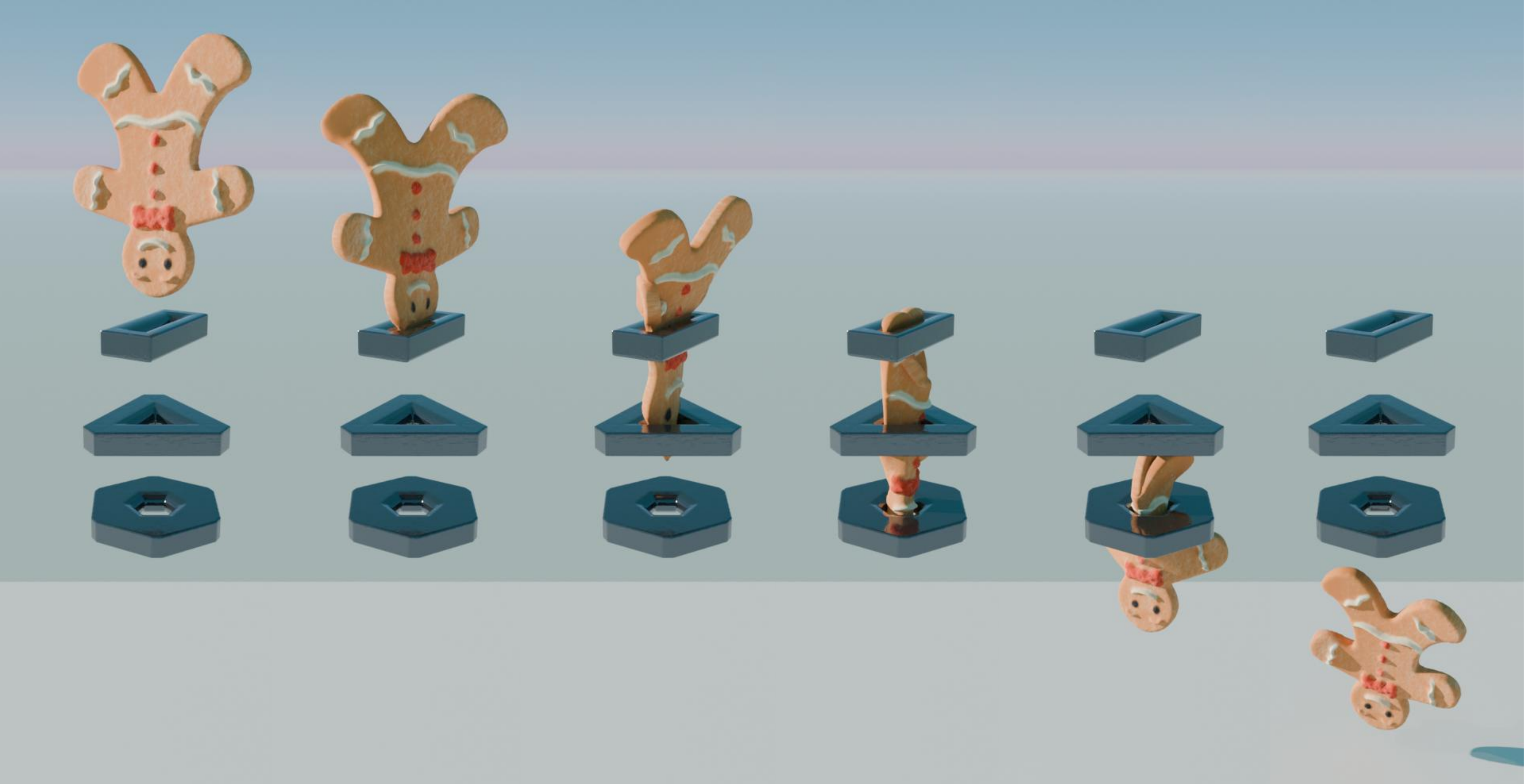}
  \caption{Crossing Gingerbread Man: 
  we propose a new framework for stable simulation of hyperelastic materials with nodes under large deformation and generic contact constraints in real-time.
  Pulling the gingerbread man ($58.5k$ DoFs for a single object) through the thin and irregular obstacles is simulated at $11.95 ms$/iteration using $5$ local-global iterations per frame when maximum contact pairs are involved ($800$ contact constraints).   
  }
  \Description{This is the teaser figure for the article.}
  \label{fig: teaser}
\end{teaserfigure}

\maketitle

\section{Introduction}


Physics-based simulation of deformable materials plays a crucial role across a variety of disciplines, such as computer graphics, robotics, and medical imaging. Within these fields, it enhances visual experiences through accurate simulations of soft objects and their interactions with the environment, often utilizing finite element discretization for complex nonlinear materials. An ideal physical simulator should possess key attributes like the capacity for \textit{generality} in simulating different materials and the \textit{robustness} to minimize failure instances. In interactive applications, achieving \textit{real-time performance} is essential, while a framework's \textit{simplicity} is highly valued for simplifying maintenance and development processes, making it widely applicable across various downstream fields.

Position-based dynamics (PBD) \cite{muller_position_2006} has gained widespread adoption in real-time simulation engines due to its simplicity, robustness, and efficiency. However, as PBD is not derived from continuum mechanics principles, its real-time performance comes at the cost of accuracy. To address this limitation, Projective Dynamics (PD) \cite{bouaziz_projective_2014} employs local-global optimization, combining localized constraint solving with global optimization for better accuracy and efficiency. 
Vertex Block Descent (VBD) \cite{chen_vertex_2024} further enhances real-time performance by optimizing matrix structures and constraints for parallelization, delivering significant computational advantages. 

In multi-body systems, addressing contact problems typically involves utilizing either penalty methods with soft constraints or Lagrange multiplier methods with hard constraints. While penalty methods are computationally efficient, they encounter difficulties in enforcing strict non-penetration and accurate friction. Incremental Potential Contact (IPC) \cite{li_incremental_2020} mitigates this using a logarithmic barrier penalty and barrier-aware line search to prevent intersections. 
Lagrange multiplier methods are robust and accurate but face challenges with convergence and parallelization in relaxation techniques like projected Gauss-Seidel \cite{Duriez2013a}.
Non-smooth Newton methods \cite{macklin_non-smooth_2019}, combined with Krylov subspace solvers and complementarity preconditioner, offer rapid convergence, high accuracy, and efficient parallelization.

We propose a framework where the local-global iterations are constrained by nonlinear complementarity conditions. Our simulator, empowered by efficient matrix operations, exhibits many favorable features:
First, our framework is \textit{efficient} and enables large-scale simulations in real-time, in the presence of nonlinear and non-smooth problems.
Within the real-time computational constraints, our solver is able to converge to a desirable \textit{accuracy}, thereby enhancing the \textit{stability}.
Second, our method is not only \textit{general} for a wide range of hyperelastic models but also preserves efficiency across materials with both low and high stiffness.
Through extensive experiments, our framework has been proven to be \textit{robust} when solving the complex contact problem with different challenges, including large deformation, non-smooth contact, and accurate friction.
Beyond these advantages, our method shines in its simplicity, as the core contributions only rely on standard matrix operations.
In summary, our contributions are listed as follows:
\begin{enumerate}
    \item Our method achieves a highly parallelizable structure while maintaining a high convergence rate by transforming the linear global system into sparse matrix multiplications.
    \item We reformulate Lagrange multiplier methods, particularly the non-smooth Newton method, to integrate seamlessly with local-global integrators.
    \item We introduce a strategy to separate non-smooth indicators, resulting in both reduced Schur-complement computations and enhanced complementarity preconditioner.
    \item Our algorithm establishes a unified, GPU-friendly system for real-time dynamics and contact resolution, with a modular design that ensures easy integration with other frameworks.
\end{enumerate}

\section{Related work}

\begin{figure*}[t!]
    \centering
	\noindent%
	\begin{subfigure}[t]{0.33\linewidth}%
		\includegraphics[width=0.95\linewidth,clip,trim=200 0 200 0]{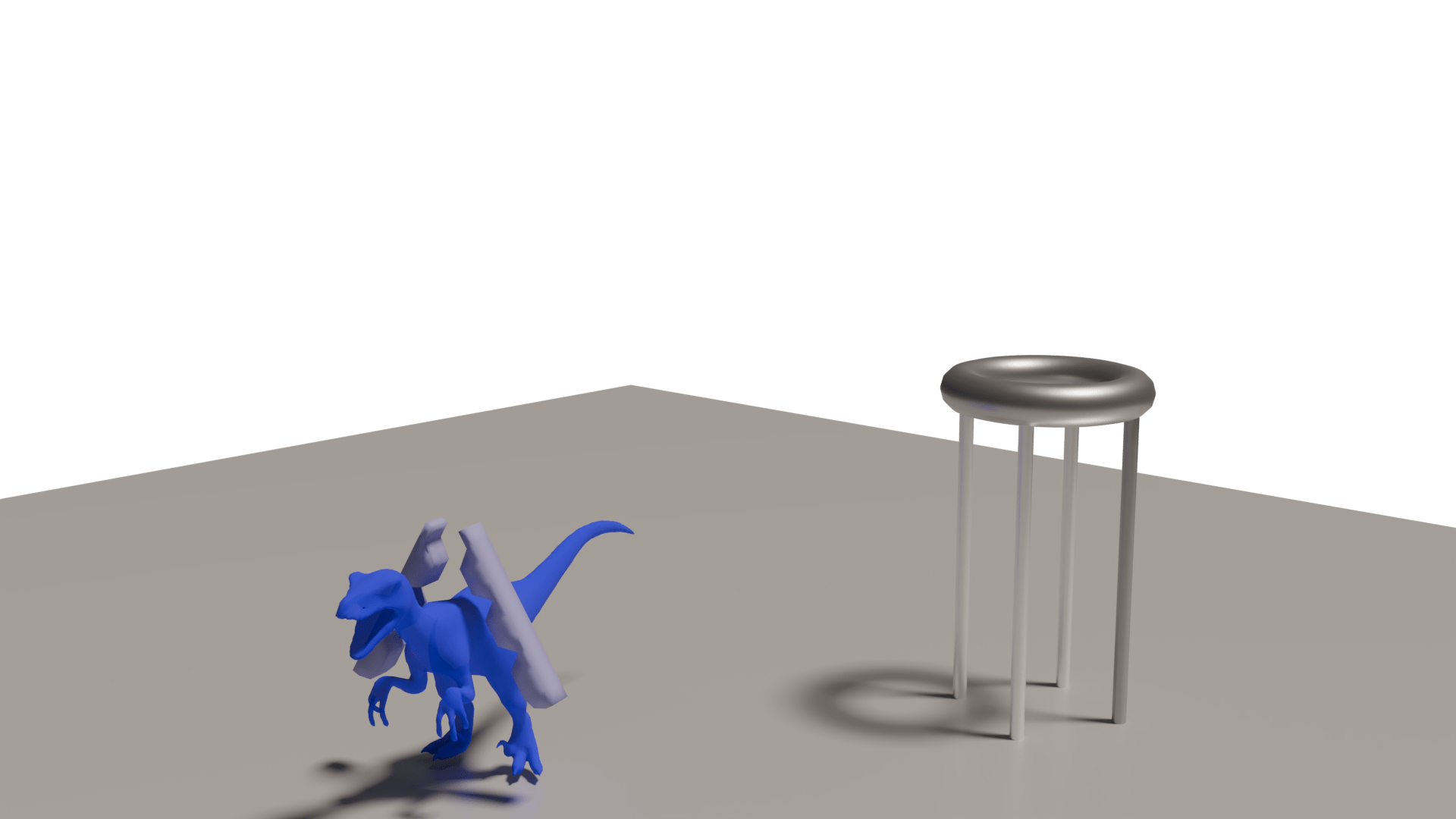}%
            \caption{Frame $0$}
	\end{subfigure}%
	\begin{subfigure}[t]{0.33\linewidth}%
		\includegraphics[width=0.95\linewidth,clip,trim=200 0 200 0]{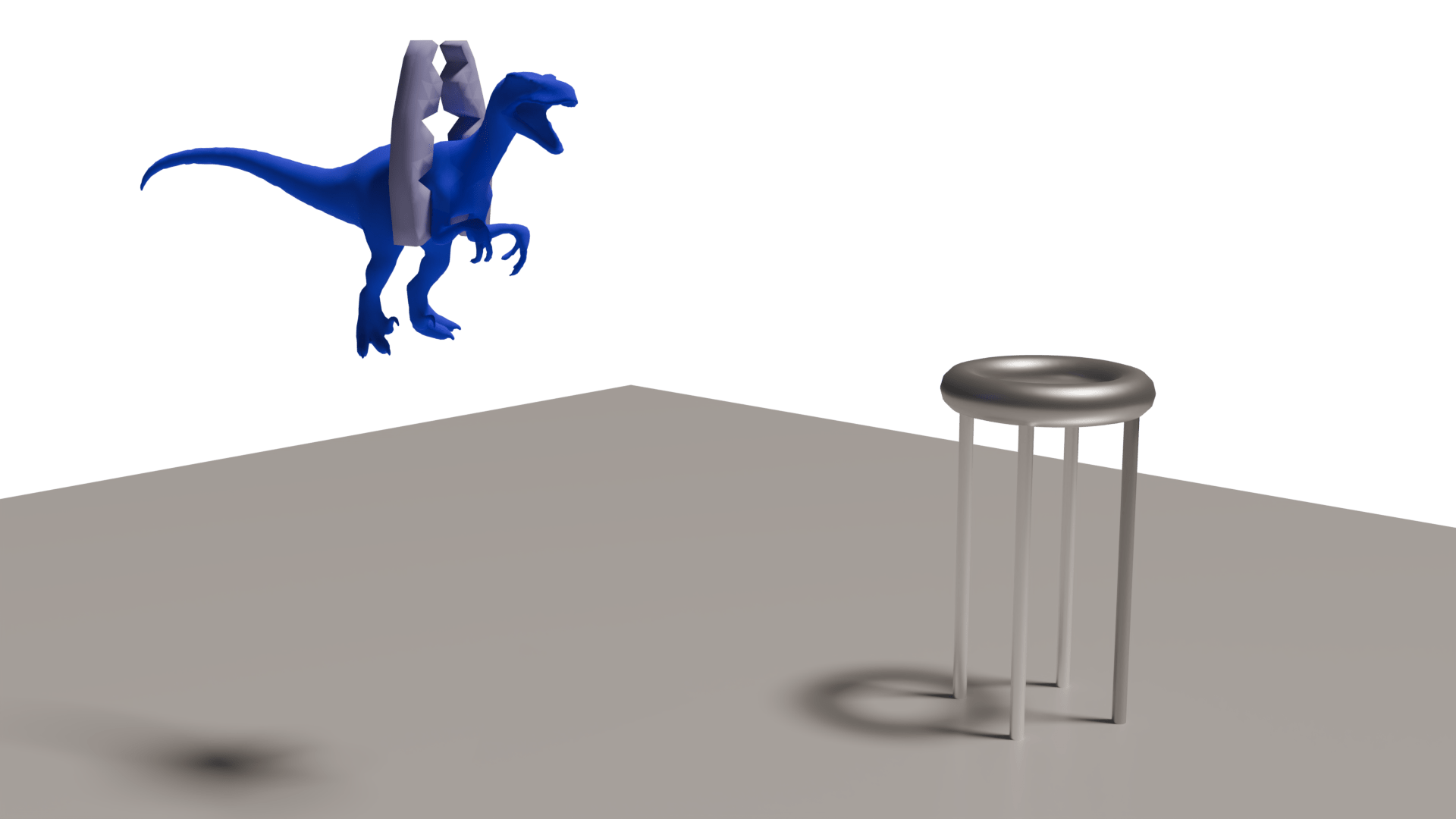}%
            \caption{Frame $1100$}
	\end{subfigure}%
	\begin{subfigure}[t]{0.33\linewidth}%
		\includegraphics[width=0.95\linewidth,clip,trim=200 0 200 0]{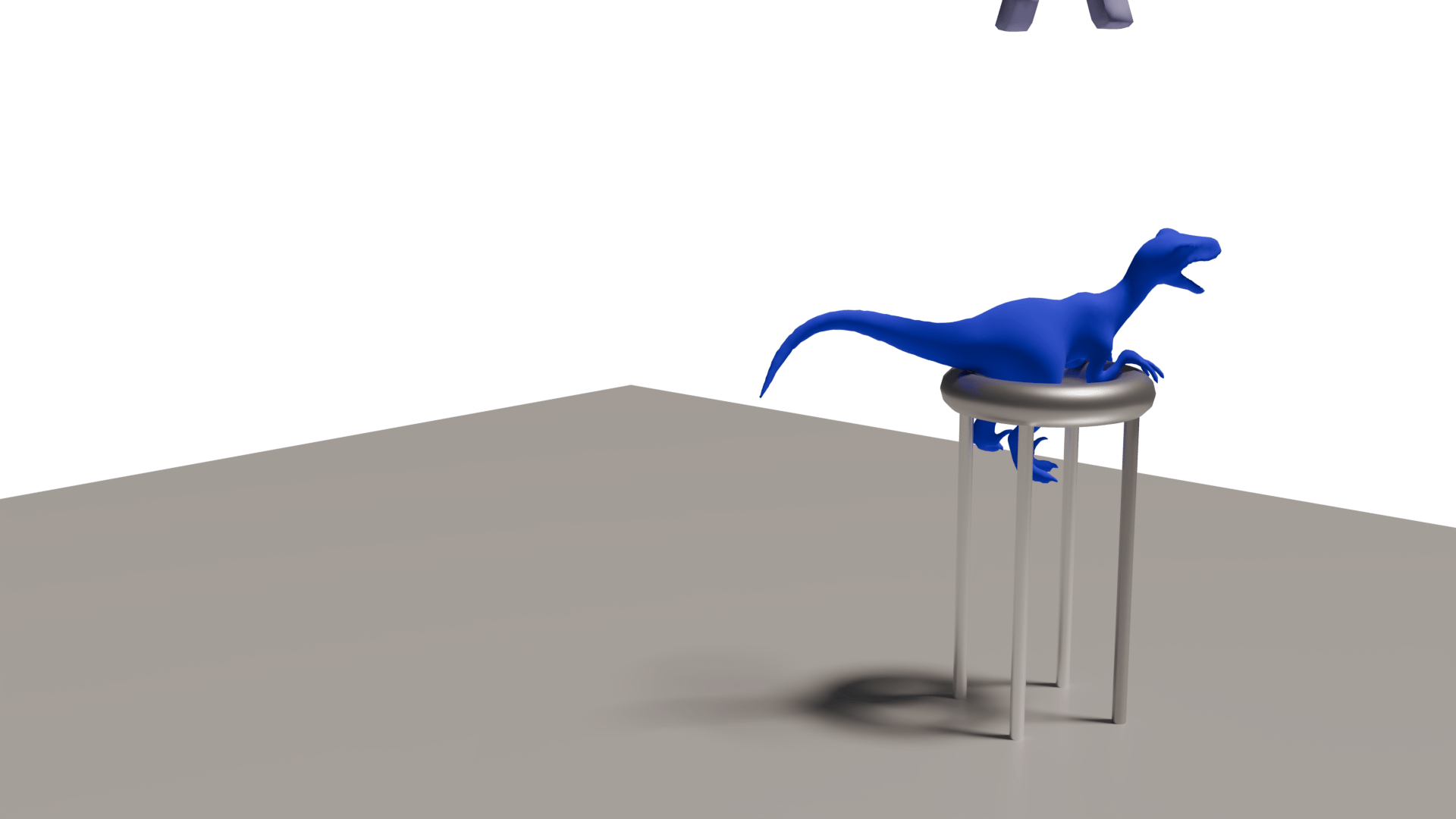}%
            \caption{Frame $3200$}
	\end{subfigure}%
    \caption{
        Grabbing Raptor: stable catching of an elastic raptor with a soft gripper actuated by cables. Lifting, rotating, and moving the raptor by the fingers are complex operations where friction constraints are necessary.
    }
    \Description{}
    \label{fig: grab raptor}
\end{figure*}

Physics-based simulation in computer graphics has been extensively studied. In this section, we focus on discussing the most recent work related to our method.

\subsection{Implicit simulation for elastic dynamics}
\label{sec: implicit simulation for elastic dynamics}

In computer graphics, implicit simulations \cite{Bro-Nielsen1996} using backward Euler integration allow for larger time steps compared to explicit methods \cite{comas_efficient_2008}, significantly improving computational stability in stiff systems.
Finite element (FE) models \cite{sifakis_fem_2012, kim_dynamic_2022} are a valuable tool for understanding the underlying mechanisms in the real world, as they provide a direct explanation of soft tissue behavior through constitutive relations.
With the rapid advancement of computational power and methods, FE models have become increasingly suitable for real-time and interactive simulations. 

In terms of integration, the traditional implicit method employs the Newton iteration technique to solve nonlinear problems.
Initially restricted to linear elastic models \cite{Bro-Nielsen1996}, the method is later extended to the co-rotational formulation \cite{Felippa2000} and hyperelastic and viscoelastic materials \cite{Marchesseau2010}.
Although the Newton's method converges fast in solving nonlinear systems, it involves re-evaluate and invert the Hessian matrix in each Newton iteration, which implies large computing costs.
In practice, Newton integrators typically perform only one iteration in real-time applications \cite{faure_sofa}.

In computer graphics, PBD \cite{muller_position_2006} is very popular in applications like cloth simulations owning to its high efficiency and stable behavior.
As an extension, the extended PBD (XBPD) \cite{macklin_xpbd_2016} addresses the limitation of stiffness dependence, providing a better approximation to the implicit Euler method. 
Recently, the VBD method \cite{chen_vertex_2024} is proposed, which constrains the positions with hyperelastic constraints.
We categorize these methods as PBD-like methods \cite{bender_survey_2017}, where the positions of the objects are iteratively adjusted to satisfy a set of local physical constraints.
As a result, the PBD-like methods iterate in a Gauss-Seidel-like manner, and their convergence rates are limited by the inefficiency of propagation.

On the other hand, PD \cite{bouaziz_projective_2014} uses a different way of integration (initially proposed in \cite{liu_fast_2013}), solving the nonlinear implicit Euler problem through iterative local and global steps.
While the PD initially only supports the as-rigid-as-possible (ARAP) model \cite{chao_simple_2010}, advanced PD methods \cite{liu_quasi-newton_2017, overby_admm_2017} extend the elastic model to generic hyperelastic models.
Moreover, \cite{liu_quasi-newton_2017} regards the local-global iterations as a Quasi-Newton method and accelerates it using the L-BFGS method, thus creating what is known as LBFGS-PD.
On the other hand, another point of view is proposed in \cite{narain_admm_2016, overby_admm_2017, brown_accurate_2018}, considering the local-global method as a special case of the ADMM method. 
This work is extended in the WRAPD \cite{brown_wrapd_2021} and the mixed variational FEM \cite{trusty_mixed_2022} to improve the convergence rate with rotation-aware global steps.
We categorize these methods as PD-like local-global methods, also briefly local-global methods in the rest of this paper.
The key difference between the PD-like methods and the PBD-like methods is how to normalize the local projection results to the global coupling.
Unlike PBD-like methods that directly adjust the positions, PD-like ones normalize the results onto the right-hand-side of the global step, via a linear mapping.
Accordingly, PD-like methods couple the local solutions through a constant system that can be pre-factorized for efficient computing.
Compared with PBD-like methods,  PD-like methods provide faster propagation of the local results, leading to much higher convergence rates.

Although local-global methods provide fast convergence and efficient global solution owning to pre-factorization, the global step is hard to be parallelized on the GPU due to the strong data dependence in forward and backward substitutions.
To tackle this issue, \cite{wang_chebyshev_2015} proposes replacing the global linear system with a single Jacobi iteration and employing the Chebyshev approach to accelerate the convergence.
The method is further extended to A-Jacobi in \cite{lan_penetration-free_2022}, which incorporates multiple Jacobian iterations to enhance the potential for parallelization. 
To address Jacobi methods' poor convergence, \cite{fratarcangeli_vivace_2016} suggests using a Gauss-Seidel approach and employing graph coloring techniques \cite{saad_iterative_2003} to improve parallelization.
Nevertheless, these methods for parallelizing the global steps only solve the global system more or less approximately, thereby largely decreasing the convergence rate.

\subsection{Muti-body Contact}

The collision modeling in computer graphics inherits from the numerical methods in constrained optimization theory \cite{nocedal_numerical_2006}.
By simplifying the non-interpenetration contact as inequality constraints, one can address the contact problems via penalty methods \cite{Kugelstadt2018, Hasegawa2004} and augmented Lagrangian methods. 
These methods handle constraints by adding penalty terms to the objective function such that the overall variational optimization becomes unconstrained.
Despite being simple and straightforward, such methods can lead to numerical instability under complex constraints.
Another general issue of these methods is the difficulty of accurate friction modeling due to the complexity in the non-smooth complementarity conditions.
To tackle these challenges, IPC \cite{li_incremental_2020} employs a logarithmic barrier penalty to create a stronger impulse for separating objects as they approach each other closely. 
Furthermore, in every Newton iteration, it utilizes a CCD-aware line search to geometrically ensure intersection-free results.
The friction modeling is also included in this pipeline through a variational approximation.

To accurately model non-smooth conditions, one can formulate frictional contact problems as complementarity problems.
In optimization theory, these problems can be solved using Quadratic Programming (QP) or Sequential QP for simulating linear elastic or nonlinear hyperelastic materials \cite{nocedal_numerical_2006, kaufman_staggered_2008, kane_finite_1999}.
Different to the penalty methods, these methods generally constrain the linearized system (e.g., at each Newton solve) with Lagrange multipliers \cite{baraff_linear_1996}, thereby known as Lagrange multiplier methods.
Early works like \cite{STEWART1996} linearize the frictional constraints along the friction cones, thereby converting the original problem to a linear complementarity problem (LCP).
The LCP can be then solved with relaxation methods such as Projected Gauss-Seidel (PGS) \cite{duriez_realistic_2006, daviet_simple_2020, li_implicit_2018}, and direct methods like pivoting methods \cite{erleben_numerical_2013}.
It has been observed in \cite{Todorov2010} that the friction cone linearization could be unnecessary since the non-smooth contact model can be simply treated as an additional set of non-linear equations in the system.
\cite{larionov_frictional_2021} proposes a smooth local implicit surface representation to accurately handle frictional contact between smooth objects.
The Newton-based approaches are successful in handling friction model \cite{Bertails-Descoubes2011}.
By transforming the Signorini-Coulomb conditions \cite{brogliato_nonsmooth_2016} to equaly nonlinear complementarity problem (NCP) functions, one can convert the origin non-smooth problem to a root-finding one.
Such approach can yield quadratic convergence, and is decoupled with the linear solver, which means that one can choose a fast linear solver like Krylov space solver (e.g., Conjugate Residual \cite{Frncu2015VIRTUALTO}) for both fast convergence and high potential for parallelization.
\cite{macklin_non-smooth_2019} extended this method to handle the soft materials and proposed an efficient complementarity preconditioner for smoothing the NCP functions.

In local-global methods, collisions can be handled either through the penalty way (i.e., adding local penalty contact energies), or with the Lagrange multipliers (i.e.,  constraining the global steps).
The penalty way follows the simple collision handing in PBD, dynamically adding contact energies while potential collisions are detected \cite{bouaziz_projective_2014, wang_optimized_2021}.
The IPC energy can also be included in this way \cite{lan_penetration-free_2022}.
However, such methods are typically only adaptable to incomplete methods that are less sensitive to changes in the system \cite{fratarcangeli_vivace_2016}.
In contrast, \cite{overby_admm_2017} proposes constraining the global steps with Lagrange multipliers when interpenetration is detected.
In their work, the non-smooth conditions are omitted, thereby simplifying the complementarity problems into linear systems.
\cite{komaritzan_fast_2019} and \cite{ly_projective_2020} continue to use this strategy while verifying the Signorini-Coulomb conditions in each global step.
To improve computational efficiency, \cite{ly_projective_2020} uses a semi-implicit approach that approximates the system inverse.


\section{Background}

\subsection{Implicit Euler scheme in elastic dynamics}
\label{sec: implicit Euler scheme in elastic dynamics}

Following \cite{Martin2011}, the implicit Euler time integration solves the following optimization problem within the time interval [$t$, $t + \dt$]:
\begin{equation}
\label{eq: implicit euler}
    \pos_{t+\dt} = \min_{\pos} \bigg( \frac{1}{2 \dt^2} || \mass^{\frac{1}{2}} (\pos -  \sn)||_F^2 + \sum_i \elasticEnergy_i(\pos)\bigg)
\end{equation}
where $\mass$ represents the mass matrix; 
$\pos$ and $\vel$ denote positions and velocity;
$\sn = \pos_t + \dt \vel_t + \dt^2 \mass^\mathrm{-1} \vecf_\mathrm{ext}$ is the predicted state with external force $\vecf_\mathrm{ext}$ when implicit forces are not considered;
$\elasticEnergy$ signifies the elastic energy for the finite elements.

\subsection{Multi-body dynamics}
\label{sec: multi-body dynamics}

\begin{figure}[tb!]
    \centering
        \noindent%
	\begin{subfigure}[t]{0.5\linewidth}%
		\includegraphics[width=1.0\linewidth,clip,trim=0 300 0 0]{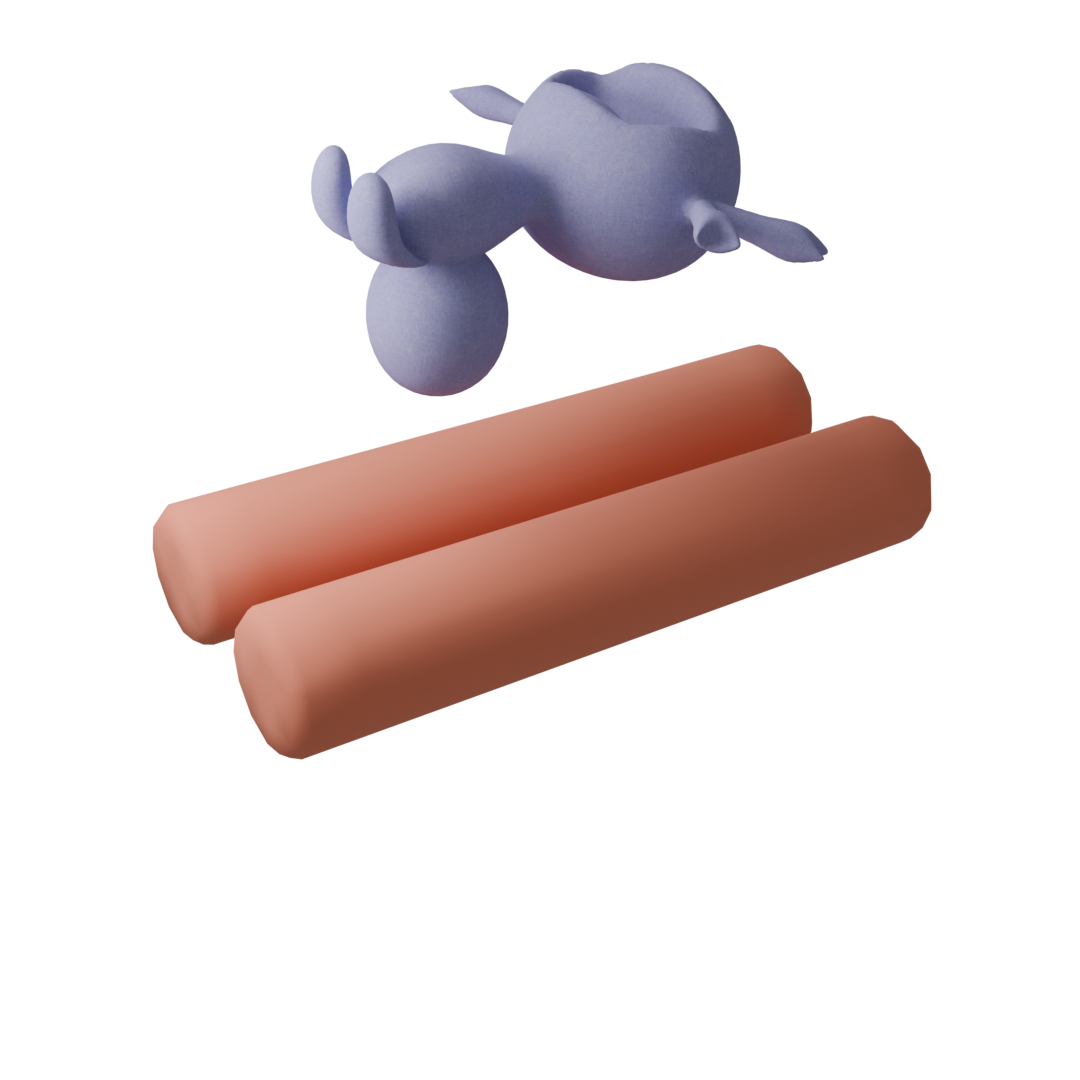}%
            \caption{Frame $0$}
	\end{subfigure}%
	\begin{subfigure}[t]{0.5\linewidth}%
		\includegraphics[width=1.0\linewidth,clip,trim=0 300 0 0]{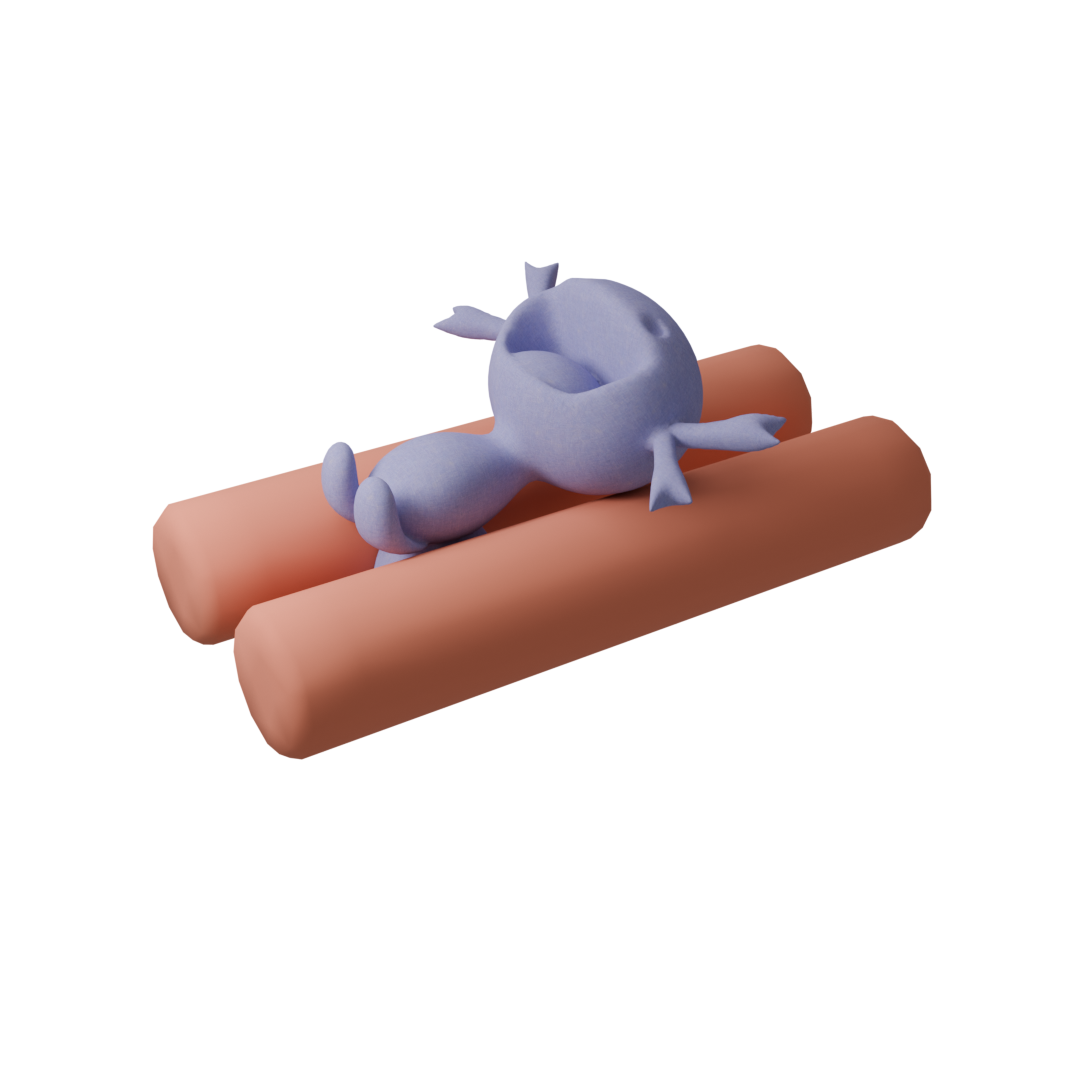}%
            \caption{Frame $240$}
	\end{subfigure}%
	\newline%
	\begin{subfigure}[t]{0.5\linewidth}%
		\includegraphics[width=1.0\linewidth,clip,trim=0 150 0 250]{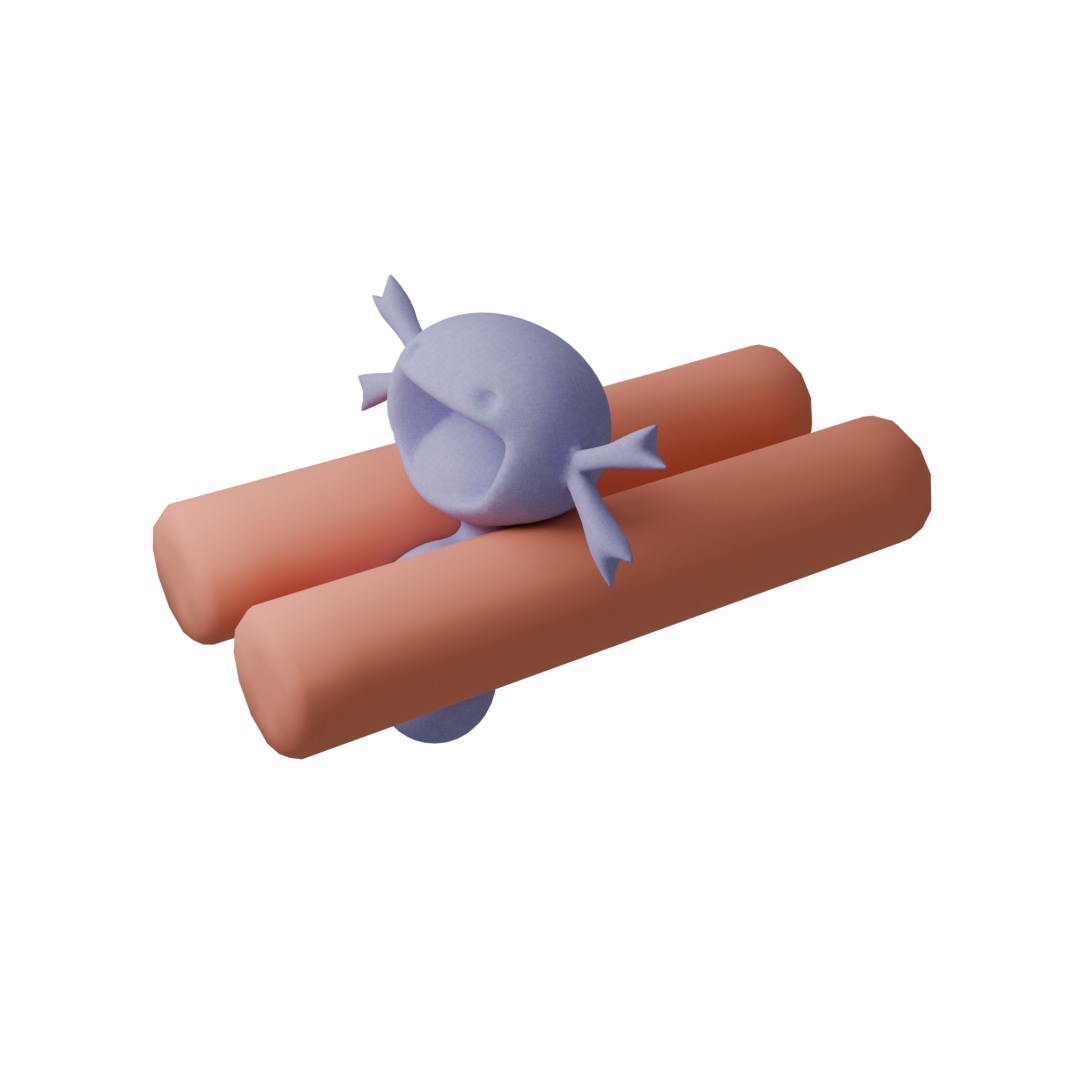}%
            \caption{Frame $340$}
	\end{subfigure}%
	\begin{subfigure}[t]{0.5\linewidth}%
		\includegraphics[width=1.0\linewidth,clip,trim=0 150 0 250]{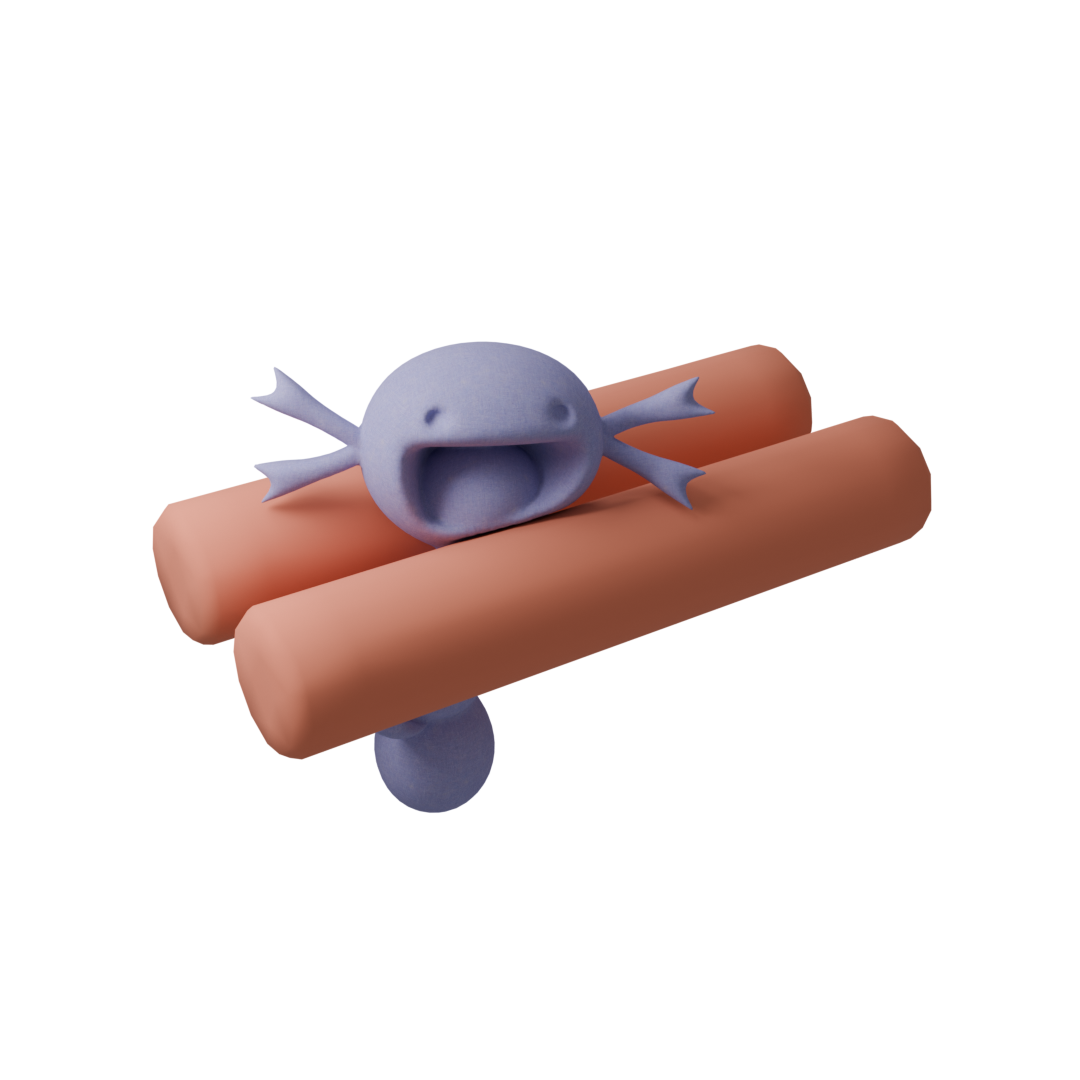}%
            \caption{Frame $420$}
	\end{subfigure}%
	\newline%
	\begin{subfigure}[t]{0.5\linewidth}%
		\includegraphics[width=1.0\linewidth,clip,trim=0 50 0 300]{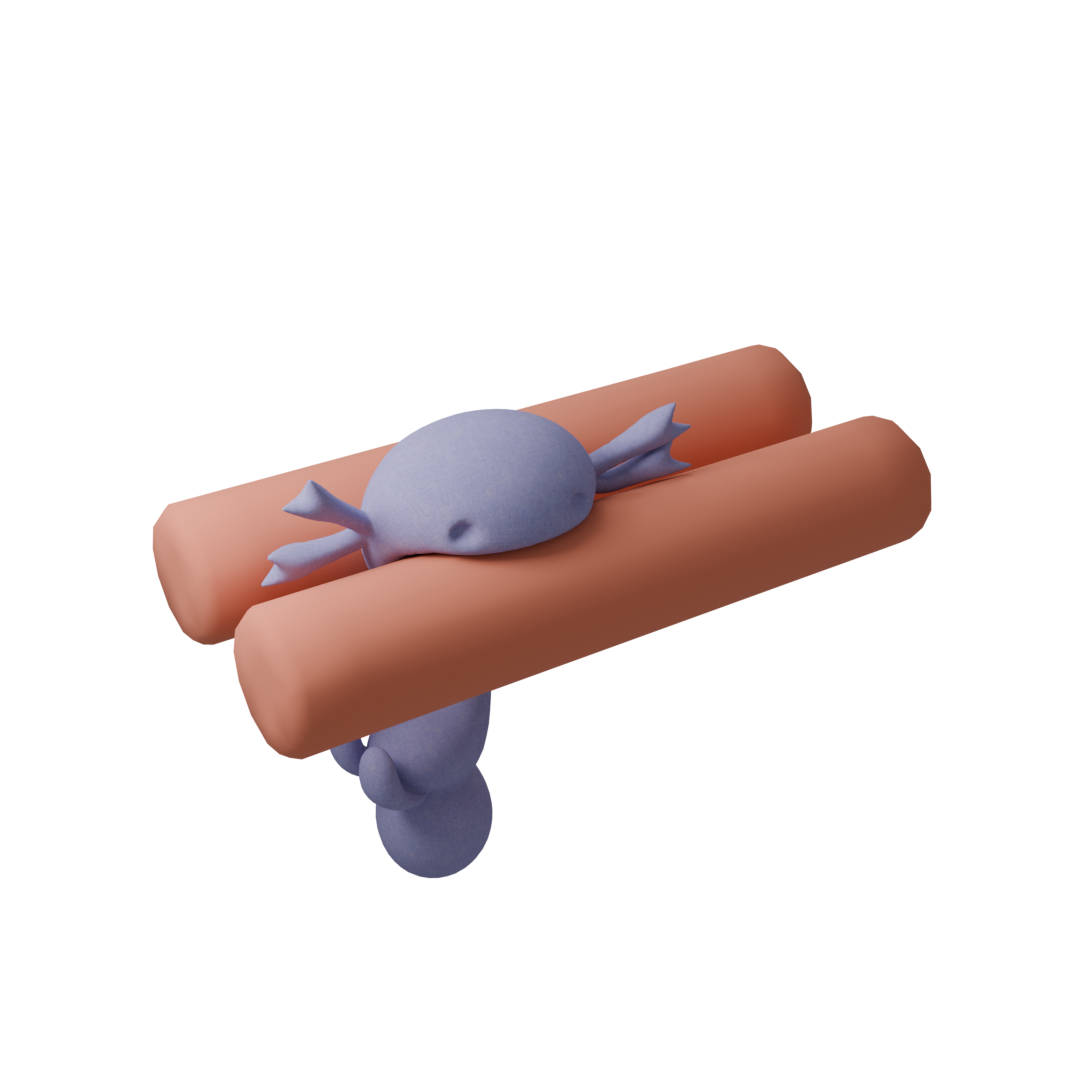}%
            \caption{Frame $500$}
	\end{subfigure}%
	\begin{subfigure}[t]{0.5\linewidth}%
		\includegraphics[width=1.0\linewidth,clip,trim=0 50 0 300]{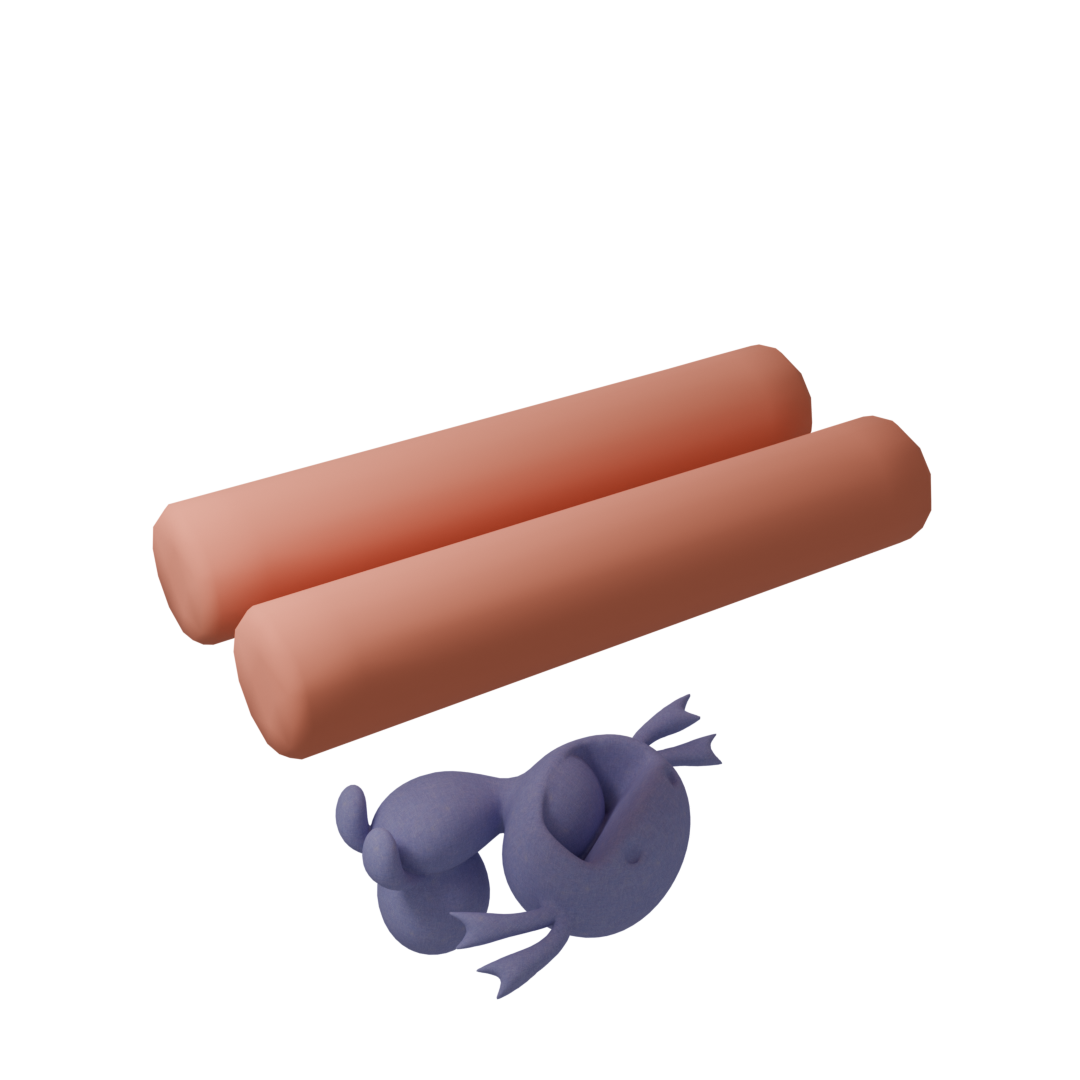}%
            \caption{Frame $600$}
	\end{subfigure}%
    \caption{
        Pulling Wooper: a moving positional constraint is applied on the tail of the wooper, pulling it through the thin gap between two cylinders.
    }
    \Description{}
    \label{fig: pulling wooper}
\end{figure}

In multi-body systems,  the typical Lagrange multiplier method incorporates contact forces $\Lam_{j}$ into the dynamic system.
Constraining the derivative of the total energy in Equation \eqref{eq: implicit euler} with contact conditions yields:
\begin{subequations}
\label{eq: constrained implicit euler}
\begin{align}
    \mass (\pos - \sn) - \dt^2 \vecf_\mathrm{int} (\pos) - \dt^2 \sum_{j \in \mathcal{L}} \contact_{j}^\mathrm{T} \Lam_{j} = \zeros \\
    \forall j \in \mathcal{L}, \quad \contact_{j}\pos - \hat{\proj}_{j}  = \Pene_{j} \\
    \forall j \in \mathcal{B}, \quad \Pene_{j} = \zeros \\
    \forall j \in \mathcal{C}, \quad \big( \Lam_{j}, \Pene_{j} \big) \in \scfunc_{\mu_j} 
\end{align}
\end{subequations}
where $\vecf_\mathrm{int}$ denotes implicit internal forces, and $\mathcal{L}$ represents indices for Lagrangian constraints which are categorized into binding pairs $\mathcal{B}$ and non-interpenetration pairs $\mathcal{C}$.
The linear mapping $\contact_{j} $ maps the mechanical state to the contact space.
The iteam $\hat{\proj}$ could be the projected state in binding constraints or a parameter of minimum separation in unilateral constraints (as in \cite{macklin_non-smooth_2019}).

For each contact pair $j \in \mathcal{L}$, interpenetration $\Pene_{j}$ and contact forces $\Lam_{j}$ must satisfy bilateral condition or Signorini-Coulomb condition $\scfunc_{\mu_j}$ \cite{brogliato_nonsmooth_2016} for accurate representation of bilateral and frictional contacts. 
Generally, to solve the problem with numerical methods, linearizing the constraints along specific directions is performed:

\paragraph*{\textbf{Binding contact}}

Bilateral constraints are prevalent in binding scenarios such as joint connections and special cases like needle constraints \cite{martin_high_2024}. 
The directions of bilateral constraints $\cbilateral \in \mathcal{R}^{1\times3}$ are defined for binding connections or special collision events.

\paragraph*{\textbf{Non-interpenetration contact}}
Non-interpenetration contacts are typically linearized along normal and tangent directions to prevent interpenetration and model friction. 
Generally, the contact normal should be defined by the collision detection.
We refer the readers to \cite{erleben_methodology_2018} for more details about the methodologies to generate the contact normal for stable numerical solution.
Given a contact normal $\cnormal \in \mathcal{R}^{1\times3}$, two tangent directions $\cfriction_1, \cfriction_2 \in \mathcal{R}^{1\times3}$ can be generated (e.g., via the Gram-Schmidt process) to represent the space perpendicular to the normal.
In the subsequent sections, friction directions are unified as $\cfriction$ for simplicity.\\

The formulation in Appendix \ref{ap: Constraint linearization} converts the governing equation \eqref{eq: constrained implicit euler} into a linearized form, as follows:
\begin{subequations}
\label{eq: constrained implicit euler linearized} 
\begin{align}
    \mass (\pos - \sn) - \dt^2 \vecf_\mathrm{int} (\pos) - \dt^2 \jacobian_b^\mathrm{T} \lam_b^{} - \dt^2 \jacobian_n^\mathrm{T} \lam_n^{} - \dt^2 \jacobian_f^\mathrm{T} \lam_f^{} &= \zeros
    \label{eq: constrained implicit euler linearized a} \\
    \jacobian_b \pos - \vecd_b  &= \pene_b 
    \label{eq: non-smooth conditions b} \\
    \forall j \in \mathcal{B}, \quad \peneval_{j} &= 0   
    \label{eq: non-smooth conditions c}\\
    \jacobian_n \pos - \vecd_n  &= \pene_n  
    \label{eq: non-smooth conditions d}\\
    \jacobian_f \vel - \vecd_f  &= \dot \pene_f   
    \label{eq: non-smooth conditions e}\\
    \forall j \in \mathcal{C}, \quad \big( \lamval_{j}, \peneval_{j}, \dot \peneval_{j} \big) &\in \scfunc_{\mu_j}   
    \label{eq: non-smooth conditions f}
\end{align}
\end{subequations}

In this formulation, $\mathcal{B} = \{b_1, ..., b_m\}$ represents bilateral constraints, and $\mathcal{C} = \{c_1, ..., c_n\}$ represents unilateral constraints along with frictional constraints. 
The Contact Jacobian matrix $\jacobian$ compiles the linearized contact mappings along specific constraint directions, exemplified by $\cbilateral$, $\cnormal$, or $\cfriction$.
The elements $\vecd$, $\pene$, and $\lam$ linearize the contact items $\hat{\proj}$ (or $\hat{\relativevel}$), $\Pene$, and $\Lam$, respectively, along these constraint directions.
To accurately capture the frictional behavior of contacts, the Signorini-Coulomb condition $\scfunc_{\mu_j}$ should be satisfied as follows: 
\begin{subequations}
\label{eq: signorini-coulomb conditions}
\begin{align}
    \forall j \in \mathcal{C}, \quad 0 \leq \peneval_{n,j} \perp \lamval_{n,j} \geq 0 \\
    \forall j \in \mathcal{A}, \quad \dot \peneval_{f,j} + \frac{|\dot \peneval_{f,j}|}{|\lamval_{f,j}|} \lamval_{f,j}= 0 \\
    \forall j \in \mathcal{A}, \quad 0 \leq |\dot \peneval_{f,j}| \perp \mu_{j} \lamval_{n,j} - |\lamval_{f,j}| \geq 0 \\
    \forall j \in \mathcal{I}, \quad \lamval_{f,j} = 0 
\end{align}
\end{subequations}
where $\mathcal{A} = \{j \in \mathcal{C} \, | \, \lamval_{n,j} > 0 \}$ is the set of all active contact indices, and $\mathcal{I} = \{j \in \mathcal{C} \, | \, \lamval_{n,j} \leq 0 \}$ is its complement in $\mathcal{C}$.
We refer readers to the Appendix \ref{ap: Frictional contact formulation} for more details on the frictional contact formulation.
The constrained system in Equation \eqref{eq: constrained implicit euler linearized} actually refers to a nonlinear complementarity problem.
Various numerical methods for solving such problems have been proposed in the literature \cite{erleben_numerical_2013, andrews_contact_2021}, such as direct methods, relaxation methods, and non-smooth Newton methods.

\subsection{Local-global iterative methods}
\label{sec: local-global iterative methods}

The PD-like local-global methods address the nonlinear optimization in Equation \eqref{eq: implicit euler} through recursive local-global iterations \cite{liu_fast_2013, bouaziz_projective_2014, overby_admm_2017}.
In the local step, the positions $\pos$ are fixed, and suitable projected local states $\proj_i$ are sought by solving local and independent sub-problems:
\begin{equation}
\label{eq: PD local}
    \proj_i^k = \min_{\proj_i} \frac{\weight_i}{2} ||\proj_i - \reduction_i \pos^k ||_F^2 
    + \localEnergy_{i}(\proj_i)
\end{equation}
where $k$ denotes the current local-global (L-G) iteration; $\weight$ is a nonnegative weight for each constraint; $\reduction$ is the linear mapping from the mechanical state $\pos$ to the projection state $\proj$. 
For elastic energy, the nonlinear energy function $\localEnergy$ could be generic hyperelastic energy $\elasticEnergy$ or an indicator function as in the Projective Dynamics \cite{bouaziz_projective_2014} ($\localEnergy = 0$ if $\proj \in SO(3)$ and $\localEnergy = +\infty$ otherwise).
Moreover, the local step formulation provides large flexibility to model other generic constraints, such as bending energy in cloth simulations and positional constraints.

After solving the independent local problems, the global step couples all the projected results in a global system:
\begin{equation}
\label{eq: PD global}
    \overbrace{ \bigg(\mass + \dt^2 \sum_{i} \weight_i \reduction_i^\mathrm{T} \reduction_i^{} \bigg)}^{\system} \pos^{k+1} =
    \overbrace{\bigg(\mass \sn + \dt^2 \sum_{i} \weight_i \reduction_i^\mathrm{T} \proj_i^k \bigg)}^{\rhs^k}
\end{equation}

In the global step, the right-hand-side (RHS) $\rhs^k$ assembles the projected state $\proj^k$ in the current L-G iteration. 
Since $\reduction$ depends only on predefined projection constraints, the global system matrix $\system$ remains invariant throughout the simulation.
Repeating the local and the global step recursively (Algorithm \ref{algo: Local-global iterations}), the integrator efficiently converges to the solution of Newton's method.

\begin{algorithm}
\caption{Local-global integrators}
\label{algo: Local-global iterations}
\SetAlgoLined
\While{\textit{simulation}}{
$\sn = \pos_t + \dt \vel_t + \dt^2 \mass^\mathrm{-1} \vecf_\mathrm{ext}$\;
\For{$k \in \{0, ..., n\}$}{
$\proj_i^k = project(\reduction_i \pos^k)$\Comment*[r]{local step} 
$\rhs^k = \mass \sn + \dt^2 \sum_i \weight_i \reduction_i^\mathrm{T} \proj_i^k$\Comment*[r]{assemble the RHS} 
$\pos^{k+1} = \system^\mathrm{-1} \rhs^k$\Comment*[r]{global step} 
}
$\pos_{t+h} = \pos^k$\Comment*[r]{integration} 
$\vel_{t+h} = \frac{1}{\dt} (\pos^k - \pos_t )$\;
}
\end{algorithm}

The success of local-global methods can be attributed to several factors:
First, the flexibility in the local step allows for simulating different constraints in a unified framework.
The local steps are parallelizable due to their independence.
Second, the global step efficiently couples the constraints through the system matrix $\system$.
Since local solutions are propagated immediately through global solving, the efficient constraint coupling leads to rapid convergence, especially in the initial iterations.
In general, local-global methods converge in significantly fewer iterations than PBD-like methods, although not as few as Newton's method. 
Most importantly, the system matrix $\system$ remains constant throughout the simulation.
Exploiting this property could significantly reduce the computational cost of the global step.
A common approach is to pre-factorize the system $\system$ once at initialization, which reduces the global steps to solving only sparse triangular systems (STS).

Despite its efficiency, the forward and backward substitutions in STS are difficult to be computed in parallel due to strong data dependencies, limiting its applicability to large-scale problems.
Moreover, it is also challenging to model accurate frictional contacts in local-global methods due to the requirement of satisfying complex non-smooth conditions described in Section \ref{sec: multi-body dynamics}.
In this paper, we seek for a method that addresses all these challenges.
For this purpose, we propose a simple yet highly efficient method for parallelizing the global step (Section \ref{sec: Sparse Inverse Solution}), and develop a unified framework where the local-global iterations are constrained by nonlinear complementarity conditions to handle frictional contacts (Section \ref{sec: multi-body dynamics with accurate frictional contact}).

\section{Sparse Inverse Solution}
\label{sec: Sparse Inverse Solution}

\subsection{Global step solution strategy}
\label{sec: global step solution strategy}

In Section \ref{sec: implicit simulation for elastic dynamics}, we review different numerical solvers used for solving the global step in the literature.
These solvers can be generally classified into complete and incomplete solutions. 

 \paragraph*{\textbf{Complete Solution}}
The traditional strategy \cite{liu_fast_2013, bouaziz_projective_2014, overby_admm_2017, brown_wrapd_2021} involves fully resolving the linear system $\pos^{k+1} = \system^{\mathrm{-1}} \rhs^k$ during each L-G iteration with pre-factorized system. 
As discussed in Section \ref{sec: local-global iterative methods}, although immediate propagation via complete global solving achieves fast convergence, data dependencies in the STS make it difficult to implement efficient parallel computation.

\paragraph*{\textbf{Incomplete Solution}}
An alternative perspective \cite{wang_chebyshev_2015} suggests that computing an exact solution is both redundant and computationally inefficient, especially considering that the local step requires modifications to the linear system in subsequent local-global iterations.
Therefore, incomplete methods \cite{wang_chebyshev_2015, fratarcangeli_vivace_2016, lan_penetration-free_2022} approximate the exact solution $\pos^{k+1} = \system^{\mathrm{-1}} \rhs^k$ with $\pos^{k+1} = \Tilde{\matP} \rhs^k + \Tilde{\matQ} \pos^k$, where $\Tilde{\matP}$ and $\Tilde{\matQ}$ depend on specific iterative algorithms.
These methods offer substantial potential for parallelization since the iterative solvers typically involve sparse matrix-vector multiplications (SpMV) that are easily parallelized on both CPU and GPU architectures.
However, approximating the global solve significantly decreases the efficiency of propagating local solutions, leading to slower convergence.
Consequently, a larger number of L-G iterations (usually one or more orders of magnitude higher) are required to reach the same level of accuracy as the complete solution.
Moreover, each additional iteration leads to an extra local step, which possibly becomes the bottleneck for real-time performance, especially in solving hyperelastic local problems.\\

Our goal is to develop a method that both efficiently couples local results with the complete solution for fast convergence and offers extensive parallelization capabilities.
A naive propose is to explicitly pre-compute the inverse of $\system$, such that the global steps simply turn to a matrix-vector product, which is straightforward to be parallelized.
However, inverting the matrix requires large precomputation cost and, more importantly, massive memory usage for large-scale simulations.

In practice, although the system matrix $\system$ is sparse, its inverse $\system^\mathrm{-1}$ is typically dense. 
For instance, a deformable object with $10k$ vertices requires more than $3GB$ of memory for storing dense matrices (as evidenced by experimental data in Tables \ref{tab: memory cost surface} and \ref{tab: memory cost volume}). 
Due to memory constraints and the overhead of computing and transferring large matrices to GPUs, this method becomes impractical for large-scale simulations.

An immediate question is: does there exist a sparse system that can accurately represent the inverse $\system^\mathrm{-1}$?
The answer is affirmative, and we give the details in the following subsection.

\subsection{Sparse inverted local-global method}
\label{sec: sparse inverted local-global method}

\begin{figure}[htb!]
    \centering 
    \includegraphics[width=1.0\linewidth,clip,trim=20 50 20 80]{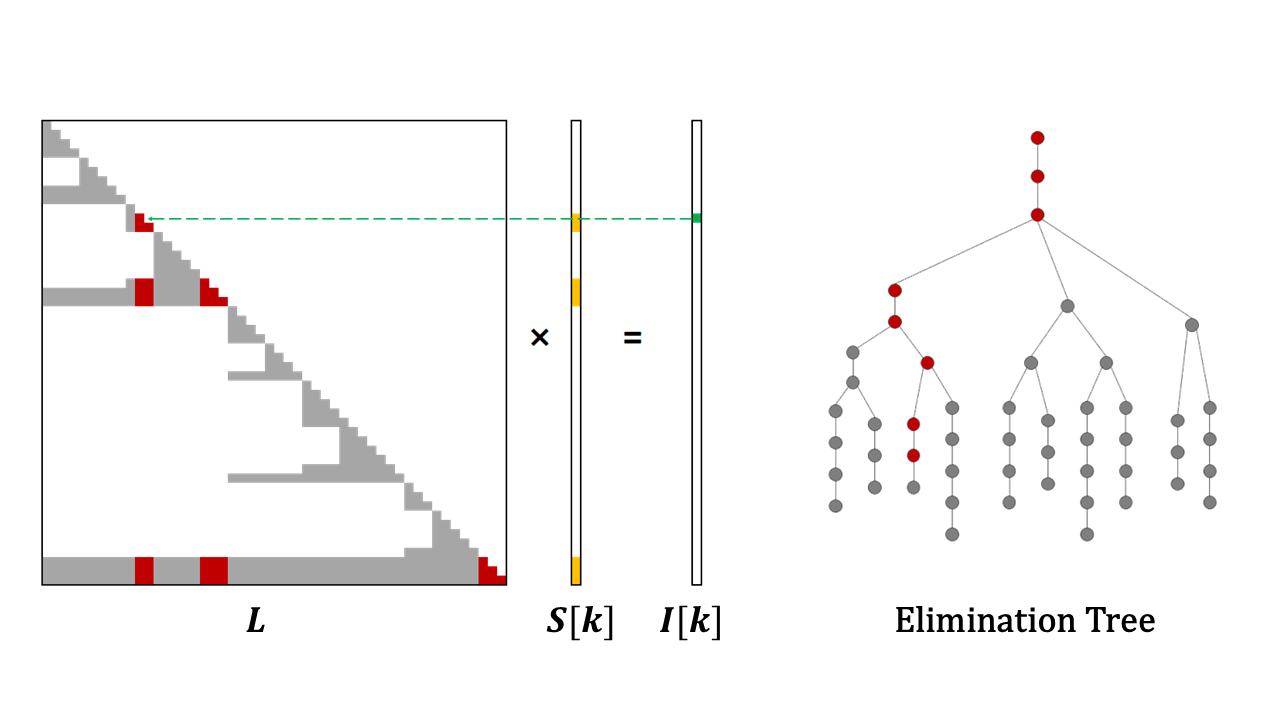}
    \caption{
        Sparse solution: after reducing the matrix pattern fill-in through nested dissection, the Cholesky factor $\choleskyfactor$ is reordered and partitioned into sub-blocks.
        For a given column $k$ in the identity matrix $\identity$, the requisite structure to be processed in $\choleskyfactor$ consists of $k$ and its ancestors (red nodes) in the elimination tree.
    }
    \Description{}
    \label{fig: sparse solution}
\end{figure}

Given the Cholesky factor $\choleskyfactor$ of the system, its inverse $\choleskyfactor^\mathrm{-1}$ is exactly the sparse system we are seeking.
Following \cite{bridson_ordering_1999, benzi_orderings_2000, scott_algorithms_2023}, we present the following theorem:
\begin{theorem}
\label{theorem: sparse cholesky inverse}
Let $\system$ be a symmetric positive definite (SPD) matrix with its Cholesky factor $\choleskyfactor$. 
The sparsity structure $\mathcal{S}\{\choleskyfactor^\mathrm{-1}\}$ is the union of all entries $(i,j)$ where $i$ is an ancestor of $j$ in the elimination tree $\mathcal{T}(\system)$.
\end{theorem}
The theorem implies that $\choleskyfactor^\mathrm{-1}$ does not need to be fully dense.
In practice, for SPD systems in local-global methods, techniques like fill-in reduction ordering (e.g., nested dissection \cite{George1973}) can efficiently reduce the number of ancestors of vertices in $\mathcal{T}(\system)$, resulting in a highly sparse $\choleskyfactor^\mathrm{-1}$.
A simple example in Figure \ref{fig: sparse solution} illustrates this concept.
The sparse scheme outperforms a full traversal of $\choleskyfactor$, yielding both efficient computation of $\sparseinverse = \choleskyfactor^\mathrm{-1}$ and high sparsity in $\sparseinverse$.
The same conclusion can be drawn from \cite{zeng_realtime_2022} concerning the STS solution in their method.
By explicitly computing $\sparseinverse$ and storing it in sparse format, we achieve the desired sparse system which is capable of representing the system inverse.
Specifically, the system inverse $\system^\mathrm{-1}$ is given by the following product:
\begin{equation}
    \system^\mathrm{-1} = \sparseinverse^\mathrm{T} \sparseinverse
\end{equation}

\begin{algorithm}
\caption{Sparse inverted local-global integrators}
\label{algo: Sparse inverted local-global iterations}
\SetAlgoLined
$\choleskyfactor = \textit{Cholesky}(\system)$\;
\textcolor{orange}{$\sparseinverse = \choleskyfactor^\mathrm{-1} \identity$}\;
\While{\textit{simulation}}{
$\sn = \pos_t + \dt \vel_t + \dt^2 \mass^\mathrm{-1} \vecf_\mathrm{ext}$\;
\For{$k \in \{0, ..., n\}$}{
$\proj_i^k = project(\reduction_i \pos^k)$\Comment*[r]{local step} 
$\rhs^k = \mass \sn + \dt^2 \sum_i \weight_i \reduction_i^\mathrm{T} \proj_i^k$\Comment*[r]{assemble the RHS} 
\textcolor{orange}{$\pos^{k+1} = \sparseinverse^\mathrm{T} \sparseinverse \rhs^k$}\Comment*[r]{global step} 
}
$\pos_{t+h} = \pos^k$\Comment*[r]{integration} 
$\vel_{t+h} = \frac{1}{\dt} (\pos^k - \pos_t )$\;
}
\end{algorithm}

\subsection{Optimized time and space efficiency}
\label{sec: optimized time and space efficiency}

Our approach, as shown in Algorithm \ref{algo: Sparse inverted local-global iterations}, includes an additional precomputation step to invert the Cholesky factor.
This step transforms the overall solution into two SpMV operations in global steps:
\begin{equation}
\pos^{k+1} = \sparseinverse^\mathrm{T} \sparseinverse \rhs^k
\end{equation}
Compared to the typical complete solutions in PD, the use of sparse inverse $\sparseinverse$ enables highly parallel computation while maintaining reasonable memory consumption.

\paragraph*{\textbf{Global step}}

Our algorithm processes complete solution, thereby maintaining the same strong convergence.
By replacing the STS with SpMV operations, we significantly enhance the potential for parallelization. 
While parallelizing STS operations on GPUs poses challenges due to data dependencies in triangular systems, SpMV is well-suited for extensive parallelization on GPU architectures. 
Optimized implementations of these operations are available in libraries such as NVIDIA's cuSPARSE.

\paragraph*{\textbf{Memory usage}}

As a trade-off, our method requires additional memory usage for $\sparseinverse$ and extra time for its computation. 
However, as detailed in Section \ref{sec: sparse inverted local-global method}, these computational and memory costs are justifiable owning to sparse solutions.
Our experiments across various meshes and element types (Section \ref{sec: results}) prove that the memory requirement for $\sparseinverse$ remains consistently reasonable.\\

Regarding the cost per global step, existing incomplete methods require two SpMV operation, $\pos^{k+1} = \Tilde{\matP} \rhs^k + \Tilde{\matQ} \pos^k$.
Our approach requires two SpMV operations as well: $\pos^{k+1} = \sparseinverse^\mathrm{T} \sparseinverse \rhs^k$, with $\sparseinverse$ typically having higher density than $\Tilde{\matP}$ and $\Tilde{\matQ}$.
Nevertheless, our sparse inverse method preserves the rapid convergence characteristics of the complete solution, requiring substantially fewer L-G iterations to reach comparable accuracy levels.
Moreover, our method avoids running too many local steps, which can be a potential bottleneck when solving nonlinear local steps. 

To summarize, our sparse inverse method combines both fast convergence and high potential for parallelization, while maintaining simplicity through its use of standard matrix operations.
However, our method heavily depends on the system invariability, which makes it challenging to incorporate penalty-based methods that dynamically modify the system during contact handling.
To address this, in the next section, we propose using the Lagrange multiplier methods to constrain the global steps where the complementarity conditions are effectively verified in each L-G iteration.

\section{Multi-body dynamics with accurate frictional contact}
\label{sec: multi-body dynamics with accurate frictional contact}

\begin{figure}[tb!]
    \centering
        \noindent%
	\begin{subfigure}[t]{0.5\linewidth}%
		\includegraphics[width=0.95\linewidth,clip,trim=500 250 500 50]{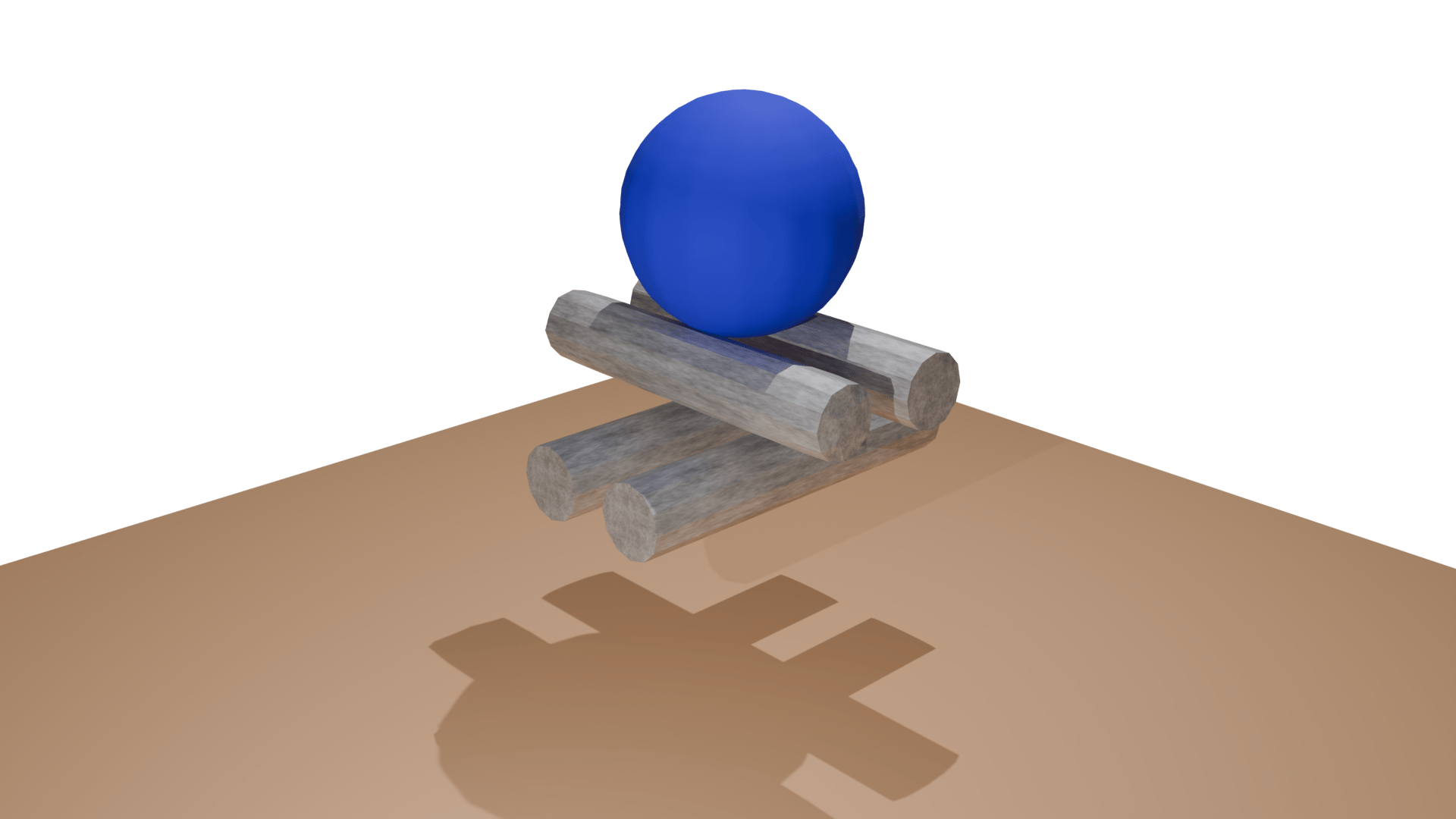}%
            \caption{Frame $0$}
	\end{subfigure}%
	\begin{subfigure}[t]{0.5\linewidth}%
		\includegraphics[width=0.95\linewidth,clip,trim=500 250 500 50]{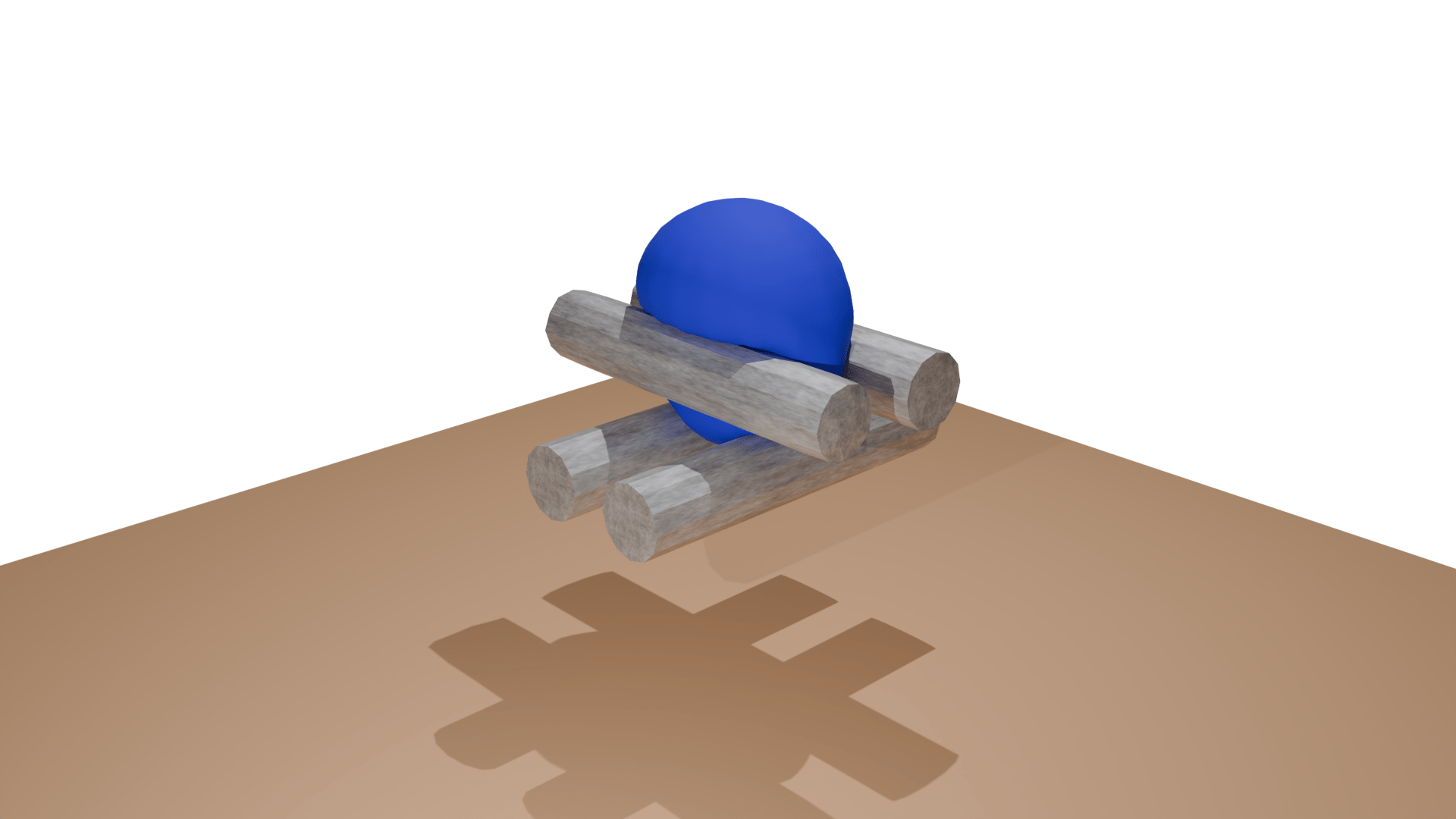}%
            \caption{Frame $200$}
	\end{subfigure}%
	\newline%
	\begin{subfigure}[t]{0.5\linewidth}%
		\includegraphics[width=0.95\linewidth,clip,trim=500 150 500 200]{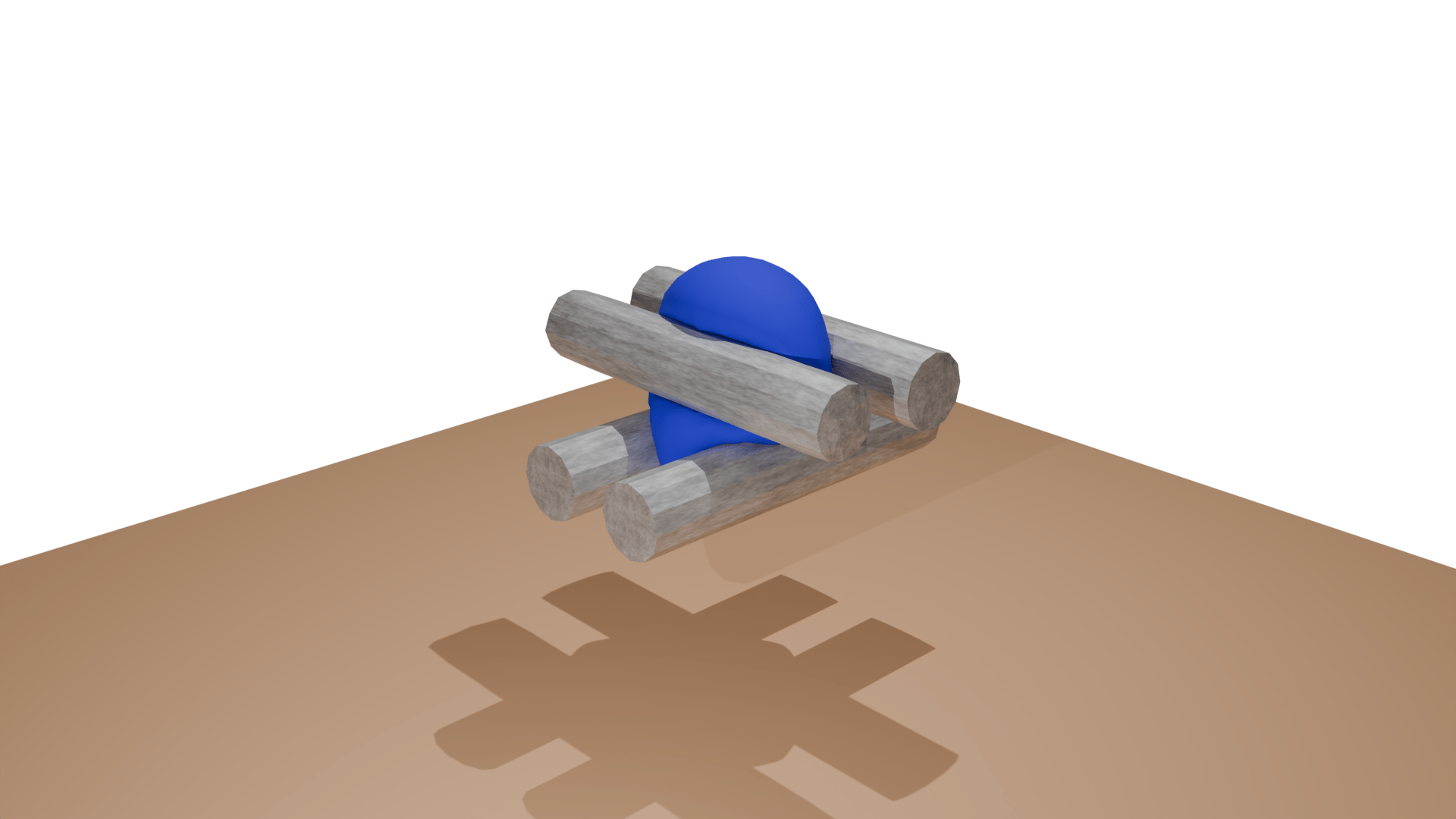}%
            \caption{Frame $300$}
	\end{subfigure}%
	\begin{subfigure}[t]{0.5\linewidth}%
		\includegraphics[width=0.95\linewidth,clip,trim=500 150 500 200]{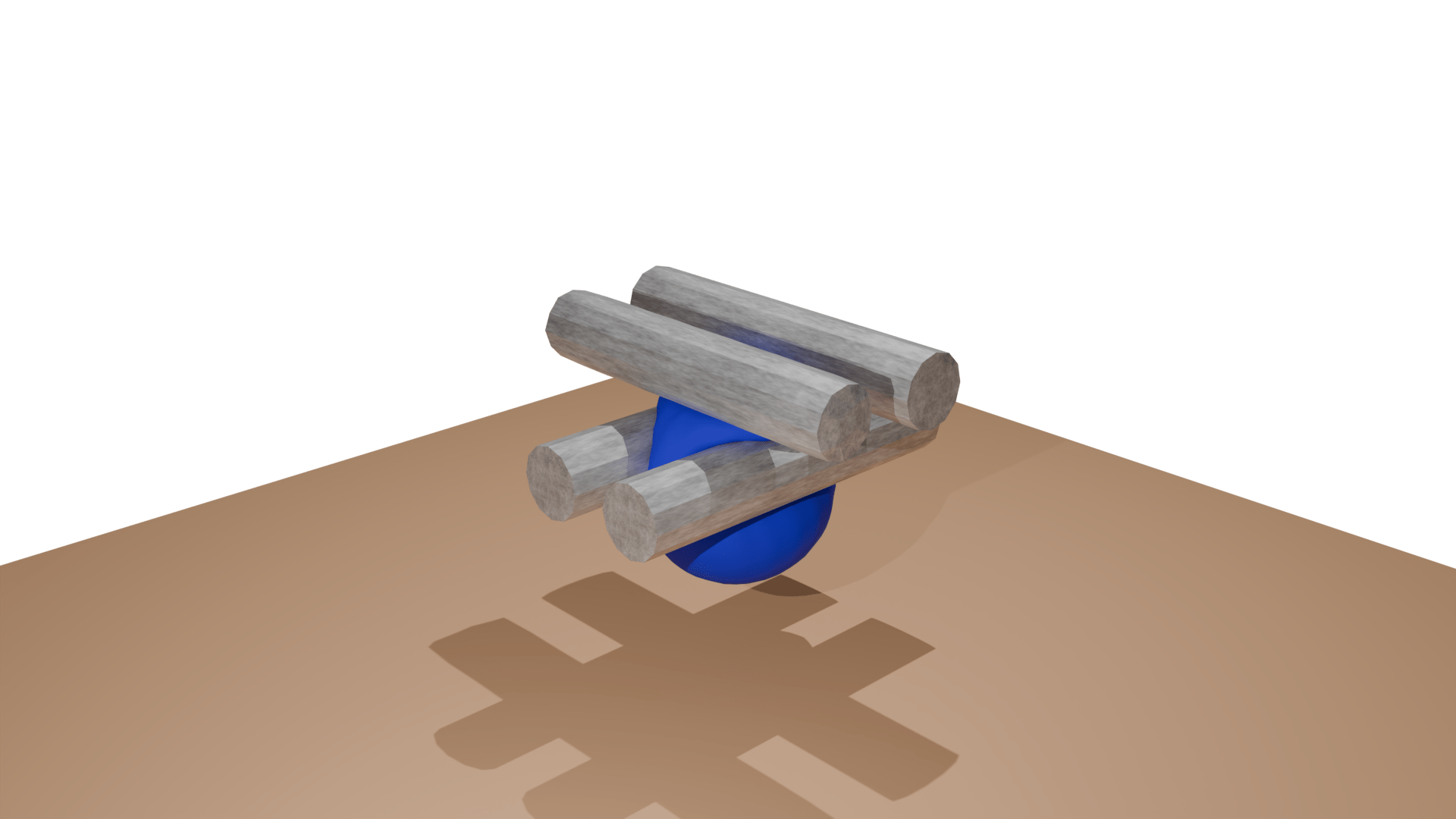}%
            \caption{Frame $400$}
	\end{subfigure}%
	\newline%
	\begin{subfigure}[t]{0.5\linewidth}%
		\includegraphics[width=0.95\linewidth,clip,trim=500 50 500 300]{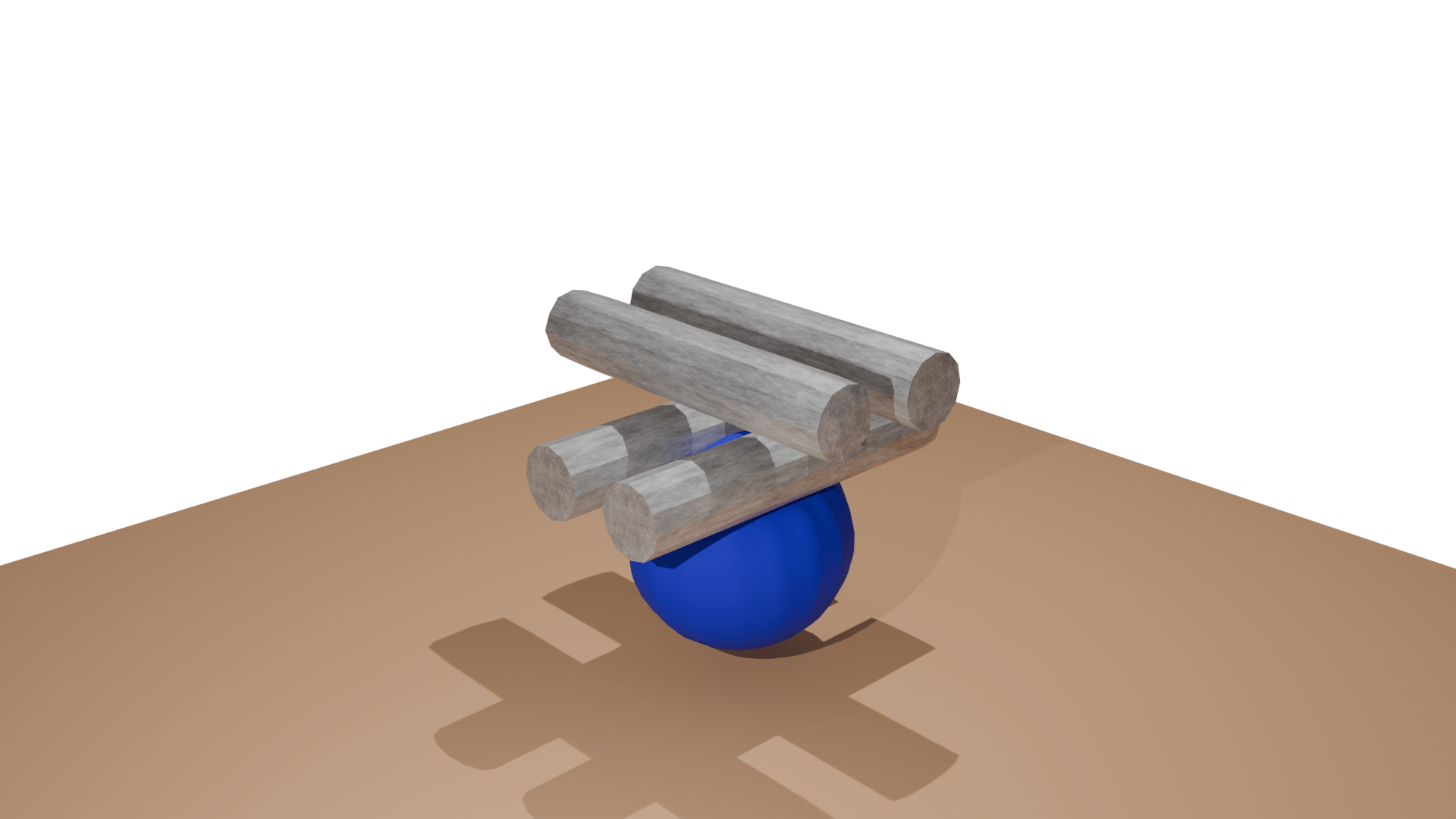}%
            \caption{Frame $500$}
	\end{subfigure}%
	\begin{subfigure}[t]{0.5\linewidth}%
		\includegraphics[width=0.95\linewidth,clip,trim=500 50 500 300]{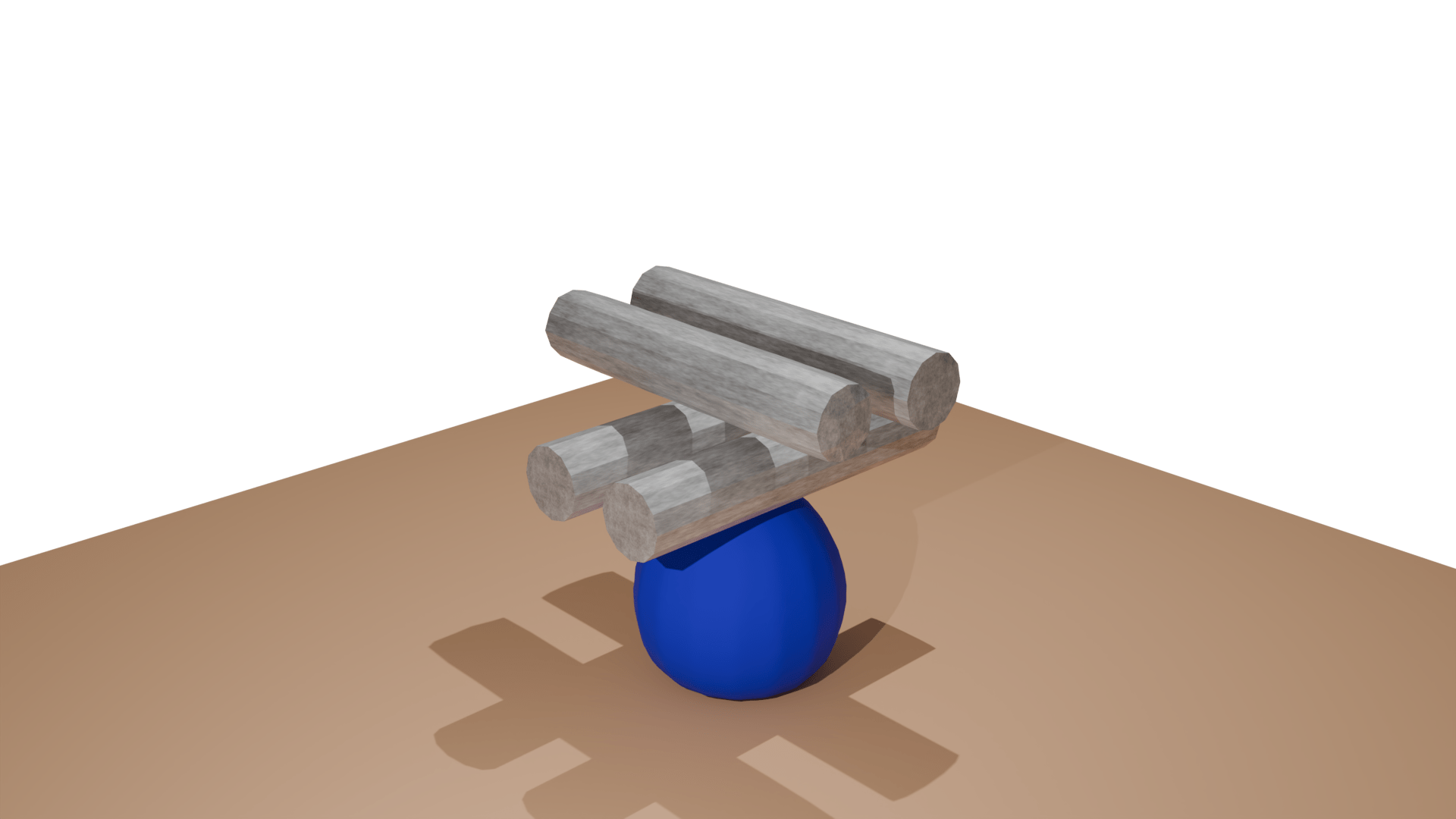}%
            \caption{Frame $600$}
	\end{subfigure}%
    \caption{
        Squeezing Ball: the rolling cylinders strongly squeeze the soft ball, thereby generating large frictional forces (with $\mu = 0.5$) that drive the ball through the gaps.
    }
    \Description{}
    \label{fig: squeezing ball}
\end{figure}

In Section \ref{sec: Sparse Inverse Solution}, we introduce the sparse inverse approach for local-global methods. 
This section now shifts its focus to addressing contact-related issues within the local-global framework. 
Although several efficient techniques have been developed to handle non-penetration contacts within local-global frameworks, none fully captures  the accurate contact and friction behaviors.
The non-smooth Newton method has proven effective for its accuracy and stability in multi-body dynamics. 
However, it is not inherently compatible with local-global methods, as existing approaches mainly combine Newton's method with impulse-based integration.

In this section, we derive a reformulation to integrate the non-smooth Newton method into the local-global framework. 
In Section \ref{sec: constrained global step}, we establish the compatibility of the Lagrange multiplier methods within the local-global framework. 
By considering the local-global iterations as quasi-Newton iterations (referencing \cite{liu_quasi-newton_2017}), we demonstrate the equivalence between position-based local-global integration and impulse-based integration, effectively merging these two methodologies within a unified framework (Section \ref{sec: position-based non-smooth Newton method}).
Subsequently, in Section \ref{sec: splitting out non-smooth indicators}, we propose a strategy that splits out the non-smooth indicators, allowing us to significantly reduce computational costs in the Schur complement calculation.
This approach also enables us to develop an effective complementarity preconditioner (Section \ref{sec: complementarity precondition}).
Finally, in Section \ref{sec: a GPU-friendly framework}, we conclude our approach as a GPU-friendly framework.

\subsection{Constrained global step}
\label{sec: constrained global step}

Local-global methods, being position-based, directly update positions $\pos$ without explicitly computing internal forces.
Therefore, it is not straightforward to integrate the Lagrange multiplier method shown in Equation \eqref{eq: constrained implicit euler} into the local-global algorithm.
To bridge this gap, we first evaluate the impulse $\Delta \vel$ in the global steps in Equation \eqref{eq: PD global} within each L-G iteration:

\begin{equation}
\begin{aligned}
     \mass \Delta \vel 
     &= \frac{1}{\dt} \mass \bigg(\pos^{k+1} - (\pos_t + \dt \vel_t) \bigg)\\
     &= \dt \vecf_\mathrm{ext} + \dt^2 \bigg( \sum_{i} \weight_i \reduction_i^\mathrm{T} \proj_i^{k} -  \sum_{i} \weight_i \reduction_i^\mathrm{T} \reduction_i^{} \pos^{k+1}\bigg)
\end{aligned}
\end{equation}

In an isolated system (excluding contact forces), the impulse $\Delta \vel$ is generated by both external and internal forces, allowing us to compute the internal forces for the current L-G iteration $k$:

\begin{equation}
\label{eq: internal forces in local-global}
    \vecf_\mathrm{int} (\pos^{k+1}) = \dt \bigg( \sum_{i} \weight_i \reduction_i^\mathrm{T} \proj_i^{k} - \sum_{i} \weight_i \reduction_i^\mathrm{T} \reduction_i^{} \pos^{k+1} \bigg)
\end{equation}
By incorporating Equation \eqref{eq: internal forces in local-global} into the constrained implicit Euler solver in Equation \eqref{eq: constrained implicit euler linearized a}, we derive the constrained global step:
\begin{equation}
\label{eq: constrained global}
\begin{aligned}
    \overbrace{ \bigg(\mass + \dt^2 \sum_{i} \weight_i \reduction_i^\mathrm{T} \reduction_i^{} \bigg)}^{\system} \pos^{k+1} -
    \overbrace{\bigg(\mass \sn + \dt^2 \sum_{i} \weight_i \reduction_i^\mathrm{T} \proj_i^k \bigg)}^{\rhs^k} \\
    - \dt^2 \jacobian^\mathrm{T}_b \lam_b^{k+1} - \dt^2 \jacobian^\mathrm{T}_n \lam_n^{k+1} - \dt^2 \jacobian^\mathrm{T}_f \lam_f^{k+1} = \zeros \\
    \text{subject to Equations \eqref{eq: non-smooth conditions b} - \eqref{eq: non-smooth conditions f}}
\end{aligned}
\end{equation}
A similar formulation can be found in \cite{overby_admm_2017}, which simplifies the NCP by disregarding the Signorini-Coulomb conditions and converting it into a linear problem.
For a thorough solution to the contact problem, we propose using the non-smooth Newton method, which will be explained in detail in the following section.

\subsection{Non-smooth constrained local-global integration}
\label{sec: position-based non-smooth Newton method}

\begin{figure}[tb!]
    \centering\noindent%
	\begin{subfigure}[t]{0.5\linewidth}%
		\includegraphics[width=0.95\linewidth,clip,trim=0 0 0 10]{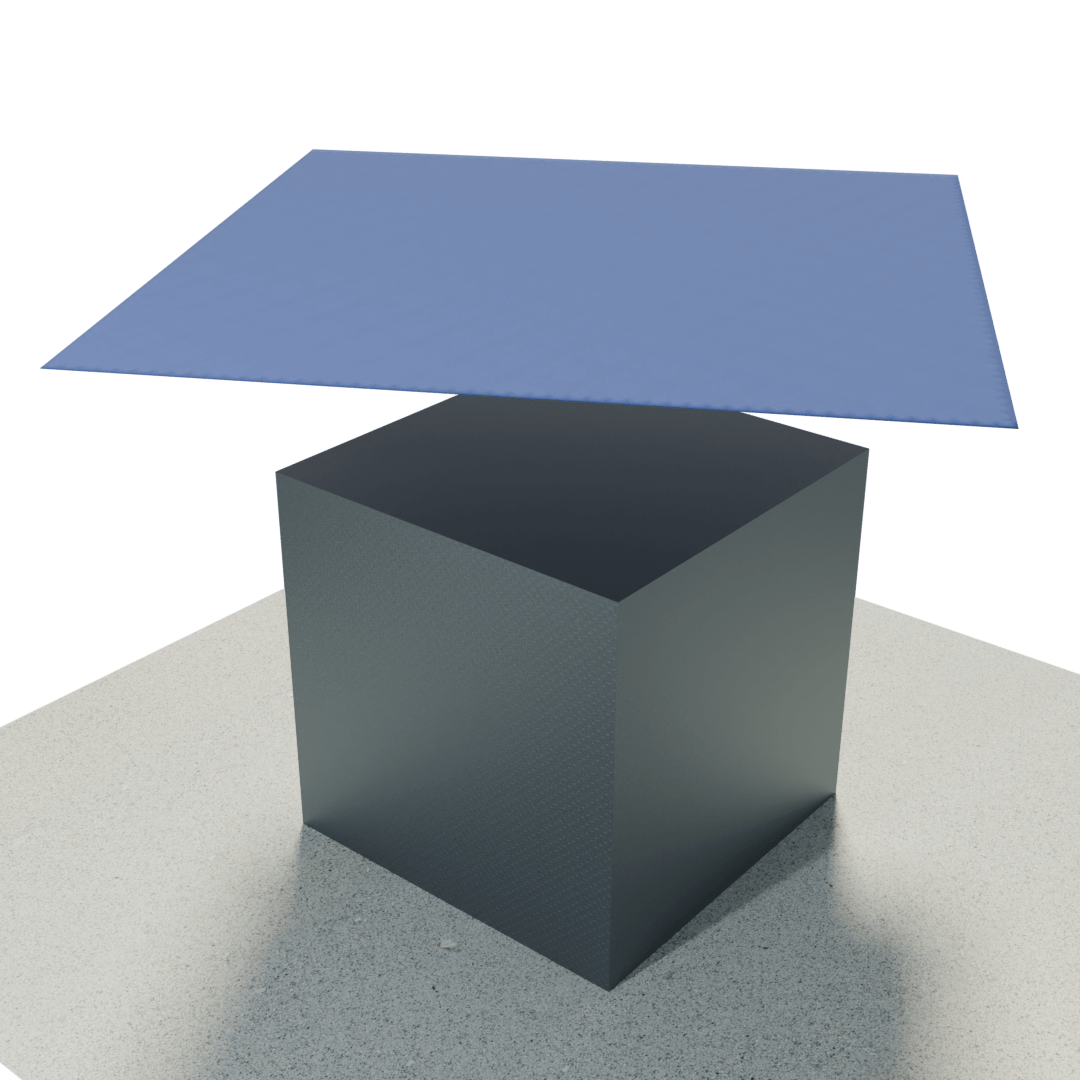}%
            \caption{Frame $0$}
	\end{subfigure}%
	\begin{subfigure}[t]{0.5\linewidth}%
		\includegraphics[width=0.95\linewidth,clip,trim=0 0 0 10]{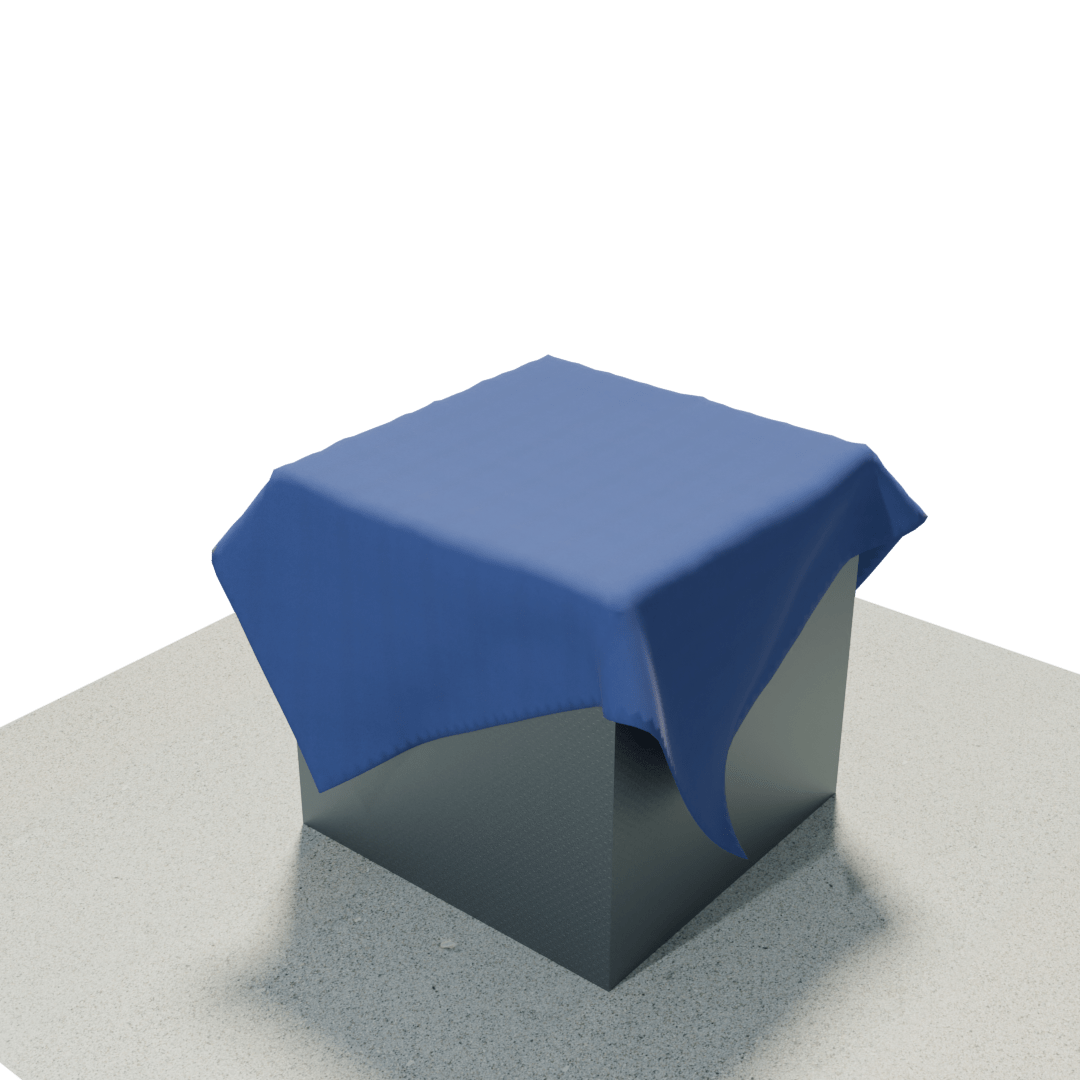}%
            \caption{Frame $120$}
	\end{subfigure}%
    \caption{
        Sharp Corner: stable simulation of cloth mesh on sharp corner obstacle, with rich and non-smooth contacts.
    }
    \Description{}
    \label{fig: sharp corner}
\end{figure}

The non-smooth Newton method reformulates the complementarity conditions into equivalent NCP functions:
\begin{equation}
    0 \leq a \perp b \geq 0 \quad \leftrightarrows \quad \ncpfunc (a, b) = 0
\end{equation}
where the NCP function $\ncpfunc$ transforms the original problem into a root-finding one.
In practice, $\ncpfunc$ could be formulated as either the \textit{minimum-map} formulation or the \textit{Fischer-Burmeister} formulation:
\begin{equation}
    \ncpfunc_{\mathrm{min}} (a, b) = \mathrm{min}(a, b)
\end{equation}
\begin{equation}
    \ncpfunc_{\mathrm{FB}} (a, b) = a + b - \sqrt{a^2 + b^2}
\end{equation}

As in \cite{macklin_non-smooth_2019}, we use complementarity preconditioners $\nsprecond$ to improve the convergence in nonlinear integration.
We refer the readers to Appendix \ref{ap: Non-smooth functions} for the detail formulation of different NCP functions and their derivatives.
By considering interpenetration $\pene$ and contact forces $\lam$ as arguments of NCP functions, the non-smooth Newton method reformulates the constrained global step (Equation \eqref{eq: constrained global}) as follows:
\begin{subequations}
\label{eq: constrained global non-smooth}
\begin{align}
    \system \pos^{k+1} - \rhs^{k} - \dt^2 \nsjacobian_b^\mathrm{T} \lam_b^{k+1} - \dt^2 \nsjacobian_n^\mathrm{T} \lam_n^{k+1}- \dt^2 \nsjacobian_f^\mathrm{T} \lam_f^{k+1} = \zeros 
    \label{eq: constrained global non-smooth a}\\
    \nsjacobian_b^{} \pos^{k+1} - \vecd_b^{} + \bcompliance \lam_b^{k+1} = \zeros  
    \label{eq: constrained global non-smooth b}\\
    \ncpfunc_n^{} (\pene_n^{k+1}, \lam_n^{k+1}) = \zeros  
    \label{eq: constrained global non-smooth c}\\
    \ncpfunc_f^{} (\dot \pene_f^{k+1}, \lam_f^{k+1}) = \zeros  
    \label{eq: constrained global non-smooth d}
\end{align}
\end{subequations}
where solving the non-smooth conditions in Equations \eqref{eq: non-smooth conditions d} - \eqref{eq: non-smooth conditions f} involves finding roots for NCP functions.

For bilateral constraints, $\nsjacobian_b^{} = \jacobian_b$ represents the non-smooth Jacobian matrix, while $\bcompliance = diag(e_{b_0}, ..., e_{b_m})$ is a diagonal compliance matrix consisting of inverse stiffness coefficients $e$ (following \cite{macklin_non-smooth_2019}).
By setting $e > 0$, we establish quadratic energy potentials with a stiffness weight of $e^{-1}$.
Besides, setting $e = 0$ is equivalent to the bilateral condition in Equations \eqref{eq: non-smooth conditions b} - \eqref{eq: non-smooth conditions c}, resulting in a hard constraint with infinite stiffness.

For unilateral and frictional constraints, the non-smooth Jacobian matrix $\nsjacobian$ is derived from the partial derivatives of $\ncpfunc$ with respect to position $\pos$ or velocity $\vel$.
The non-smooth compliance matrix $\compliance$ is derived from the partial derivatives of $\ncpfunc$ with respect to $\lam_n$ and $\lam_f$:
\begin{equation}
\begin{aligned}
\label{eq: non-smooth contact jacobian}
    &\nsjacobian_n^{} = \frac{\partial \ncpfunc_n}{\partial \pos}, \qquad 
    &\ncompliance = \frac{\partial \ncpfunc_n}{\partial \lam_n} \\
    &\nsjacobian_f^{} = \frac{\partial \ncpfunc_f}{\partial \vel}, \qquad 
    &\fcompliance = \frac{\partial \ncpfunc_f}{\partial \lam_f} 
\end{aligned}
\end{equation}

With $\Delta \pos = \pos^{k+1} - \pos^{k}$ and $\Delta \vel = \vel^{k+1} - \vel^{k} = \frac{1}{\dt} (\pos^{k+1} - \pos^{k})$, we write the conditions in Equations \eqref{eq: constrained global non-smooth b} - \eqref{eq: constrained global non-smooth d} into the position-based form using a first-order Taylor expansion:
\begin{subequations}
\label{eq: ncp position-based}
\begin{align}
    \begin{split}
    \zeros & = \nsjacobian_b^{} \pos^{k+1} - \vecd_b^{} + \bcompliance \lam_b^{k+1} \\
    & = \nsjacobian_b^{} \pos^{k+1} + \bcompliance \Delta \lam_b^{} - \vecd_b^{}  + \bcompliance \lam_b^k  
    \end{split} \\
    \begin{split}
    \zeros & = \ncpfunc_n^{} (\pene_n^{k+1}, \lam_n^{k+1}) \\
    & = \ncpfunc_n^{} (\pene_n^k, \lam_n^k) + \nsjacobian_n^{} \Delta \pos + \ncompliance \Delta \lam_n^{} \\
    & = \nsjacobian_n^{} \pos^{k+1} + \ncompliance \Delta \lam_n^{} + \ncpfunc_n(\pene_n^k, \lam_n^k) - \nsjacobian_n^{} \pos^k 
    \end{split} \\
    \begin{split}
    \zeros & = \ncpfunc_f^{} (\dot \pene_f^{k+1}, \lam_f^{k+1}) \\
    & = \ncpfunc_f^{} (\dot \pene_f^k, \lam_f^k, \lam_n^k) + \nsjacobian_f^{} \Delta \vel + \fcompliance \Delta \lam_f^{} \\
    & = \frac{1}{\dt} \bigg( \nsjacobian_f^{} \pos^{k+1} + \dt \fcompliance \Delta \lam_f^{}  + \dt \ncpfunc_f^{} (\dot \pene_f^k, \lam_f^k, \lam_n^k) - \nsjacobian_f^{} \pos^k \bigg)  
    \end{split}
\end{align}
\end{subequations}
Assembling the system for Equations \eqref{eq: constrained global non-smooth a} and \eqref{eq: ncp position-based}, we have:
\begin{equation}
\label{eq: non-smooth newton system}
\begin{aligned}
    \begin{bmatrix}
    \system & -\nsjacobian_b^\mathrm{T} & -\nsjacobian_n^\mathrm{T} & -\nsjacobian_f^\mathrm{T} \\
    \nsjacobian_b^{} & \frac{1}{\dt^2} \bcompliance & \zeros & \zeros \\
    \nsjacobian_n^{} & \zeros & \frac{1}{\dt^2} \ncompliance & \zeros \\
    \nsjacobian_f^{} & \zeros & \zeros & \frac{1}{\dt} \fcompliance
    \end{bmatrix} 
    \begin{bmatrix}
    \pos \\ \dt^2 \Delta \lam_b^{} \\ \dt^2 \Delta \lam_n^{} \\ \dt^2 \Delta \lam_f^{} 
    \end{bmatrix}
    = 
    \begin{bmatrix}
    \vecg \\ \vech_b^{} \\ \vech_n^{} \\ \vech_f^{}
    \end{bmatrix}
\end{aligned}
\end{equation}
where
\begin{subequations}
\begin{align}
    &\vecg = \rhs^k + \dt^2 \nsjacobian_b^\mathrm{T} \lam_b^k + \dt^2 \nsjacobian_n^\mathrm{T} \lam_n^k + \dt^2 \nsjacobian_f^\mathrm{T} \lam_f^k \\
    &\vech_b^{} = \vecd_b^{} - \bcompliance \lam_b^k \\
    &\vech_n^{} = - \ncpfunc_n^{} (\pene_n^k, \lam_n^k) + \nsjacobian_n^{} \pos^k \\
    &\vech_f^{} = - \dt \ncpfunc_f^{} (\dot \pene_f^k, \lam_f^k, \lam_n^k) + \nsjacobian_f^{} \pos^k
\end{align}
\end{subequations}

Grouping the contact items such that
\begin{equation}
\begin{aligned}
    &\nsjacobian = 
    \begin{bmatrix}
    \nsjacobian_b^{} \\\nsjacobian_n^{} \\ \nsjacobian_f^{}
    \end{bmatrix}, \qquad
    \lam = 
    \begin{bmatrix}
    \lam_b^{} \\ \lam_n^{} \\ \lam_f^{}
    \end{bmatrix}, \qquad
    \vech = 
    \begin{bmatrix}
    \vech_b^{} \\ \vech_n^{} \\ \vech_f^{}
    \end{bmatrix}, \\
    &\text{and} \quad \compliance = 
    \begin{bmatrix}
    \frac{1}{\dt^2} \bcompliance & \zeros & \zeros\\ 
    \zeros & \frac{1}{\dt^2} \ncompliance & \zeros\\ 
    \zeros & \zeros & \frac{1}{\dt}\fcompliance
    \end{bmatrix}, 
\end{aligned}
\end{equation}
we can rewrite Equation \eqref{eq: non-smooth newton system} as
\begin{equation}
\label{eq: non-smooth newton problem matrix form}
\begin{aligned}
    \begin{bmatrix}
    \system & -\nsjacobian^\mathrm{T} \\
    \nsjacobian & \compliance
    \end{bmatrix} 
    \begin{bmatrix}
    \pos \\ \dt^2 \Delta \lam
    \end{bmatrix}
    = 
    \begin{bmatrix}
    \vecg \\ \vech
    \end{bmatrix}\\
\end{aligned}
\end{equation}
which eventually leads to a saddle point problem after a Schur-complement:
\begin{equation}
\label{eq: saddle point problem}
    \bigg( \nsjacobian \system^\mathrm{-1} \nsjacobian^\mathrm{T} + \compliance  \bigg) \Delta \lam 
    = \frac{1}{\dt^2} \bigg( \nsjacobian \system^\mathrm{-1} \vecg - \vech \bigg)
\end{equation}

A key advantage of the non-smooth Newton method is its independence from the linear solver used for the saddle point system in Equation \eqref{eq: saddle point problem}.
This flexibility allows for various solver choices: decomposition methods can be used for precise solutions in small-scale problems, while Krylov subspace methods offer both good convergence and parallelization potential for large-scale problems.

After computing $\Delta \lam$, we solve the global step by integrating the contact forces $\lam^{k+1}$, with corrections to the right-hand side:
\begin{subequations}
\label{eq: rhs correction}
\begin{align}
    &\lam^{k+1} = \lam^{k} + \Delta \lam\\
    &\pos^{k+1} = \system^\mathrm{-1} \Big(\rhs^k + \dt^2 \nsjacobian^\mathrm{T} \lam^{k+1} \Big)
\end{align}
\end{subequations}

The local-global iteration constrained with non-smooth functions is outlined in Algorithm \ref{algo: Local-global iterations with non-smooth functions}.
Implementing the non-smooth constraints into the local-global methods brings two major additional computing cost within each L-G iteration: the Schur-complement $\nsjacobian \system^\mathrm{-1} \nsjacobian^\mathrm{T}$ and the linear system solution in Equation \eqref{eq: saddle point problem}.
As discussed earlier, the flexibility in choosing linear solvers enables us to use Krylov subspace methods, such as Conjugate Gradient (CG) and Conjugate Residual (CR), which are well-suited for parallel implementation on both CPUs and GPUs for large-scale problems.

In contrast, efficiently computing the Schur-complement is challenging as it involves solving a large linear system (scale of DOFs $n$) with multiple right-hand sides (scale of contact constraints $c$).
Such computation can easily become the bottleneck in case of a detailed mesh discretization or large amount of contacts.
The solution in \cite{macklin_non-smooth_2019} simplifies the computing by replacing the Hessian by a diagonal approximation computed in each Newton iteration.
In general, this should cause slower convergence since propagating the impact of constraints through the non-diagonal entries is eliminated, leading to weak coupling between the constraints.
To address this problem, we present effective enhancements in subsequent sections.

\begin{algorithm}[tb!]
\caption{Non-smooth constrained local-global method}
\label{algo: Local-global iterations with non-smooth functions}
\SetAlgoLined
\While{\textit{simulation}}{
\text{Perform collision detection}\;
$\sn^k = \pos_t + \dt \vel_t + \dt^2 \mass^\mathrm{-1} \vecf_\mathrm{ext}$\;
\For{$k \in \{0, ..., n\}$}{
$\proj_i^k = project(\reduction_i \pos^k)$\Comment*[r]{local step} 
$\rhs = \mass \sn + \dt^2 \sum_i \weight_i \reduction_i^\mathrm{T} \proj_i^k$\;
Evaluate $\nsjacobian$, $\compliance$, $\vecg$, $\vech$\;
$ \Delta \lam 
= \frac{1}{\dt^2} \bigg( \nsjacobian \system^\mathrm{-1} \nsjacobian^\mathrm{T} + \compliance  \bigg)^\mathrm{-1} \bigg( \vech - \nsjacobian \system^\mathrm{-1} \vecg \bigg)$ \Comment*{solve constraint linear system} 
$\lam^{k+1} = \lam^{k} + \Delta \lam$\;
$\pos^{k+1} = \system^\mathrm{-1} \Big(\rhs + \dt^2 \nsjacobian^\mathrm{T} \lam^{k+1} \Big)$\Comment*[r]{apply constraint correction} 
}
$\pos_{t+h} = \pos^{k+1} $\Comment*[r]{integration} 
$\vel_{t+h} = \frac{1}{\dt} \bigg(\pos^{k+1} - \pos_t \bigg)$\;
}
\end{algorithm}

\subsection{Splitting out non-smooth indicators}
\label{sec: splitting out non-smooth indicators}

We propose a splitting strategy to address the expensive computation of the Schur-complement $\nsjacobian \system^\mathrm{-1} \nsjacobian^\mathrm{T}$ in each L-G iteration.
Recalling the formulations of non-smooth Jacobian in Equations \eqref{eq: non-smooth jacobian in minimum-map for contact} \eqref{eq: non-smooth jacobian in minimum-map for friction} \eqref{eq: non-smooth jacobian in FB for contact} \eqref{eq: non-smooth jacobian in FB for friction} for different NCP functions, we unify these formulations in following way:  

\begin{subequations}
\label{eq: spliting non-smooth indicator}
\begin{align}
    &\nsjacobian_{n_{\mathrm{min}}} = 
    \weightns_{n_{\mathrm{min}}} \jacobian_n
    &&\text{with} \quad 
    \weightns_{n_{\mathrm{min}}} = 
    \begin{cases} 
    \vecones \qquad &\text{if} \; \pene_n \leq \nsprecond \lam_n\\
    \veczeros \qquad &\text{otherwise}
    \end{cases} \\
    &\nsjacobian_{f_{\mathrm{min}}} = 
    \weightns_{f_{\mathrm{min}}} \jacobian_f
    &&\text{with} \quad 
    \weightns_{f_{\mathrm{min}}} = 
    \begin{cases} 
    \vecones \qquad &\text{if} \; \lam_n > \veczeros \\
    \veczeros \qquad &\text{otherwise}
    \end{cases} \\
    &\nsjacobian_{n_{\mathrm{FB}}} = 
    \weightns_{n_{\mathrm{FB}}} \jacobian_n
    &&\text{with} \quad 
    \weightns_{n_{\mathrm{FB}}} = 
    \vecones - \frac{\pene_n}{\sqrt{\pene_n^2 + \nsprecond^2 \lam_n^2}} \\
    &\nsjacobian_{f_{\mathrm{FB}}} = 
    \weightns_{f_{\mathrm{FB}}} \jacobian_f
    &&\text{with} \quad 
    \weightns_{f_{\mathrm{FB}}} = 
    \begin{cases} 
    \vecones \qquad &\text{if} \; \lam_n > \veczeros \\
    \veczeros \qquad &\text{otherwise}
    \end{cases}
\end{align}
\end{subequations}
where we define the non-smooth indicators $\weightns$ as the weighting parameters for $\jacobian$.

In matrix form, we can reformulate the non-smooth Jacobian as follows:
\begin{equation}
\label{eq: spliting non-smooth indicator matrix form}
    \nsjacobian = 
    \begin{bmatrix}
        \weightns_0 & & \\
        & \ddots & \\
        & & \weightns_c
    \end{bmatrix} \jacobian
    = \weightnsjacobian \jacobian
\end{equation}
where $\weightnsjacobian$ is a diagonal matrix containing the non-smooth indicators.

A key finding is that the non-smooth indicators $\weightns$ dominate the dynamic behavior of the non-smooth Jacobian $\nsjacobian$, while $\jacobian$ remains static throughout the current time step.
This is because $\jacobian$ only depends on the output of the collision detection and the constraint linearization which are performed only once at the beginning of the time step.
With Equation \eqref{eq: spliting non-smooth indicator matrix form}, the Schur-complement in each Newton iteration becomes:
\begin{equation}
\label{eq: reduce schur complement}
    \nsjacobian \system^{-1} \nsjacobian^\mathrm{T}
    = \weightnsjacobian \jacobian \system^{-1} \jacobian^\mathrm{T} \weightnsjacobian^\mathrm{T}   
    = \weightnsjacobian \delasus \weightnsjacobian^\mathrm{T} 
\end{equation}
where $\delasus = \jacobian \system^{-1} \jacobian^\mathrm{T}$ is called \textit{delasus operator} in multi-body dynamics, and it only needs to be computed once per time step.
Consequently, this converts the original $m$ Schur-complement computations (where $m$ is the number of Newton or local-global iterations) into a single Schur-complement computation at the beginning of the time step, plus $2 \times m$ diagonal matrix - matrix multiplications, which are typically much more efficient.

The splitting strategy can be applied for non-smooth Newton methods with generic linear solvers.
Moreover, referring back to the sparse inverse resolution outlined in Section \ref{sec: Sparse Inverse Solution}, we can efficiently compute the Schur-complement through matrix product operations as follows:
\begin{equation}
\label{eq: schur-complement}
    \delasus = \jacobian \system^\mathrm{-1} \jacobian^\mathrm{T} = \jacobian \sparseinverse^\mathrm{T} \sparseinverse \jacobian^\mathrm{T}
\end{equation}
In contrast to the diagonal approximation detailed in \cite{macklin_non-smooth_2019}, our exact solution can efficiently couple the constraints using sparse matrix - sparse matrix multiplication (SpGEMM) operations which are fundamental operations for sparse matrices.

To accelerate the computation of Equation \eqref{eq: schur-complement}, one can utilize the reuse strategy described in \cite{zeng_realtime_2022} to exploit shared contact data between consecutive time steps.

\subsection{Complementarity precondition}
\label{sec: complementarity precondition}

Another advantage of explicitly computing $\delasus$ is that it provides an effective complementarity preconditioner.
As discussed in \cite{macklin_non-smooth_2019}, applying a preconditioner $r$ to NCP functions dose not change the solution of the original problem.
However, using the preconditioner significantly affects the convergence rate of the non-smooth Newton method and our local-global method constrained with NCP functions.

The strategy in \cite{macklin_non-smooth_2019} for choosing the preconditioner is to ensure both sides of the complementarity equation have the same slop: 
For a unilateral constraint, $\nsprecond_j = \dt^2 \big[ \nsjacobian \mass^{\mathrm{-1}}  \nsjacobian^{\mathrm{T}} \big]_{jj}$ is used to achieve appropriate scaling in the position-force relationship between $\pene_j$ and $\lam_j$.
For a frictional constraint, the scaling factor should be $\nsprecond_j = \dt \big[ \nsjacobian \mass^{\mathrm{-1}}  \nsjacobian^{\mathrm{T}} \big]_{jj}$ to follow the velocity-force relationship between $\dot \pene_j$ and $\lam_j$.

However, there are two problems to use such preconditioner.
First, the preconditioner $\nsprecond$ is computed by $\nsjacobian$ while $\nsjacobian$ actually depends on $\nsprecond$, leading to a circular dependency issue.
Second, using the mass inverse will decrease the efficiency of the preconditioner.
This is particularly obvious in soft body simulations with high-stiffness materials, where the stiffness matrix, rather than the mass matrix, dominates the system's diagonal elements.

\begin{figure}[b!]
        \noindent%
	\begin{subfigure}[t]{1.0\linewidth}%
            \centering
		  \includegraphics[width=0.65\linewidth,clip,trim=0 105 0 175]{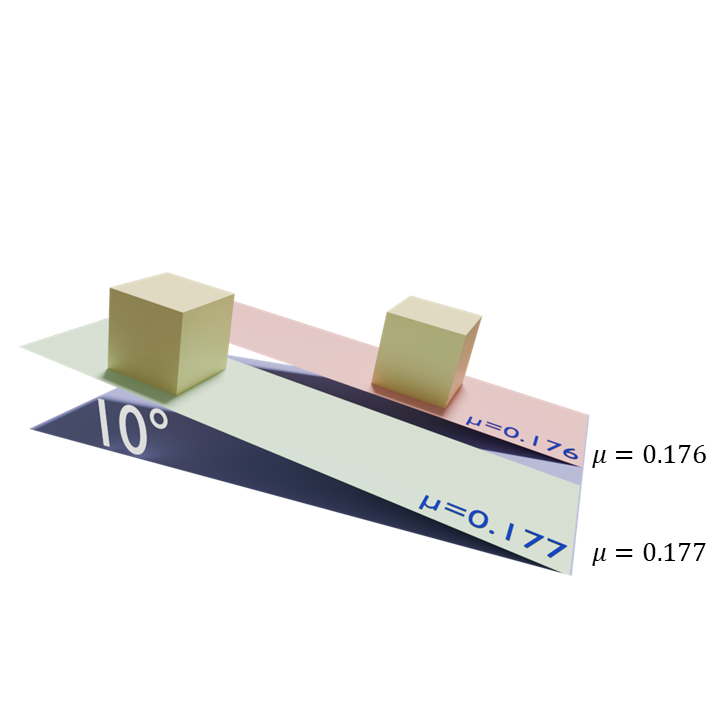}%
            \caption{Our system inverse $\big[ \jacobian \system^{\mathrm{-1}}  \jacobian^{\mathrm{T}} \big]_{jj}$ complementarity preconditioner.}
            \label{fig: accurate friction a}
	\end{subfigure}%
        \newline%
	\begin{subfigure}[t]{1.0\linewidth}%
            \centering
		  \includegraphics[width=0.65\linewidth,clip,trim=0 105 0 175]{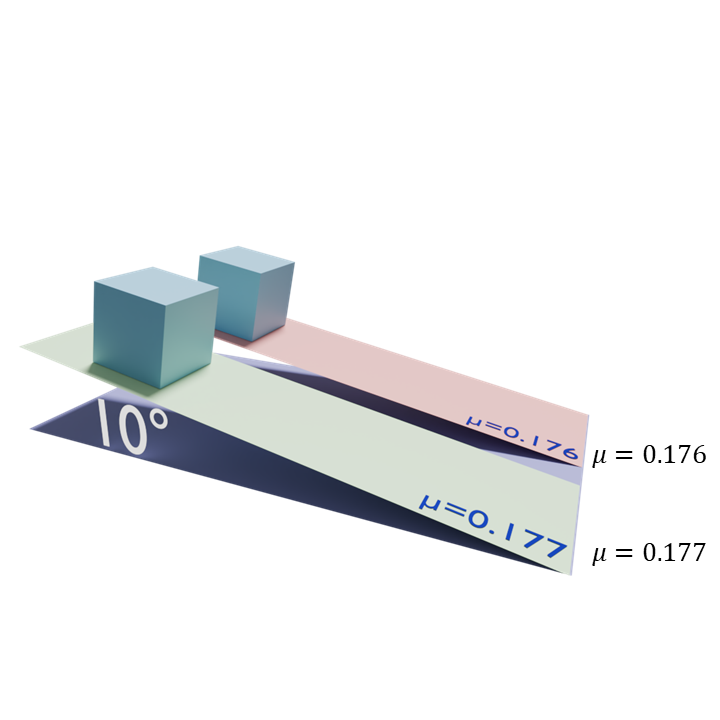}%
	\end{subfigure}%
        \newline%
	\begin{subfigure}[t]{1.0\linewidth}%
            \centering
		  \includegraphics[width=0.65\linewidth,clip,trim=0 105 0 175]{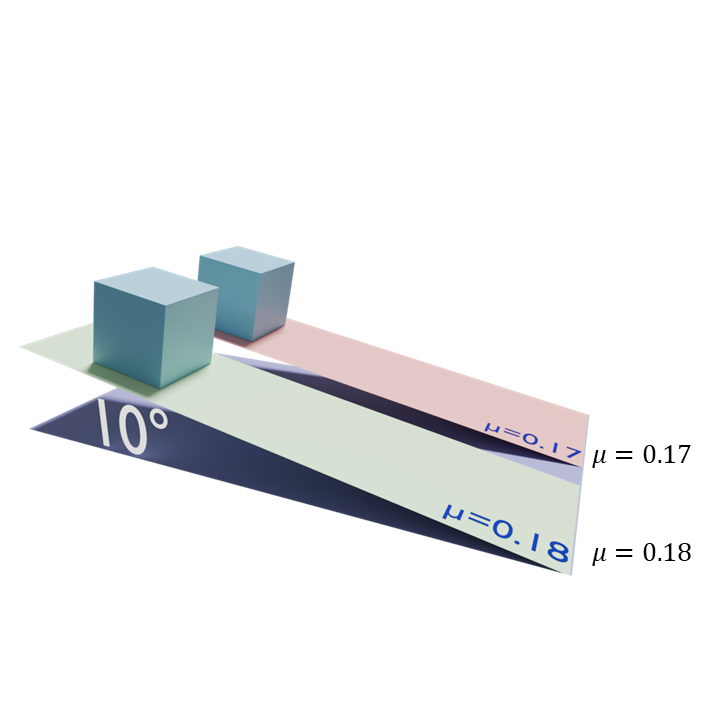}%
	\end{subfigure}%
        \newline%
	\begin{subfigure}[t]{1.0\linewidth}%
            \centering
		  \includegraphics[width=0.65\linewidth,clip,trim=0 105 0 175]{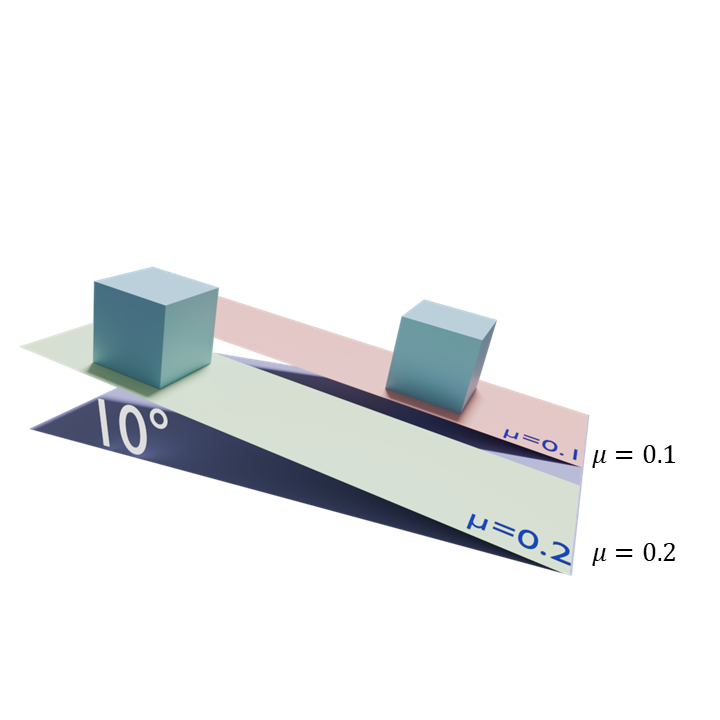}%
            \caption{Mass inverse $\big[ \jacobian \mass^{\mathrm{-1}}  \jacobian^{\mathrm{T}} \big]_{jj}$ complementarity preconditioner.}
            \label{fig: accurate friction b}
	\end{subfigure}%
    \caption{
        Stick-Sliding test: simulating a high-stiff block on a slope.
        With the analytical stick-sliding discontinuity $\mu^* = tan(10\pi/180) = 0.17632698$, the ideal behavior is that the box sticks to the plane when $\mu > \mu^*$, and slides when $\mu \leq \mu^*$.
        (a) Our complementarity preconditioner with the system inverse achieves the accuracy level at $0.001$.
        (b) The complementarity preconditioner computed with the mass inverse can only capture the discontinuity for stick-sliding behavior with an accuracy of $0.1$. 
    }
    \Description{}
    \label{fig: accurate friction}
\end{figure}

To address these problems, we propose computing the preconditioner using the diagonal elements in the \textit{delasus operator}:
\begin{subequations}
\begin{align}
    &\text{Unilateral constraint:} \quad \nsprecond_j = \dt^2 \delasus_{jj} \\
    &\text{Frictional constraint:} \quad \nsprecond_j = \dt \delasus_{jj}
\end{align}
\label{eq: complementarity preconditioner}
\end{subequations}

We give several reasons for using this strategy:
First, our preconditioner uses $\system^\mathrm{-1}$, instead of $\mass^\mathrm{-1}$ to couple the constraints, efficiently taking the stiffness into consideration.
Second, the \textit{delasus operator} $\delasus$ is computed at the beginning of each time step, thereby avoiding the circular dependency issue.
Finally, as shown in previous studies of complementarity problems \cite{erleben_numerical_2013}, the \textit{delasus operator} $\delasus$ inherently acts as a scaling factor by establishing the connection between interpenetration and contact force (e.g., $\pene = \dt^2 \delasus \lam$ for LCP \cite{duriez_realistic_2006, zeng_realtime_2022}).

Consequently, our complementarity preconditioning strategy provides effective scaling for the NCP parameters.
In Section, \ref{sec: complementarity precondition and friction accuracy} we present experimental results demonstrating our method's efficiency in accurately capturing the discontinuities between sticking and sliding behavior.

\section{A GPU-friendly framework and its implementation}
\label{sec: a GPU-friendly framework}

\begin{algorithm}[tb!]
\caption{A unified and GPU-friendly pipeline for solving elastic dynamics involving fricitonal contacts}
\label{algo: Enhanced local-global iterations with non-smooth functions}
\SetAlgoLined
$\choleskyfactor = \textit{Cholesky}(\system)$\;
$\sparseinverse = \choleskyfactor^\mathrm{-1} \identity$\;
\While{\textit{simulation}}{
\text{Perform collision detection}\;
\textcolor{orange}{$\delasus = \jacobian \sparseinverse^\mathrm{T} \sparseinverse \jacobian^\mathrm{T}$\Comment*[r]{computing delasus operator} }
$\sn = \pos_t + \dt \vel_t + \dt^2 \mass^\mathrm{-1} \vecf_\mathrm{ext}$\;
\For{$k \in \{0, ..., n\}$}{
$\proj_i^k = project(\reduction_i \pos^k)$\Comment*[r]{local step} 
$\rhs = \mass \sn + \dt^2 \sum_i \weight_i \reduction_i^\mathrm{T} \proj_i^k$\;
Evaluate \textcolor{orange}{$\weightnsjacobian$}, $\nsjacobian$, $\compliance$, $\vecg$, $\vech$\;
$ \Delta \lam 
= \frac{1}{\dt^2} \bigg( \textcolor{orange}{\weightnsjacobian \delasus \weightnsjacobian^\mathrm{T}} + \compliance  \bigg)^\mathrm{-1} \bigg( \vech - \textcolor{orange}{\nsjacobian \sparseinverse^\mathrm{T} \sparseinverse \vecg} \bigg)$\Comment*[r]{solve constraint linear system}
$\lam^{k+1} = \lam^{k} + \Delta \lam$\;
$\pos^{k+1} = \textcolor{orange}{\sparseinverse^\mathrm{T} \sparseinverse} \Big(\rhs + \dt^2 \nsjacobian^\mathrm{T} \lam^{k+1} \Big)$\Comment*[r]{apply constraint correction}
}
$\pos_{t+h} = \pos^{k+1} $\Comment*[r]{integration}  
$\vel_{t+h} = \frac{1}{\dt} \bigg(\pos^{k+1} - \pos_t \bigg)$\;
}
\end{algorithm}

\begin{table*}[b!]
\centering
\caption{Memory usage (MB) for storing matrices inverse (in double floating) with triangular surface meshes.}
\begin{tabular}{ccccccccccccc}
\toprule
Shape & \multicolumn{4}{c}{Square cloth (no bending)} &\multicolumn{4}{c}{Square cloth (isometric bending)} &\multicolumn{4}{c}{Square cloth (Laplace-Beltrami bending)} \\
\cmidrule(lr){2-5} \cmidrule(lr){6-9} \cmidrule(lr){10-13} 
Vertex & 5.1k & 10.2k & 15.3k & 20.2k & 5.1k & 10.2k & 15.3k & 20.2k & 5.1k & 10.2k & 15.3k & 20.2k \\
Tri. & 10.0k & 20.2k & 30.3k & 40.0k & 10.0k & 20.2k & 30.3k & 40.0k & 10.0k & 20.2k & 30.3k & 40.0k \\
\cmidrule(lr){2-5} \cmidrule(lr){6-9} \cmidrule(lr){10-13} 
$\system^\mathrm{-1}$ & 893.33 & 3589.46 & 8050.51 & 14010.30 & 893.33 & 3589.46 & 8050.51 & 14010.30 & 893.33 & 3589.46 & 8050.51 & 14010.30 \\
$\sparseinverse$ & 34.80 & 102.51 & 183.79 & 280.41 & 62.37 & 183.82 & 342.68 & 523.36 & 87.08 & 258.85 & 487.14 & 749.10 \\
\bottomrule
\end{tabular}
\label{tab: memory cost surface}
\end{table*}

\begin{table*}[b!]
\centering
\caption{Memory usage (MB) for matrices inverse (in double floating) with tetrahedral volume meshes}
\begin{tabular}{ccccccccccccc}
\toprule
Shape & \multicolumn{3}{c}{Raptor} & \multicolumn{2}{c}{Ball} & \multicolumn{3}{c}{Bar} & \multicolumn{2}{c}{Wooper} & \multicolumn{2}{c}{Gingerbreadman}\\
\cmidrule(lr){2-4} \cmidrule(lr){5-6} \cmidrule(lr){7-9} \cmidrule(lr){10-11} \cmidrule(lr){12-13}
Vertex & 10.0k & 15.0k & 20.1k & 7.1k & 14.7k & 5.3k & 11.3k & 20.8k & 5.3k & 11.1k & 11.8k & 19.5k \\
Tetra. & 33.9k & 54.4k & 77.0k & 37.5k & 81.7k & 26.4k & 59.0k & 111.1k & 20.4k & 52.5k & 48.2k & 93.6k \\
\cmidrule(lr){2-4} \cmidrule(lr){5-6} \cmidrule(lr){7-9} \cmidrule(lr){10-11} \cmidrule(lr){12-13}
$\system^\mathrm{-1}$ & 3661.43 & 8071.55 & 14308.70 & 1744.86 & 7452.21 & 961.48 & 4414.98 & 14870.70 & 973.51 & 4784.48 & 4255.27 & 13045.50 \\
$\sparseinverse$ & 120.00 & 260.85 & 467.58 & 266.87 & 946.97 & 125.02 & 463.03 & 1310.19 & 67.14 & 295.09 & 227.30 & 619.41 \\
\bottomrule
\end{tabular}
\label{tab: memory cost volume}
\end{table*}

Our final integrated algorithm is presented in Algorithm \ref{algo: Enhanced local-global iterations with non-smooth functions}, 
where the modifications in orange highlight our core contributions with efficient forward matrix operations.

In conclude, we elaborate on the synergistic efficiency of combining the sparse inverse local-global method with the non-smooth Newton method:

The sparse inverse local-global method relies on system invariability, precluding penalty-based contact methods that would modify the system matrix.
As a Lagrange multiplier approach, the non-smooth Newton method operates without such alterations, effectively ensuring the fulfillment of complementarity conditions during global steps.

In contrast, complete-solution-based local-global methods facilitate achieving the desired precision with a limited number of iterations. 
This characteristic aligns well with the preferences of the non-smooth Newton method, as it helps reduce significant computational cost.
Notably, the sparse inverse method provides efficient computation of the Schur-complement by transforming the linear problem into matrix multiplication operations.

This unified approach enables parallelization across all major computational stages, making it well-suited for GPU acceleration. The implementation details are as follows.

\paragraph*{\textbf{Matrix operations}}
As discussed in Section \ref{sec: optimized time and space efficiency}, the matrix operations are highly parallelizable on both the CPU and the GPU.
We use the NVIDIA's cuSPARSE library to implement SpMV operations in global steps.
The Schur-complement (Equation \eqref{eq: schur-complement}) can be implemented with SpGEMM operations.
To accelerate this process, we implement the reusing strategy in \cite{zeng_realtime_2022} to reduce computational costs.

\begin{figure}[tb!]
    \centering
        \noindent%
	\begin{subfigure}[t]{0.5\linewidth}%
		\includegraphics[width=0.95\linewidth,clip,trim=0 0 0 130]{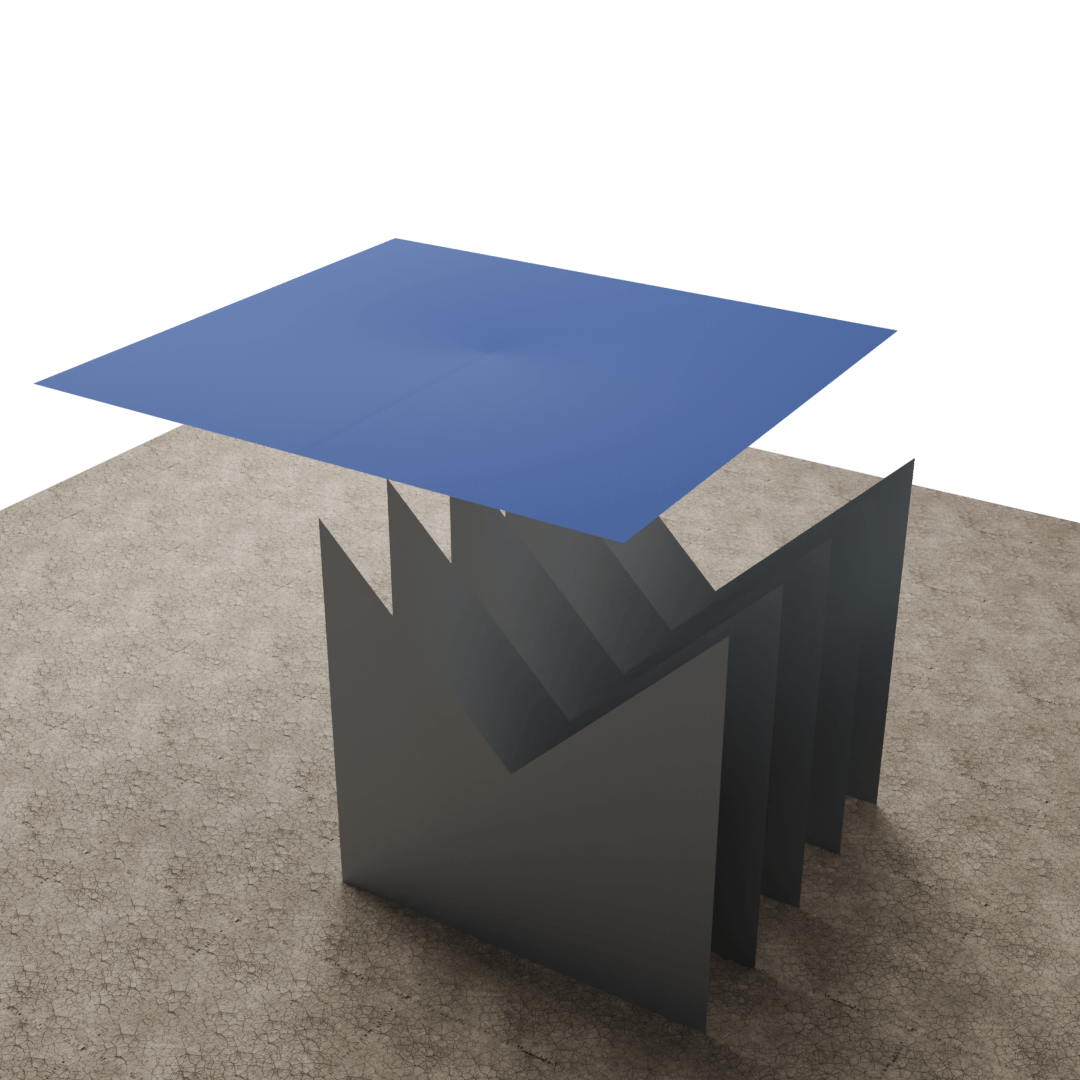}%
            \caption{Frame $0$}
	\end{subfigure}%
	\begin{subfigure}[t]{0.5\linewidth}%
		\includegraphics[width=0.95\linewidth,clip,trim=0 0 0 130]{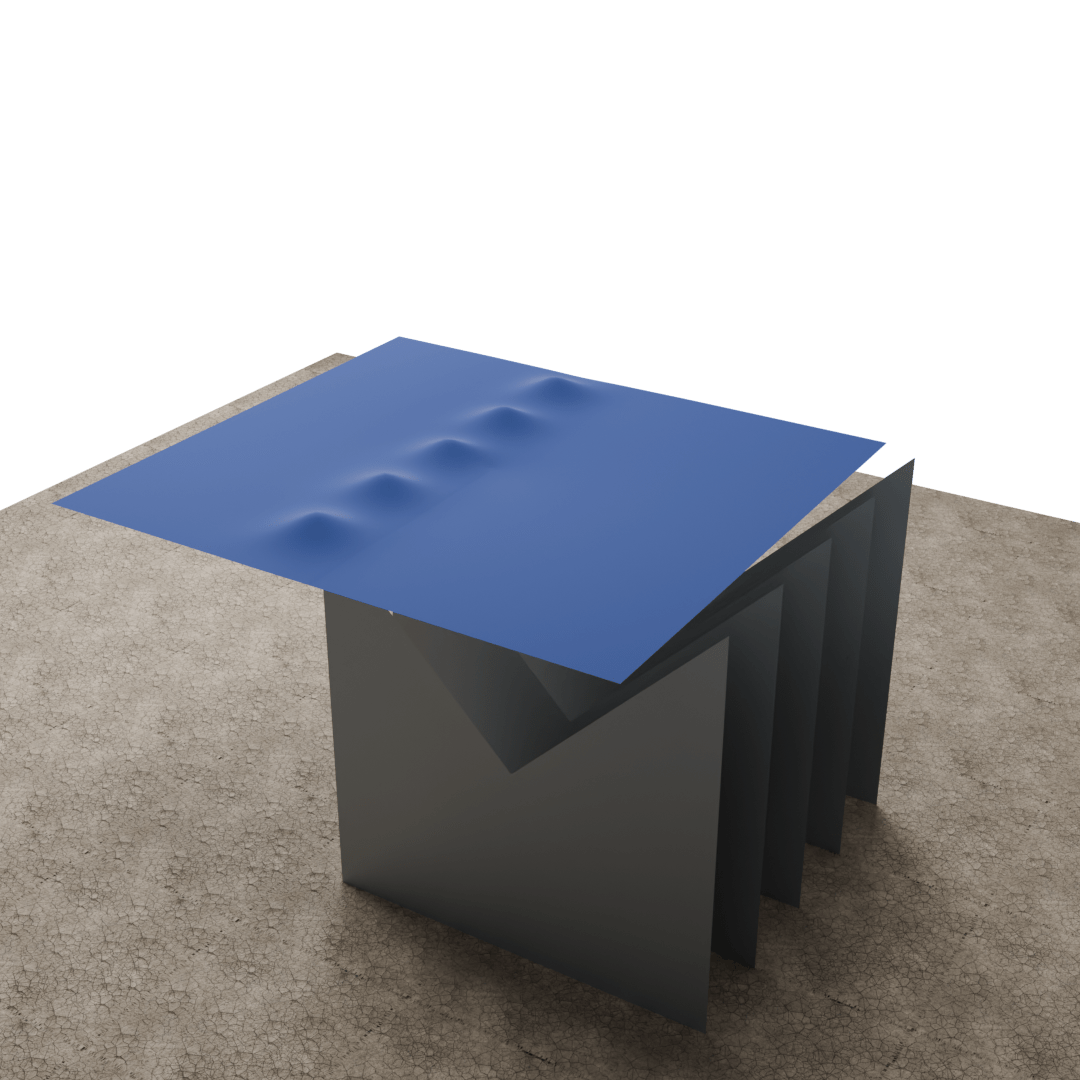}%
            \caption{Frame $60$}
	\end{subfigure}%
	\newline%
	\begin{subfigure}[t]{0.5\linewidth}%
		\includegraphics[width=0.95\linewidth,clip,trim=0 0 0 230]{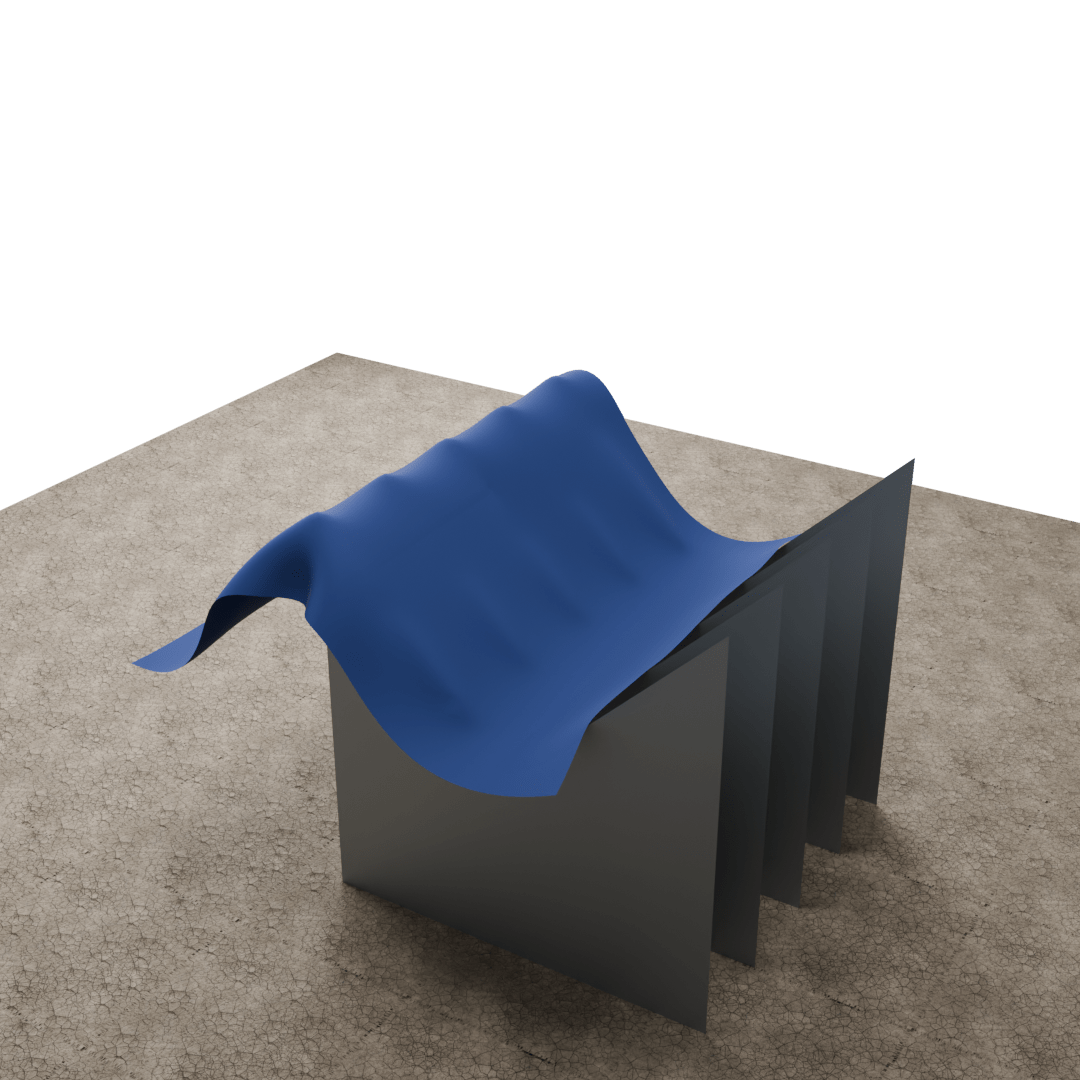}%
            \caption{Frame $90$}
	\end{subfigure}%
	\begin{subfigure}[t]{0.5\linewidth}%
		\includegraphics[width=0.95\linewidth,clip,trim=0 0 0 230]{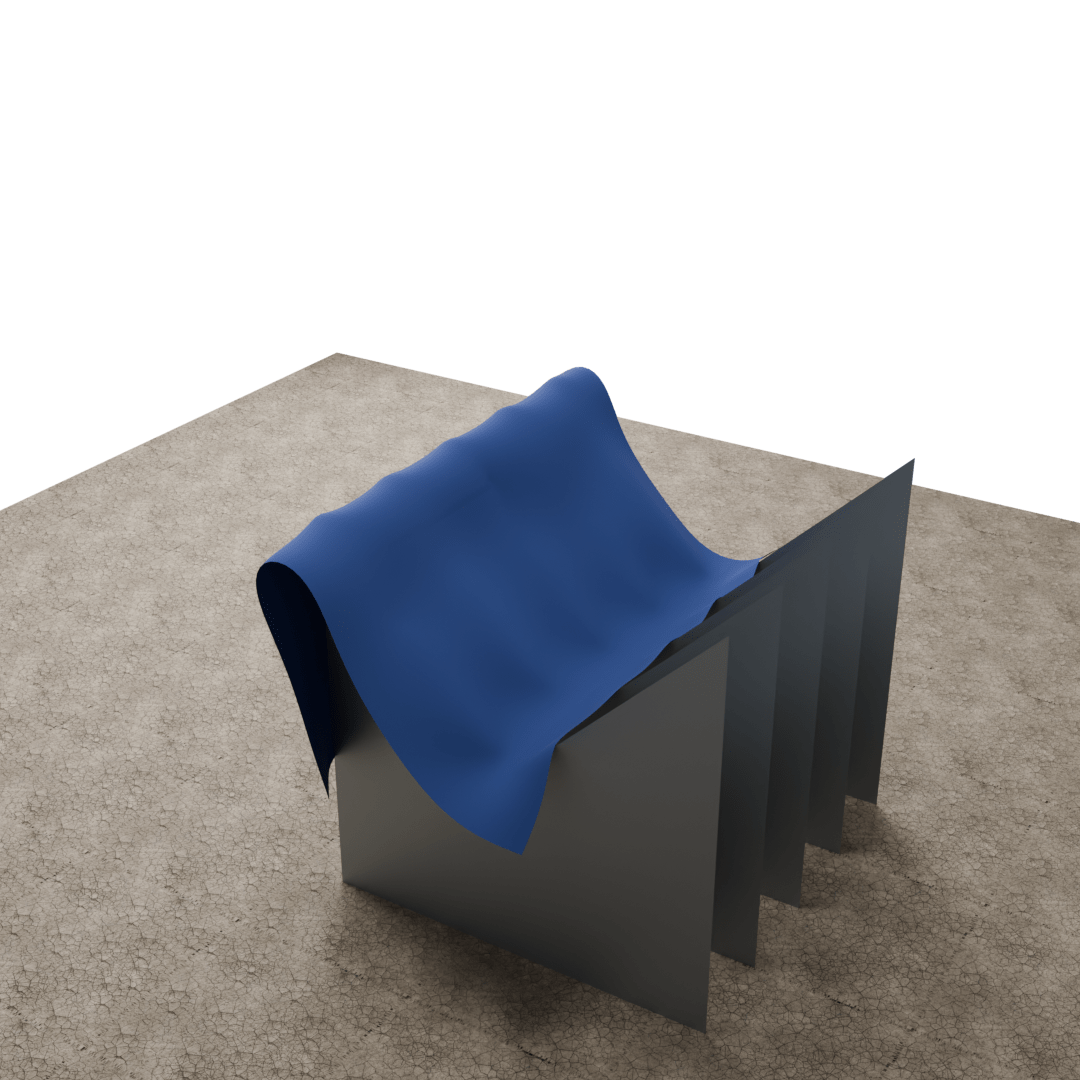}%
            \caption{Frame $300$}
	\end{subfigure}%
    \caption{
        Cloth on Knives: stable cloth simulation with non-smooth and codimensional contacts.
    }
    \Description{}
    \label{fig: cloth on knives}
\end{figure}

\paragraph*{\textbf{Constraint linear solver and NCP functions}}
We choose the Conjugate Residual (CR) as our linear solver in the constraint space, and the Fischer-Burmeister function as our NCP function, since their efficiency has been proven in \cite{macklin_non-smooth_2019}.
We implement a GPU version of the CR solver using NVIDIA cuBLAS library.

\paragraph*{\textbf{Local step}}
Following \cite{overby_admm_2017}, we implement the singular value decomposition (SVD) and a small-scale L-BFGS quasi-Newton method to solve the local nonlinear problems in Equation \eqref{eq: PD local}.
Such process is naturally parallelizable owning to the independency of the local problems to each other.
Since the local step is not the core of our methods, we implement a CPU-based parallelization for it.
The migration of this process to create a fully GPU-based simulator remains future work. 
Our current implementation utilizes a hybrid CPU-GPU framework.

\paragraph*{\textbf{Collision detection}}
Our contact method, based on the non-smooth Newton method, is decoupled with the collision detection.
It is compatible with any detection method that provides accurate collision pair information.
Like in \cite{ly_projective_2020} and \cite{macklin_non-smooth_2019}, we implement collision detection using simple proximity queries with basic parallelization strategies.
For simplicity, we do not activate the self-collision handling in our scenarios. 
However, our method supports self-collisions in principle, as the contact Jacobian $\jacobian$ can effectively couple the DOFs within the same object.

\paragraph*{\textbf{Chebyshev acceleration}}
As indicated in \cite{wang_chebyshev_2015}, local-global methods can be accelerated with the Chebyshev method.
However, since the complete solution converges fast in the first iterations where the Chebyshev method shows minimal impact, we do not use the Chebyshev acceleration for our examples which generally achieve real-time performance with just a few local-global iterations.

\section{Results}
\label{sec: results}

\begin{figure*}[htb]
    \centering
    \noindent%
	\begin{subfigure}[t]{0.25\linewidth}%
        \centering
            \includegraphics[width=0.95\linewidth,clip,trim=0 110 690 130]{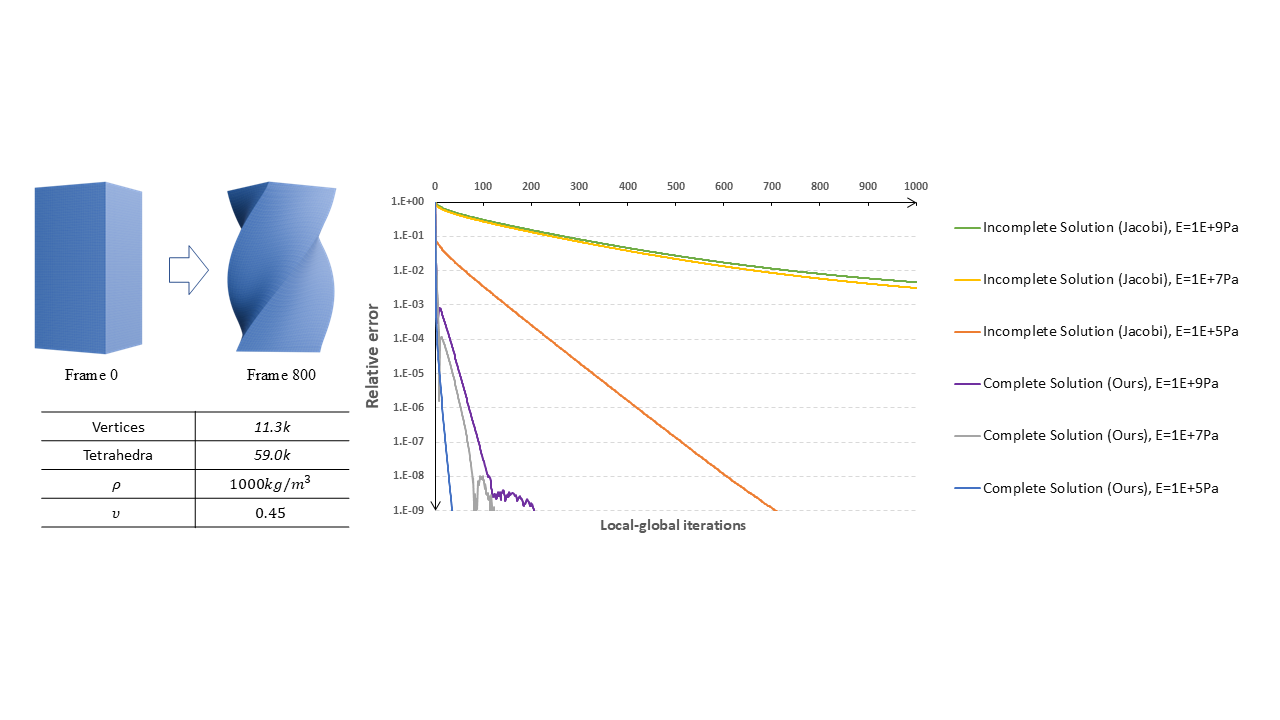}%
            \caption{Test scenario: twisting bar.}
            \label{fig: convergence comparison 00}
	\end{subfigure}%
	\begin{subfigure}[t]{0.75\linewidth}%
        \centering
            \includegraphics[width=0.95\linewidth,clip,trim=270 130 0 120]{figure/Plot_compare_01.PNG}%
            \caption{Comparison of relative error with respect to L-G iterations.}
            \label{fig: convergence comparison 01}
	\end{subfigure}%
    \newline%
	\begin{subfigure}[t]{1.0\linewidth}%
            \includegraphics[width=0.95\linewidth,clip,trim=0 120 0 120]{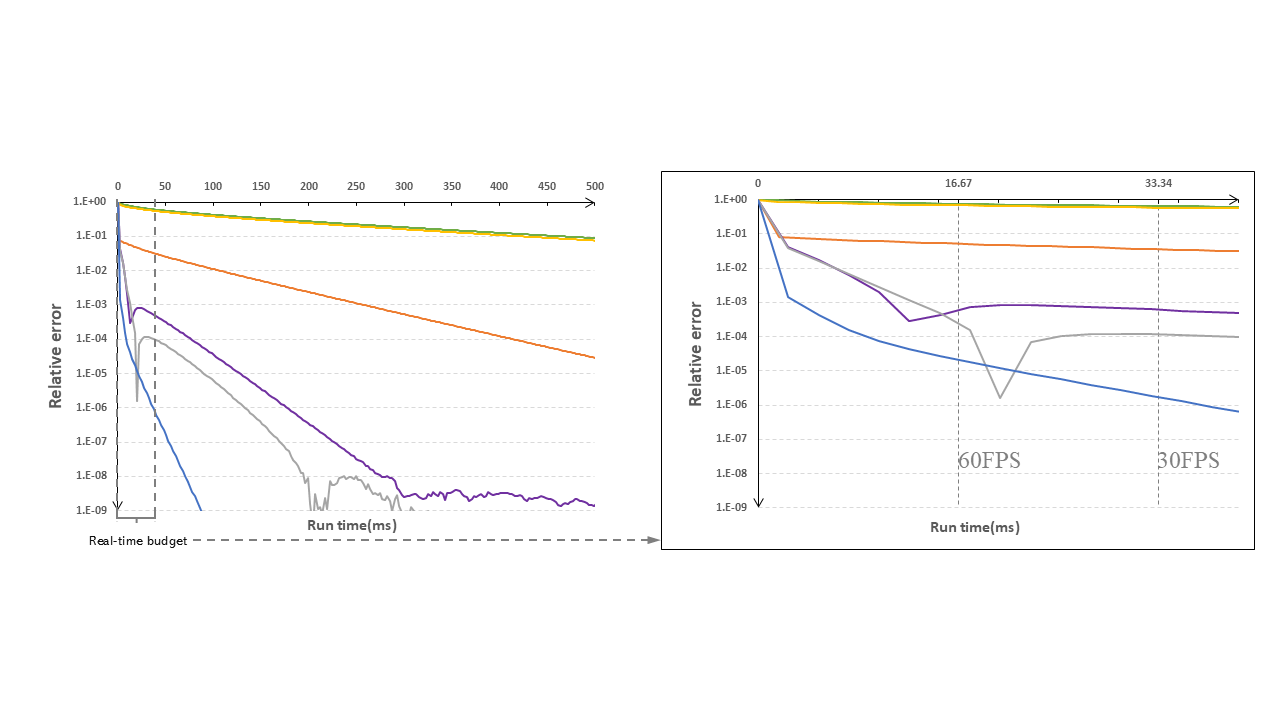}%
            \caption{
            Comparison of relative error with respect to computational time (ms): convergence rate (left) and performance at real-time frame rates of $30$ FPS and $60$ FPS (right).}
            \label{fig: convergence comparison 02}
	\end{subfigure}%
        \caption{Performance and accuracy evaluation in scenario under large rotational deformation: 
        The relative error is defined as ($\epsilon(\pos^k) - \epsilon(\pos^*))/ (\epsilon(\pos^0) - \epsilon(\pos^*)$) with $\epsilon$ the system energy and $\pos^*$ the ground truth (the converged solution in Newton's method). 
        (a) At Frame $800$ where the bar is largely deformed, we compare the convergence rate in different stiffness settings.
        (b) We plot the same comparison of convergence behavior with particular emphasis on the accuracy achieved within real-time computational constraints.}
    \Description{}
    \label{fig: convergence comparison}
\end{figure*}

\subsection{Experimental setup}

In this section, we evaluate the performance and accuracy of our proposed framework.
The simulation tests are performed on a computer equipped with an Intel@ i9-13900KF CPU 24-Core at 5.5GHz with 128GB RAM, and a GPU NVIDIA GeForce RTX 4090 with 24GB RAM.

Our tests cover various challenging scenarios, including large-scale systems, large deformation, non-smooth and massive contacts, and accurate frictions. 
Each simulation typically runs with 5 local-global iterations, and $10$ Conjugate Residual iterations for the constraint solver if contacts are involved.
The detailed statistics for the tests are summarized in Table \ref{tab: simulation statics}.

\subsection{Memory usage}
\label{sec: memory usage}

We evaluate the memory usage for storing the inverted Cholesky factor $\sparseinverse = \choleskyfactor^\mathrm{-1} \identity$ for objects of varying shapes, discretizations, and element types.
Tables \ref{tab: memory cost surface} and \ref{tab: memory cost volume} present memory usage comparisons for surface and volume meshes in our test scenarios.
As discussed in Section \ref{sec: Sparse Inverse Solution}, the system inverse $\system^\mathrm{-1}$ shows a high density, and storing such matrix in the memory becomes soon inhibitive as the problem size increases.
In contrast, our method requires reasonable memory usage for the sparse inverse of the Cholesky factor.
For large objects with $20k$ nodes ($60k$ DOFs), the memory cost for storing $\sparseinverse$ generally remains below $1$GB, with the volumetric bar being the only exception.

The topology connection strongly affects matrix sparsity, as demonstrated in Table \ref{tab: memory cost volume}.
Generally, objects with more complex shapes have fewer connections (fewer elements), resulting in sparser matrices and lower memory requirements for $\sparseinverse$.
This conclusion is also evident in the triangle meshes shown in Table \ref{tab: memory cost surface} where the bending constraints introduce virtual topology connections and increase the matrix fill-in.

\subsection{Performance and accuracy evaluation}

In Figure \ref{fig: convergence comparison}, we compare the convergence rate and the performance between our method and incomplete solutions. 
In the test, a bar is twisted with rotational boundary conditions applied to its ends (Figures \ref{fig: convergence comparison 00}). 
We take the Jacobi's method \cite{wang_chebyshev_2015} as a baseline and compute the relative error against the ground truth $\pos^*$. 
Since both complete and incomplete solutions can be accelerated via the Chebyshev method \cite{wang_chebyshev_2015}, we exclude the Chebyshev method from our comparison.

In Figure \ref{fig: convergence comparison 01}, we plot the convergence rate and observe that the complete solution quickly reduces the errors in the first iterations, for both low and high stiffness systems.
In contrast, the incomplete solution, approximating the global step with a single Jacobi iteration, converges slowly due to the inefficiency in propagating the local results. 
Furthermore, the incomplete method exhibits severely limited convergence for highly stiff systems. For example, when applied to an elastic bar with Young's modulus $E\geq10^{7}$, the method fails to achieve an accuracy of $10^{-3}$ even after $1000$ iterations.

In Figure \ref{fig: convergence comparison 02}, we compare the relative error with respect to computational time.
Combining the effect of computational efficiency (both local and global costs) and convergence rates, our method substantially outperforms the incomplete approach across all stiffness levels.
This gap will be more significant when solving nonlinear local problems, such as simulating Neo-Hookean materials, which requires more computational costs in local steps. 
This highlights the inherent simplicity of our method: even with a basic implementation of the local step, our method achieves a high accuracy within real-time computational constraints.
Conversely, while the incomplete method may achieve favorable results with optimized local steps, its performance exhibits high sensitivity to implementation details.




The magnified plot (Figure \ref{fig: convergence comparison 02}, right) illustrates the rapid convergence of our method within real-time computational constraints, achieving high accuracy (error < $10^{-3}$) for both low and high stiffness elastic materials.
In contrast, for low-stiffness systems, the incomplete method achieves a limited accuracy (error=$0.1$) within real-time constraints. 
The performance deteriorates further in high-stiffness scenarios, where solutions deviate significantly from the ground truth.

\subsection{Hyperelastic material}

\begin{figure}[htb!]
    \centering        
    \noindent%
	\begin{subfigure}[t]{1.0\linewidth}%
        \centering    
		\includegraphics[width=0.95\linewidth,clip,trim=0 350 0 300]{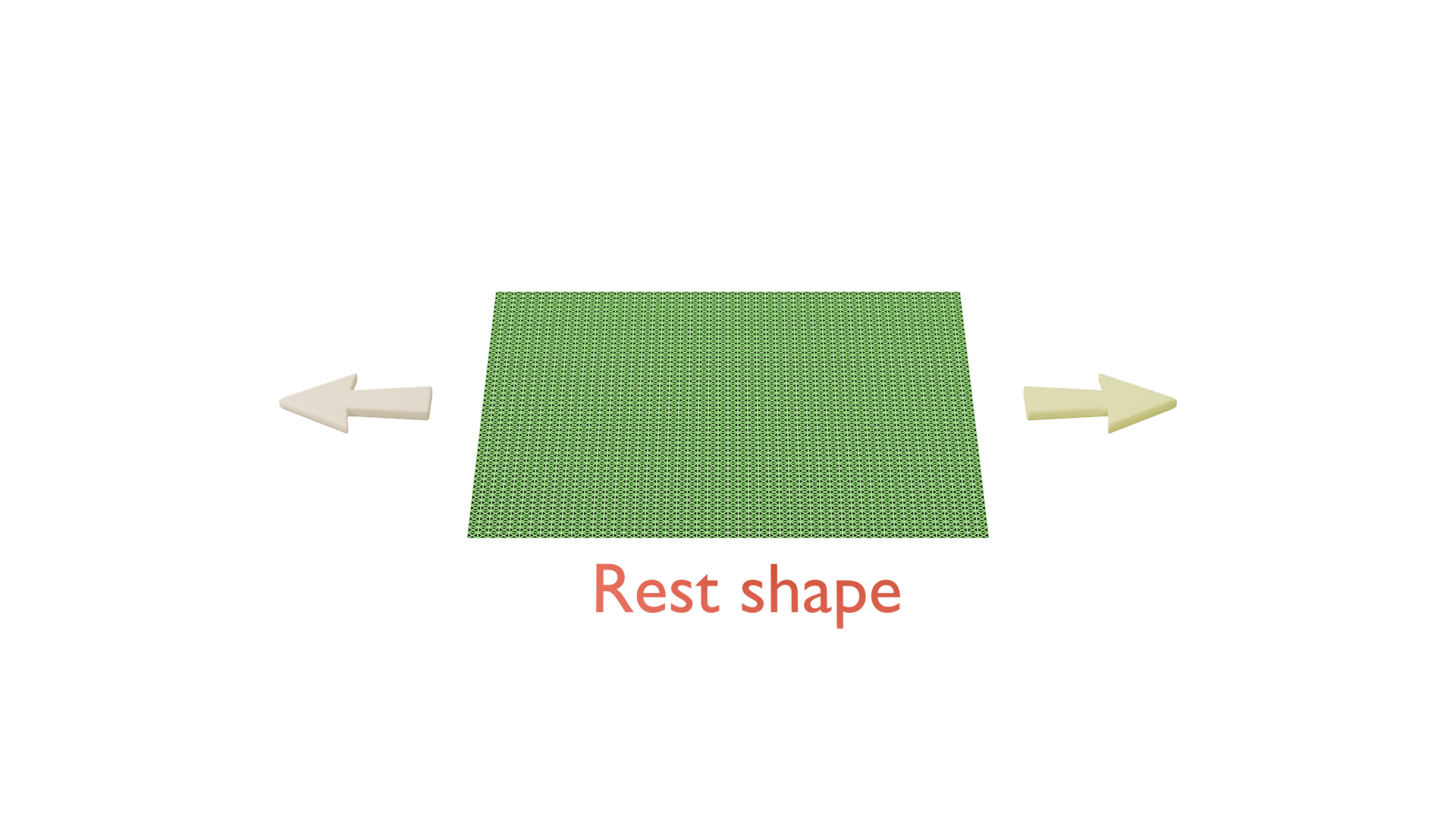}%
            \caption{Rest shape}
	\end{subfigure}%
	\newline%
	\begin{subfigure}[t]{1.0\linewidth}%
        \centering
		\includegraphics[width=0.95\linewidth,clip,trim=0 350 0 300]{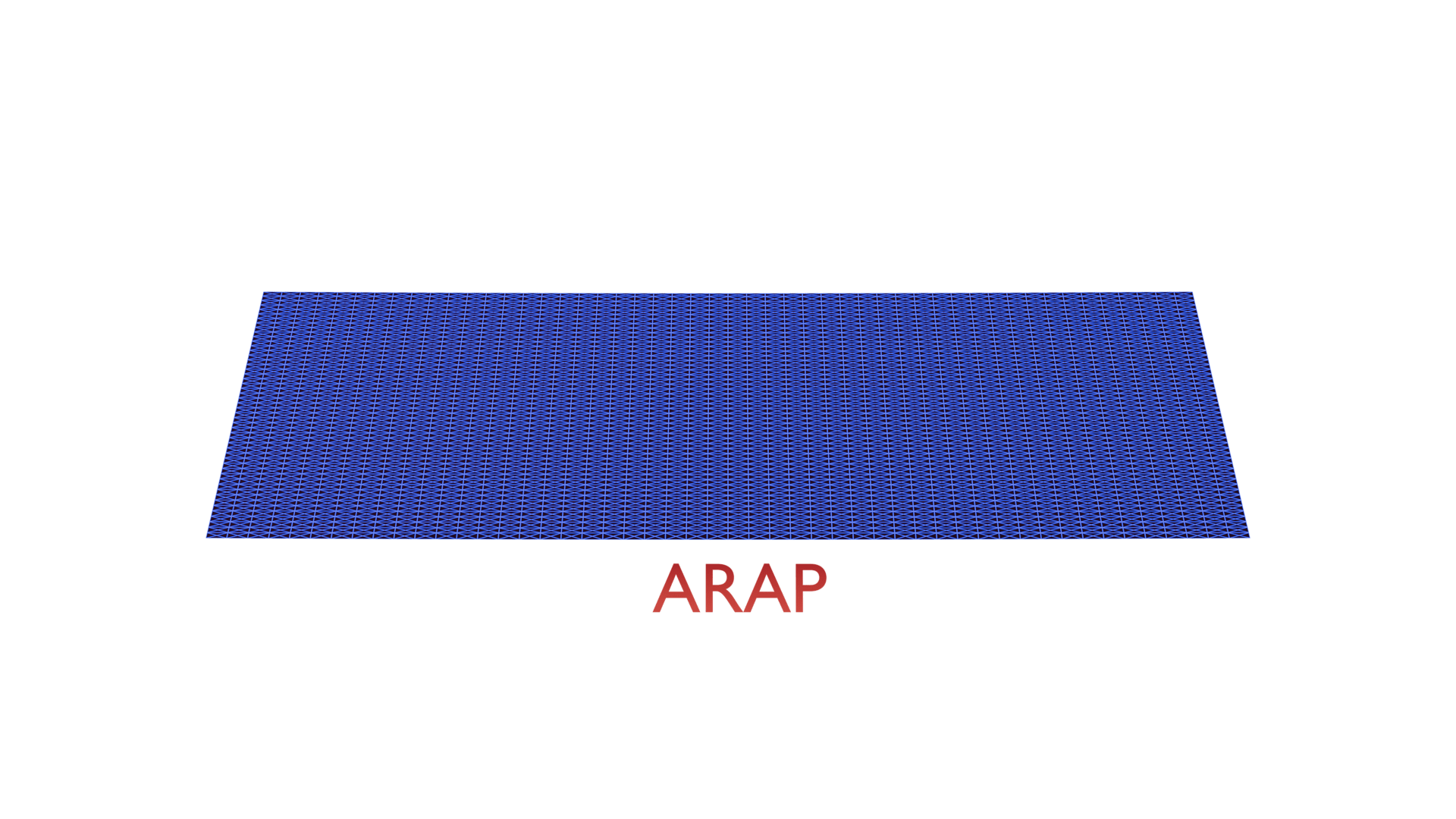}%
            \caption{ARAP material after stretching}
	\end{subfigure}%
	\newline%
	\begin{subfigure}[t]{1.0\linewidth}%
        \centering
		\includegraphics[width=0.95\linewidth,clip,trim=0 350 0 300]{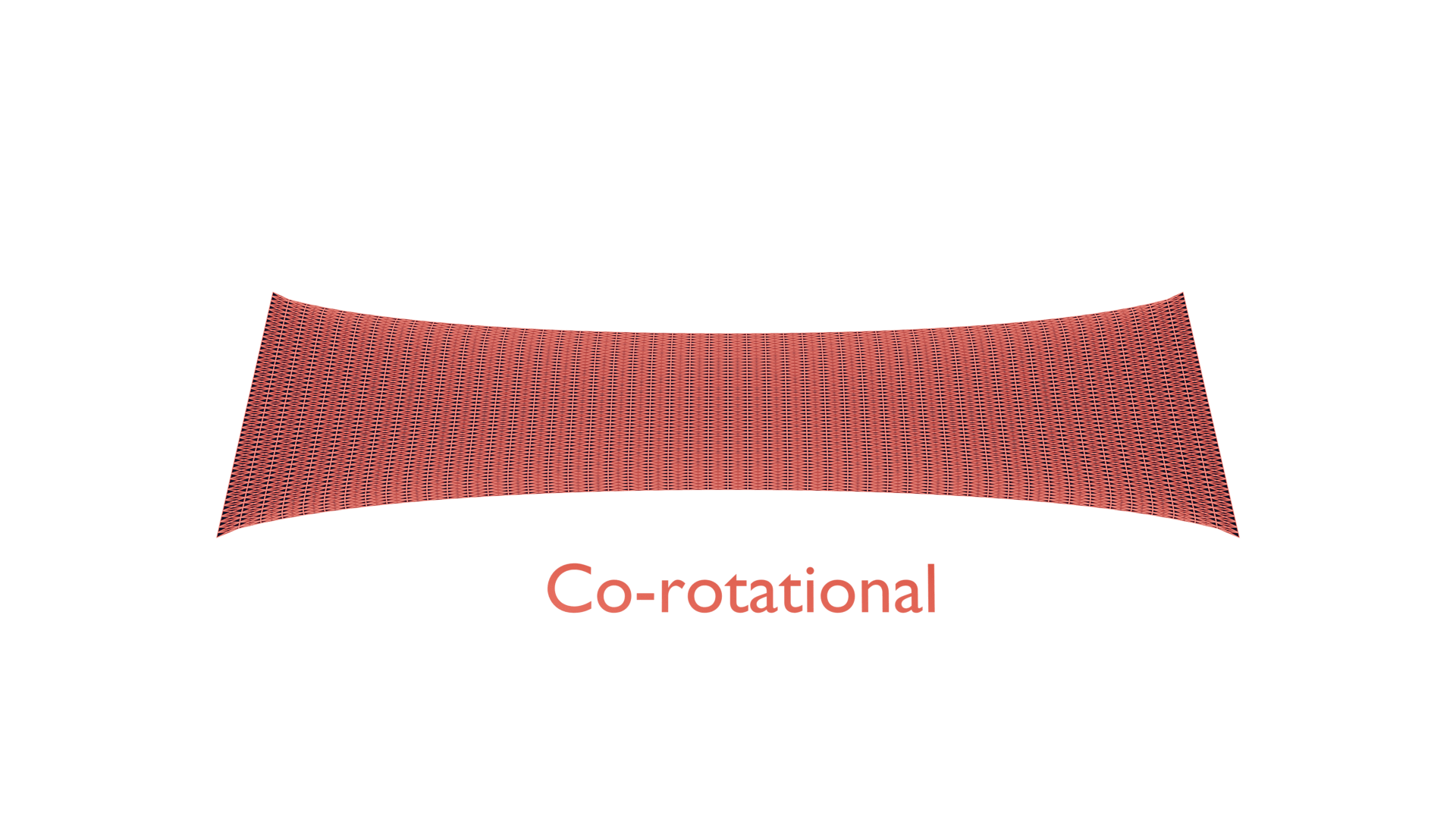}%
            \caption{Co-rotational material after stretching}
	\end{subfigure}%
	\newline%
	\begin{subfigure}[t]{1.0\linewidth}%
        \centering    
		\includegraphics[width=0.95\linewidth,clip,trim=0 350 0 300]{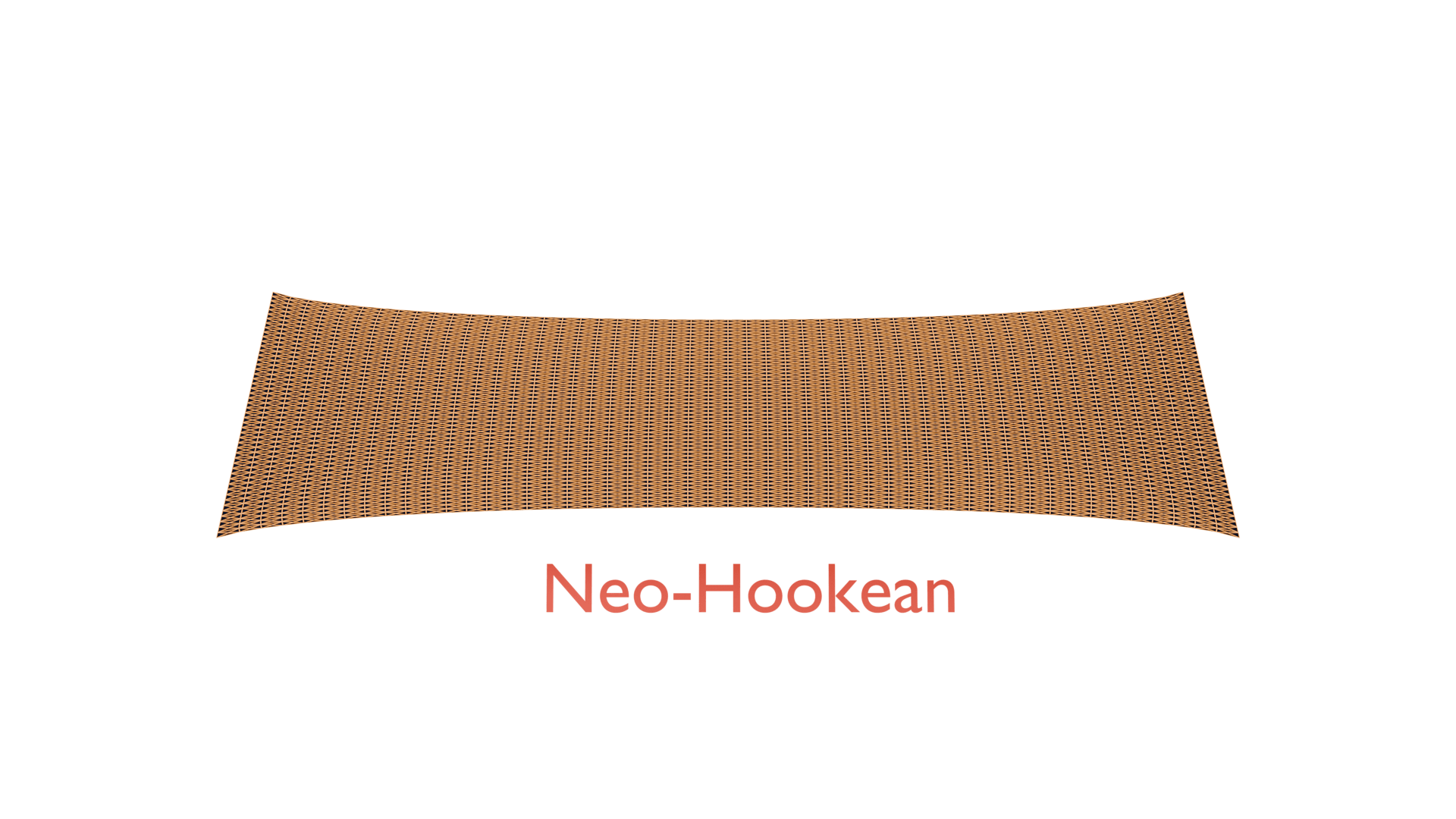}%
            \caption{Neo-Hookean material after stretching}
	\end{subfigure}%
    \caption{
        Cloth Extension: capability of volume preservation with different elastic materials.
    }
    \Description{}
    \label{fig: stretching cloth}
\end{figure}

To show the compatibility with different hyperelastic materials, we make extension tests similar to those in \cite{trusty_mixed_2022} and \cite{macklin_non-smooth_2019}.
Figure \ref{fig: stretching cloth} compares different elastic materials (Neo-Hookean, co-rotational, and ARAP) applied to an identical square cloth.
We observe different volume preservation capability:
The linear co-rotational model exibits high volume loss compared to the Neo-Hookean model, while the ARAP material only deforms orthogonally due to the omission of the preservation item in its local energy formulation.

\subsection{Complementarity precondition and friction accuracy}
\label{sec: complementarity precondition and friction accuracy}

To evaluate the accuracy of the complementarity preconditioner proposed in Section \ref{sec: complementarity precondition}, we make a stick-sliding test similar to that in \cite{larionov_frictional_2021}.
In Figure \ref{fig: accurate friction}, we simulate an elastic square box with high stiffness ($\rho=1000kg/m^3$, $E=10^8Pa$) on a slope inclined at $10$ degrees to the horizontal.
The analytical stick-sliding discontinuity occurs at $\mu^* = tan(10\pi/180) = 0.17632698$.
In this test, the problems are solved with $10$ L-G iterations and $24$ CR iterations for constraint resolution.

While the mass-inverse approach \cite{macklin_non-smooth_2019} achieves a sliding accuracy of only $0.1$, our system-inverse method based on Equation \eqref{eq: complementarity preconditioner} captures such discontinuity with a precision of $0.001$.
This aligns with our analysis in Section \ref{sec: complementarity precondition}: For systems with high stiffness, it is generally the stiffness item that dominates the system diagonal.
In this case, the mass inverse elements are generally significantly larger than those of the actual system inverse.
As a result, the non-smooth indicators frequently fall within the range $|\dot \pene_f| \leq \nsprecond (\mu \lam_n - |\lam_f|)$.
In the phase of solving linear system, this forces the numerical solver to reduce $\vel$ to zeros, resulting in undesired sticking behavior (Figure \ref{fig: accurate friction b}).
Consequently, this gap makes it hard to capture the discontinuities within a few iterations.
In contrast, our method uses exactly the system inverse, resulting in efficient capture of the discontinuities and high accuracy of friction behavior.

\subsection{Non-smooth and codimensional contacts}

We apply our methods in a scenario similar to that in \cite{li_incremental_2020}.
In Figure \ref{fig: cloth on knives}, a square cloth falls onto codimensional triangle-shaped obstacles.
Unlike the IPC test, we use a triangle surface mesh instead of a thin volumetric mat.
Collision handling for triangle meshes is generally more challenging than for volumetric meshes with finite thickness, as infinitely thin structures can generate large motions that typically lead to instabilities.
Owning to the fast convergence of the CR solver, our method can efficiently reach a desired accuracy in a few iterations, thereby maintaining a good stability.

Another scenario with non-smooth contacts is demonstrated in Figure \ref{fig: sharp corner}, where a cloth falls on to a static cube obstacle.
Rich contact is the key difference between this example and the previous one.
For a cloth containing $10.2k$ vertices, such collision event generates $8.62k$ contact constraints (see Table \ref{tab: simulation statics}), creating a challenging large-scale contact problem.
The CR solver efficiently reduces errors within a few iterations, resulting in stable behavior.
Computing the Schur complement is computationally intensive even with acceleration techniques.
Owing to our splitting method for non-smooth indicators in Section \ref{eq: spliting non-smooth indicator}, such computation should only be performed once per frame (unlike per Newton iteration in \cite{macklin_non-smooth_2019}), significantly reducing computational costs.

\subsection{Large deformation with contacts}

We reproduce another example from \cite{li_incremental_2020}. 
In Figure \ref{fig: squeezing ball}, a soft ball drops onto a set of rollers. 
While being strongly squeezed by the rolling cylinders, the ball is under large compressions and therefore generates strong resistant contact forces with the Neo-Hookean material.
This further generates large frictional forces that pulls the ball through the rollers.
After passing through the obstacles, the soft ball is able to recover its shape, exhibiting a strong stability.
In Table \ref{tab: simulation statics}, our method efficiently simulates such example at near real-time rates ($46.15 ms$ per frame for $7k$ vertices), significantly outperforming IPC technique ($60 s$ per frame with same number of vertices).

We further test contact handling for highly stiff objects undergoing large deformations.
In Figure \ref{fig: pulling wooper}, a complex-shaped soft wooper ($E = 10^7 Pa$) is pulled with a moving positional constraint.
Another similar example is in Figure \ref{fig: teaser} where a gingerbread man ($E = 10^6 Pa$) is being pulled though several obstacles.
In both cases, the elastic materials show an interesting behavior to automatically deform and fit with the obstacle shape.
Our method is proven to be stable to simulate these high-stiffness objects, being a key difference with the incomplete solutions and PBD-like methods (e.g., VBD \cite{chen_vertex_2024}).
This is owning to the high efficiency on propagating the local results through the complete solution, as discussed in Section \ref{sec: global step solution strategy}.

\begin{table*}[htb!]
\centering
\caption{Simulation statics: we report the detail of examples in Section \ref{sec: results}.
All simulations employ \textbf{5 L-G iterations}, \textbf{10 CR iterations} (if contacts are involved), and a fixed time interval $\dt = 0.01s$.
For each example, we choose several discretizations (from $5k$ to $20k$ vertices) for the soft objects to compare the performance in different scales of problem.
We detail the memory usage for storing the sparse inverse $\sparseinverse$, and maximum number of contact constraints during simulation.
Owning to our splitting non-smooth indicator strategy, the Schur-complement (Schur.) is performed once per frame and the other steps are performed once per LG iteration.
The SpMV operations in the global step (Global), the assembly phase (Assem.), the constraint linear solution (Cst. Solve), the constraint correction (Corr.), and the total constrained global step (Cst. Global) are presented respectively.
The total cost of each frame without the collision detection (Frame max*) is measured when maximum contact pairs are presented.
*The local step (Local*) is parallelized on the CPU, while the other steps are on the GPU, therefore the total frame (Frame max*) being the measure of a hybrid implementation.
}
\resizebox{\textwidth}{!}{%
\begin{tabular}{ccccccccccccccc}
\toprule
\multirow{2}{*}{Example} & \multicolumn{2}{c}{Mesh} & \multicolumn{2}{c}{Elasticity} & Memory  & Contact & \multicolumn{8}{c}{Time cost (ms)} \\ 
\cmidrule(lr){2-5} \cmidrule(lr){8-15}
& Vert. & Elem. & Constitutive & $\rho(kg/m^3)$, $E(Pa)$, $\nu$ & $\sparseinverse$ (MB) & max. & Schur. & Local* & Global & Assem. & Cst. Solve & Corr. & Cst. Global & Frame max* \\ 
\cmidrule(lr){1-15}
\multirow{3}{*}{Twisting Bar} 
& 5.3k & 26.4k & ARAP & 1000, 1E+9, 0.45 & 125.02 & N/A & N/A & 0.87 & 0.27 & N/A & N/A & N/A & N/A & 6.23 \\
& 11.3k & 59.0k & ARAP & 1000, 1E+9, 0.45 & 463.03 & N/A & N/A & 1.67 & 0.86 & N/A & N/A & N/A & N/A & 13.55 \\
& 20.8k & 111.1k & ARAP & 1000, 1E+9, 0.45 & 1310.19 & N/A & N/A & 2.99 & 2.41 & N/A & N/A & N/A & N/A & 28.55 \\
\cmidrule(lr){1-15}
\multirow{6}{*}{Cloth Extension} 
& 10.2k & 20.2k & ARAP & 1000, 1E+5, 0.45 & 
102.51 & N/A & N/A & 0.39 & 0.33 & N/A & N/A & N/A & N/A & 4.44 \\
& 10.2k & 20.2k & Co-rotation & 1000, 1E+5, 0.45 & 102.51 & N/A & N/A & 0.87 & 0.23 & N/A & N/A & N/A & N/A & 6.43 \\
& 10.2k & 20.2k & Neo-Hookean & 1000, 1E+5, 0.45 & 102.51 & N/A & N/A & 0.93 & 0.23 & N/A & N/A & N/A & N/A & 6.65 \\
\cmidrule(lr){2-15}
& 20.2k & 40.0k & ARAP & 1000, 1E+5, 0.45 & 280.41 & N/A & N/A & 0.59 & 0.58 & N/A & N/A & N/A & N/A & 7.53 \\
& 20.2k & 40.0k & Co-rotation & 1000, 1E+5, 0.45 & 280.41 & N/A & N/A & 1.59 & 0.51 & N/A & N/A & N/A & N/A & 12.11 \\
& 20.2k & 40.0k & Neo-Hookean & 1000, 1E+5, 0.45 & 280.41 & N/A & N/A & 1.73 & 0.52 & N/A & N/A & N/A & N/A & 12.88 \\
\cmidrule(lr){1-15}
\multirow{3}{*}{Cloth on Knives} 
& 5.1k & 10.0k & Neo-Hookean & 1000, 1E+5, 0.4 & 62.37 & 0.76k & 3.45 & 0.47 & 0.13 & 0.25 & 0.54 & 0.18 & 0.98 & 14.73 \\
& 10.2k & 20.2k & Neo-Hookean & 1000, 1E+5, 0.4 & 183.82 & 1.19k & 5.37 & 0.79 & 0.33 & 0.46 & 0.52 & 0.36 & 1.35 & 21.42 \\
& 15.3k & 30.3k & Neo-Hookean & 1000, 1E+5, 0.4 & 342.68 & 1.78k & 11.07 & 1.09 & 0.59 & 0.80 & 0.60 & 0.62 & 2.02 & 33.02 \\
\cmidrule(lr){1-15}
\multirow{2}{*}{Sharp Corner} 
& 5.1k & 10.0k & Neo-Hookean & 1000, 1E+5, 0.4 & 62.37 & 4.57k & 41.45 & 0.34 & 0.14 & 0.87 & 2.79 & 0.15 & 3.82 & 70.77 \\
& 10.2k & 20.2k & Neo-Hookean & 1000, 1E+5, 0.4 & 183.82 & 8.62k & 200.08 & 0.49 & 0.37 & 2.66 & 9.13 & 0.37 & 12.15 & 277.36 \\
\cmidrule(lr){1-15}
\multirow{2}{*}{Squeezing Ball} 
& 7.1k & 37.5k & Neo-Hookean & 1000, 1E+4, 0.4 & 266.87 & 3.18k & 11.60 & 2.75 & 0.50 & 0.99 & 1.39 & 0.51 & 2.89 & 46.15 \\
& 14.7k & 81.7k & Neo-Hookean & 1000, 1E+4, 0.4 & 946.97 & 4.62k & 27.91 & 6.09 & 1.61 & 2.46 & 2.80 & 1.67 & 6.92 & 102.01 \\
\cmidrule(lr){1-15}
\multirow{2}{*}{Pulling Wooper} 
& 5.3k & 20.4k & Neo-Hookean & 1000, 1E+7, 0.3 & 67.14 & 0.42k & 4.76 & 1.25 & 0.14 & 0.26 & 0.54 & 0.19 & 0.98 & 19.95 \\
& 11.8k & 52.5k & Neo-Hookean & 1000, 1E+7, 0.3 & 295.09 & 0.73k & 16.30 & 3.06 & 0.54 & 0.77 & 0.51 & 0.59 & 1.87 & 46.20 \\
\cmidrule(lr){1-15}
\multirow{2}{*}{Gingerbread Man} 
& 11.1k & 48.2k & Neo-Hookean & 1000, 1E+6, 0.3 & 227.30 & 0.74k & 8.37 & 2.77 & 0.41 & 0.59 & 0.48 & 0.45 & 1.52 & 34.75 \\
& 19.5k & 93.6k & Neo-Hookean & 1000, 1E+6, 0.3 & 619.41 & 0.80k & 14.16 & 5.03 & 1.05 & 1.30 & 0.50 & 1.15 & 2.95 & 59.76 \\
\cmidrule(lr){1-15}
\multirow{3}{*}{Grabbing Raptor} 
& 10.3k & 34.7k & Neo-Hookean & 10, 2E+4, 0.1 & 120.00 & 0.14k & 2.42 & 1.72 & 0.23 & 0.37 & 0.62 & 0.28 & 1.27 & 21.49 \\
& 15.3k & 55.2k & Neo-Hookean & 10, 2E+4, 0.1 & 260.85 & 0.13k & 2.43 & 2.63 & 0.45 & 0.62 & 0.71 & 0.51 & 1.84 & 29.68 \\
& 20.4k & 77.8k & Neo-Hookean & 10, 2E+4, 0.1 & 
467.58 & 0.14k & 2.62 & 3.51 & 0.80 & 1.08 & 0.63 & 0.89 & 2.61 & 38.62 \\
\bottomrule
\end{tabular}%
}
\label{tab: simulation statics}
\end{table*}

\subsection{Soft manipulator}

\begin{figure}[htb!]
    \centering
    \includegraphics[width=0.8\linewidth,clip,trim=300 250 300 250]{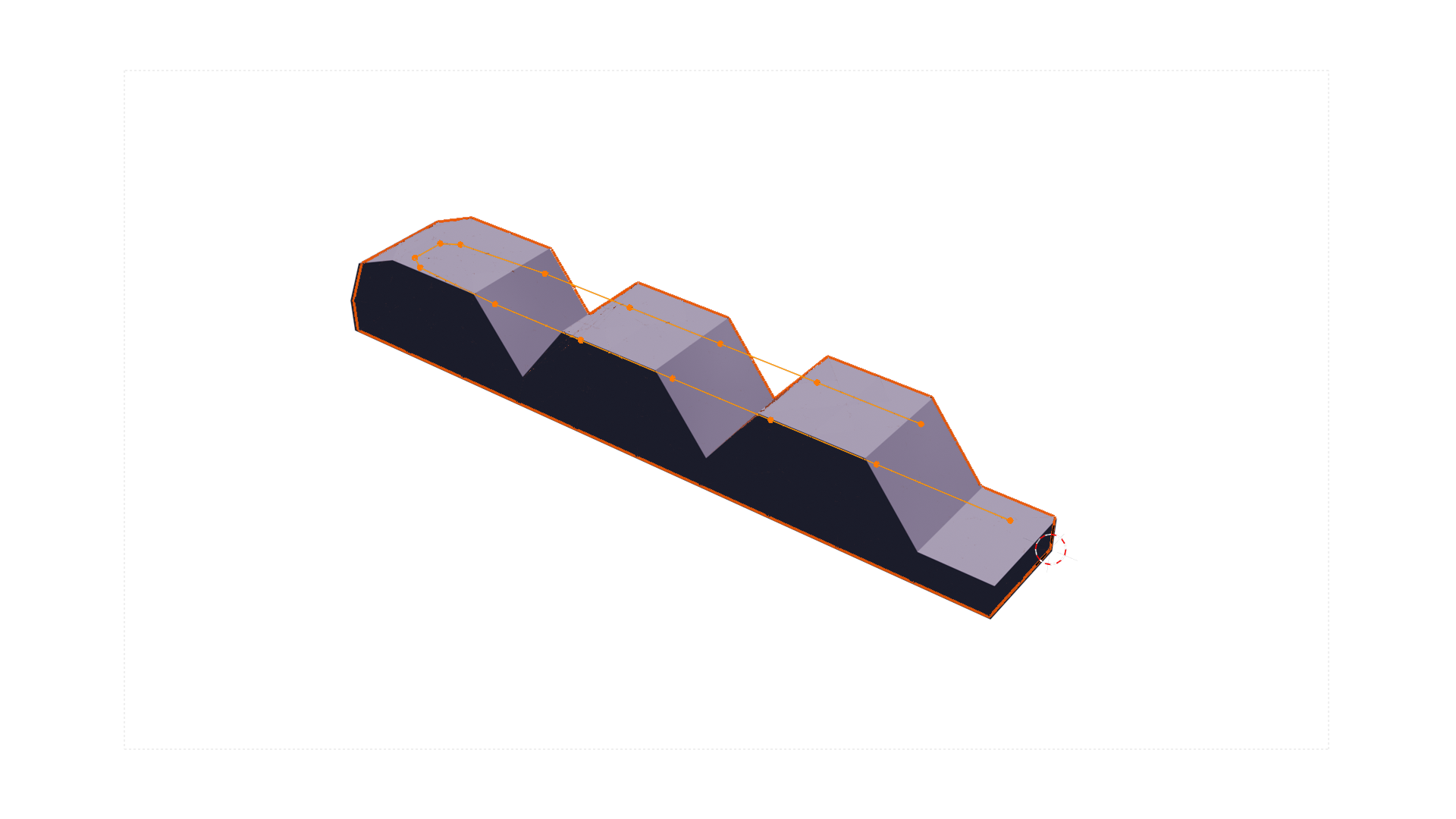}%
    \caption{
        Soft gripper driven by cables (simulated with bilateral constraints, highlighted with orange lines).
    }
    \Description{}
    \label{fig: cable gripper}
\end{figure}

We apply our methods on a complex gripping task.
In Figure \ref{fig: grab raptor}, a soft gripper \cite{coevoet_software_2017} performs a pick-and-place operation with a soft raptor.
The gripper consists of two soft fingers (detailed in Figure \ref{fig: cable gripper}), where bilateral constraints are used to simulate the cable constraints.
By pulling and pushing the cable, the finger performs bending and straightening behavior.

While gripping the raptor, both the fingers and raptor deform to fit their contact surfaces, generating numbers of contact constraints.
Lifting, rotating, and moving the raptor by the fingers are complex operations which require precise handling of frictional constraints.
In this scenario, all objects and obstacles are coupled within a unified system through Equation \eqref{eq: non-smooth newton system} during each L-G iteration.
Our pipeline efficiently simulates this multi-object system at real-time rates (see Table \ref{tab: simulation statics}).
Unlike the other scenarios, the contact constraints remain at the same level when we change the discretization of the soft raptor.
This is because the gripper mesh remains unchanged, and the collision detection iterates on the finger mesh elements.

\section{Limitations and future work}

\paragraph*{\textbf{Generic local-global methods}}
Our sparse inverse method efficiently accelerates the global steps in general local-global integrators.
Although our implementations are mainly based on PD and ADMM-PD, the sparse inverse strategy applicable to various advanced techniques such as LBFGS-PD \cite{liu_quasi-newton_2017}, WRAPD \cite{brown_wrapd_2021}, and Mixed-FEM \cite{trusty_mixed_2022}.
While these methods generally improve the L-G convergence through L-BFGS with line search, they all rely on a constant and pre-factorized system as defined in Equation \eqref{eq: PD global} for efficient computing, thereby can effectively benefit from our sparse inverse method.
We leave the parallelization of these methods as a future work.

\paragraph*{\textbf{Intersection-free collision}}
The collision response in our pipeline is penetration-based.
Although the Signorini-Coulomb condition in Equation \eqref{eq: signorini-coulomb conditions in detail} theoretically ensures intersection-free contacts, such condition is generally never exactly met due to numerical reasons. 
Therefore, unlike IPC \cite{li_incremental_2020}, our method cannot guarantee intersection-free.
Exploring such intersection-free collision could be an interesting topic in future work.

\paragraph*{\textbf{Topology changes}}
Since the efficiency of our sparse inverse method strongly relies on the pre-computation of $\sparseinverse$, it is challenging to to handle topology-altering events such as cutting and tearing.
A possibly valuable exploration is to efficiently update $\sparseinverse$ with the progressively updated Cholesky factor proposed in \cite{herholz_sparse_2020, zhang_progressive_2022}.

\paragraph*{\textbf{Hyper-rich contacts}}
As shown in Table \ref{tab: simulation statics}, the Sharp Corner example exhibits significant computational costs in both the Schur-complement and the constraint linear solver when contact constraints reach a massive scale ($8.62k$).
Although our splitting strategy reduces the Schur-complement computation to once per time step, such computing process, plus the large-scale linear solution process, still prevent a real-time performance.
We suggest simplifying contact constraints by reducing the contact space dimension, such as grouping the contacts in specific areas \cite{allard_volume_2010}.

\section{Conclusion}

In this manuscript, we introduce a unified, GPU-friendly framework for simulating elastic objects in the presence of general contacts.
Through our proposed reformulation, we adeptly address nonlinear and non-smooth challenges by integrating nonlinear complementarity conditions into the local-global iterations.
To optimize performance, we introduce two crucial strategies: a sparse inverse method for parallelizing the local-global integrators while maintaining a rapid convergence rate, and a splitting strategy for non-smooth indicators, which largely reduces Schur-complement computation while refining the complementarity preconditioner.
With the experimental results in various examples, we effectively demonstrate the generality, efficiency, accuracy, and robustness of our method in handling hyperelasticity and frictional contacts.
Notably, our approach emphasizes simplicity, as the core operations rely on standard sparse matrix operations.
We believe this simplicity makes our framework applicable across various downstream fields.

\settopmatter{printacmref=false}
\bibliographystyle{ACM-Reference-Format}
\bibliography{main.bbl}

\appendix
\clearpage

\section{Frictional contact constraints}

\subsection{Constraint linearization}
\label{ap: Constraint linearization}

For each contact pair $j$, we unify the constraint directions ($\cbilateral$, $\cnormal$, and $\cfriction$) as a general form $\cgeneral$:
\begin{equation}
\label{eq: contact linearizations}
\begin{aligned}
    &\peneval_{c,j} =\cgeneral_{j} \cdot \Pene_{j} \\
    &\lamval_{c,j} = \cgeneral_{j} \cdot \Lam_{j} 
\end{aligned}
\end{equation}

\subsubsection*{Interpenetration linearization}
For the interpenetration measure along $\cbilateral$ and $\cnormal$, we have:
\begin{equation}
\label{eq: interpenetration linearizations in pos}
\begin{aligned}
    \peneval_{c,j} = \cgeneral_{j} \cdot \Pene_{j} = \cgeneral_{j} \cdot (\contact_{j} \pos - \hat{\proj}_{j}) = \cgeneral_{j} \cdot \contact_{j} \pos - \cgeneral_{j} \cdot \hat{\proj}_{j}
\end{aligned}
\end{equation}
while the relative velocity in the tangent space (which is needed in the Coulomb's law for friction formulation, see \ref{ap: Frictional contact formulation}) is given as follows:
\begin{equation}
\label{eq: interpenetration linearizations in vel}
    \dot \peneval_{f,j} = \cfriction_{j} \cdot (\frac{\partial \Pene_{j}}{\partial t} - \hat{\relativevel}_{j})
    = \cfriction_{j} \cdot (\frac{\partial (\contact_{j} \pos - \hat \proj_i)}{\partial t} - \hat{\relativevel}_{j})
    = \cfriction_{j} \cdot \contact_{j} \vel - \cfriction_{j} \cdot \hat{\relativevel}_{j}
\end{equation}
where $\hat{\relativevel}$ is the velocity of the moving obstacle if there exist one within the contact pair (see details in \cite{ly_projective_2020}).

\subsubsection*{Contact forces linearization}
We define the force $\vec{\lam}_{c,j} = \lamval_{c,j} \cdot \cgeneral_{j} ^\mathrm{T} = \cgeneral_{j} \cdot \lam_{j} \cdot \cgeneral_{j}^\mathrm{T}$ that is the component of $\lam_{j}$ along $\cgeneral_{j}$.
For a bilateral contact pair $j \in \mathcal{B}$, we have:
\begin{equation}
\label{eq: bilateral forces linearization}
\begin{aligned}
    \contact_{j}^\mathrm{T} \Lam_{j} = (\cbilateral \cdot \contact_{j})^\mathrm{T} \lamval_{b,j}
\end{aligned}
\end{equation}
For a non-interpenetration contact pair $j \in \mathcal{C}$, since the the basis $\cnormal$ and $\cfriction$ are orthonormal basis in euclidean space, we have:
\begin{equation}
\label{eq: components in euclidean space}
\begin{aligned}
    \Lam_{j} 
    = \cnormal_{j} \cdot \Lam_{j} \cdot \cnormal_{j}^\mathrm{T} + \cfriction_{j} \cdot \Lam_{j} \cdot \cfriction_{j}^\mathrm{T} 
    = \lamval_{n,j} \cdot \cnormal_{j} ^\mathrm{T} + \lamval_{f,j} \cdot \cfriction_{j} ^\mathrm{T}
\end{aligned}
\end{equation}
Consequently, the contact items in the constrained implicit Euler \eqref{eq: constrained implicit euler} can be separated into contact and friction items:
\begin{equation}
\label{eq: separated contact and friction forces linearization}
\begin{aligned}
    \contact_{j}^\mathrm{T} \Lam_{j} = \contact_{j}^\mathrm{T} (\cnormal^\mathrm{T}\lamval_{n,j} + \cfriction^\mathrm{T}\lamval_{f,j} )
    = (\cnormal \cdot \contact_{j})^\mathrm{T} \lamval_{n,j} +  (\cfriction \cdot \contact_{j})^\mathrm{T} \lamval_{f,j} 
\end{aligned}
\end{equation}

\subsubsection*{Grouping linearized items}
By grouping the contact terms in \eqref{eq: interpenetration linearizations in pos} \eqref{eq: interpenetration linearizations in vel} \eqref{eq: bilateral forces linearization} \eqref{eq: separated contact and friction forces linearization}, we defining the following contact items for different constraints $\cbilateral$, $\cnormal$, and $\cfriction$:

\begin{equation}
\label{eq: bilateral items}
\begin{aligned}
    &\jacobian_b = 
    \begin{bmatrix}
    \cbilateral_{b_1} \cdot \contact_{b_1} \\ \textrm{...} \\ \cbilateral_{b_m} \cdot \contact_{b_m}
    \end{bmatrix}, \qquad
    &&\lam_b = 
    \begin{bmatrix}
    \lamval_{b_1} \\ \textrm{...} \\ \lamval_{b_m}
    \end{bmatrix}, \\
    &\vecd_b = 
    \begin{bmatrix}
    \cbilateral_{b_1} \cdot \hat{\proj}_{b_1} \\ \textrm{...} \\ \cbilateral_{b_m} \cdot \hat{\proj}_{b_m}
    \end{bmatrix}, \qquad
    &&\pene_b = 
    \begin{bmatrix}
    \peneval_{b_1} \\ \textrm{...} \\ \peneval_{b_m}
    \end{bmatrix}
\end{aligned}
\end{equation}

\begin{equation}
\label{eq: contact items}
\begin{aligned}
    &\jacobian_n = 
    \begin{bmatrix}
    \cnormal_{c_1} \cdot \contact_{c_1} \\ \textrm{...} \\ \cnormal_{c_n} \cdot \contact_{c_n}
    \end{bmatrix}, \qquad
    &&\lam_n = 
    \begin{bmatrix}
    \lamval_{n,c_1} \\ \textrm{...} \\ \lamval_{n,c_n}
    \end{bmatrix}, \\
    &\vecd_n = 
    \begin{bmatrix}
    \cnormal_{c_1} \cdot \hat{\proj}_{c_1} \\ \textrm{...} \\ \cnormal_{c_n} \cdot \hat{\proj}_{c_n}
    \end{bmatrix}, \qquad
    &&\pene_n = 
    \begin{bmatrix}
    \peneval_{n,c_1} \\ \textrm{...} \\ \peneval_{n,c_n}
    \end{bmatrix}
\end{aligned}
\end{equation}

\begin{equation}
\label{eq: friction items}
\begin{aligned}
    &\jacobian_f = 
    \begin{bmatrix}
    \cfriction_{c_1} \cdot \contact_{c_1} \\ \textrm{...} \\ \cfriction_{c_n}  \cdot \contact_{c_n}
    \end{bmatrix}, \qquad
    &&\lam_f = 
    \begin{bmatrix}
    \lamval_{f,c_1} \\ \textrm{...} \\ \lamval_{f,c_n}
    \end{bmatrix}, \\
    &\vecd_f = 
    \begin{bmatrix}
    \cfriction_{c_1} \cdot \hat{\relativevel}_{c_1} \\ \textrm{...} \\ \cfriction_{c_n} \cdot \hat{\relativevel}_{c_n}
    \end{bmatrix}, \qquad
    &&\dot \pene_f = 
    \begin{bmatrix}
    \dot \peneval_{f,c_1} \\ \textrm{...} \\ \dot \peneval_{f,c_n}
    \end{bmatrix}
\end{aligned}
\end{equation}
which allows us to convert the governing Equation \ref{eq: constrained implicit euler} to linearized form:
\begin{equation}
\begin{aligned}
    \pene_b = \jacobian_b \pos - \vecd_b \\
    \pene_n = \jacobian_n \pos - \vecd_n \\
    \dot \pene_f = \jacobian_f \vel - \vecd_f \\
\end{aligned}
\end{equation}
\begin{equation}
\begin{aligned}
    \sum_{j \in \mathcal{L}} \contact_{j}^\mathrm{T} \Lam_{j} 
    &= \sum_{j \in \mathcal{B}} \jacobian_{b,j}^\mathrm{T} \lamval_{b,j} + \sum_{j \in \mathcal{C}} \jacobian_{n,j}^\mathrm{T} \lamval_{n,j} + \sum_{j \in \mathcal{C}} \jacobian_{f,j}^\mathrm{T} \lamval_{f,j} \\
    &= \jacobian_{b}^\mathrm{T} \lam_{b} + \jacobian_n^\mathrm{T} \lam_n + \jacobian_f^\mathrm{T} \lam_f
\end{aligned}
\end{equation}

\subsubsection*{Computing interpenetration}
Note that we do not need to explicitly compute the velocity in each L-G iteration. 
In practice, we can efficiently compute the interpenetration as follows:
\begin{equation}
    \begin{bmatrix}
        \pene_{b}^k \\
        \pene_{n}^k \\
        \dt \dot \pene_{f}^k
    \end{bmatrix} = 
    \begin{bmatrix}
        \jacobian_{b} \pos^k - \vecd_{b} \\
        \jacobian_{n} \pos^k - \vecd_{n} \\
        \jacobian_{f} (\pos^k - \pos_t) - \dt \vecd_{f}
    \end{bmatrix} = 
    \begin{bmatrix}
        \jacobian_{b}  \\
        \jacobian_{n} \\
        \jacobian_{f} 
    \end{bmatrix} \pos^k -
    \begin{bmatrix}
        \vecd_{b} \\
        \vecd_{n} \\
        \jacobian_{f} \pos_t + \dt \vecd_{f}
    \end{bmatrix}
\end{equation}
where the initial interpenetration $\begin{bmatrix} \vecd_{b} \\\vecd_{n} \\ \jacobian_{f} \pos_t + \dt \vecd_{f} \end{bmatrix}$  can be computed at the beginning of the time step.

\subsection{Frictional contact formulation}
\label{ap: Frictional contact formulation}

While the bilateral contact has a simple formula that eliminates the interpenetration $\peneval_{b,j} = 0$, the non-interpenetration contact is a more complicated case that is usually formulated with the Signorini's law for the contact normal constraints:
\begin{equation}
\label{eq: signorini's law}
    0 \leq \peneval_{n,j} \perp \lamval_{n,j} \geq 0
\end{equation}

Following the principle of maximal dissipation \cite{stewart_rigid-body_2000}, the frictional forces $\vec{\lam}_{f,j}$ remove the maximum amount of energy from the system while having their magnitude bounded by the normal forces:
\begin{equation}
\label{eq: maximal dissipation}
\begin{aligned}
    \min_{\vec{\lam}_{f,j}} \quad & \dot{\vec{\pene}}_{f,j}^\mathrm{T} \vec{\lam}_{f,j} \\
    \textrm{s.t.} \quad & \lamval_{f,j} \leq \mu_{j} \lamval_{n,j}
\end{aligned}
\end{equation}
where $\dot{\vec{\pene}}_{f,j}$ measures the relative velocity in the contact space.
When the contact is active ($\lamval_n > 0$), the first-order KKT conditions for Equation \eqref{eq: maximal dissipation} is given by:
\begin{equation}
\label{eq: coulomb's law 1}
    \dot \peneval_{f,j} + \frac{|\dot \peneval_{f,j}|}{|\lamval_{f,j}|} \lamval_{f,j} = 0
\end{equation}
\begin{equation}
\label{eq: coulomb's law 2}
    0 \leq |\dot \peneval_{f,j}| \perp \mu_{j} \lamval_{n,j} - |\lamval_{f,j}| \geq 0
\end{equation}
The first condition \eqref{eq: coulomb's law 1} defines the direction of the frictional force $\cfriction$ as the opposite of velocity direction.
The second one \eqref{eq: coulomb's law 2} gives the complementarity conditions for the cases of sliding ($|\dot \peneval_f| > 0$) and stick ($|\dot \peneval_f| = 0$).
On the other hand, when the contact is inactive ($\lamval_n = 0$), the frictional forces should be set as 0.

Combining \eqref{eq: signorini's law}, \eqref{eq: coulomb's law 1}, \eqref{eq: coulomb's law 2}, \eqref{eq: contact items}, \eqref{eq: friction items}, we assemble the bilateral condition and Signorini-Coulomb condition:
\begin{equation}
\label{eq: signorini-coulomb conditions in detail}
\begin{aligned}
    \forall j \in \mathcal{C}, \quad 0 \leq \peneval_{n,j} \perp \lamval_{n,j} \geq 0 \\
    \forall j \in \mathcal{A}, \quad \dot \peneval_{f,j} + \frac{|\dot \peneval_{f,j}|}{|\lamval_{f,j}|} \lamval_{f,j}= 0 \\
    \forall j \in \mathcal{A}, \quad 0 \leq |\dot \peneval_{f,j}| \perp \mu_{j} \lamval_{n,j} - |\lamval_{f,j}| \geq 0 \\
    \forall j \in \mathcal{I}, \quad \lamval_{f,j} = 0 
\end{aligned}
\end{equation}
where $\mathcal{A} = \{j \in \mathcal{C} \, | \, \lamval_{n,j} > 0 \}$ is the set of all active contact indices, and $\mathcal{I} = \{j \in \mathcal{C} \, | \, \lamval_{n,j} \leq 0 \}$ is its complement in $\mathcal{C}$.

\section{Non-smooth functions}
\label{ap: Non-smooth functions}

\subsection{Minimum-Map formulation}

For unilateral constraints:
\begin{equation}
    \ncpfunc_n = 
    \begin{cases}
    \pene_n \qquad &\text{if} \; \pene_n \leq \nsprecond \lam_n\\
    \nsprecond \lam_n \qquad &\text{if} \; \pene_n > \nsprecond \lam_n
    \end{cases}
\end{equation}

\begin{equation}
\label{eq: non-smooth jacobian in minimum-map for contact}
    \nsjacobian_n = 
    \frac{\partial \ncpfunc_n}{\partial \pos} = 
    \begin{cases}
    \jacobian_n \qquad &\text{if} \; \pene_n \leq \nsprecond \lam_n\\
    \veczeros \qquad &\text{if} \; \pene_n > \nsprecond \lam_n
    \end{cases}
\end{equation}

\begin{equation}
    \ncompliance = 
    \frac{\partial \ncpfunc_n}{\partial \lam_n} = 
    \begin{cases}
    \veczeros \qquad &\text{if} \; \pene_n \leq \nsprecond \lam_n\\
    \nsprecond \qquad &\text{if} \; \pene_n > \nsprecond \lam_n
    \end{cases}
\end{equation}
For friction constraints:

\begin{equation}
    \ncpfunc_f = 
    \nsjacobian_f \vel + \fcompliance \lam_f 
\end{equation}

\begin{equation}
\label{eq: non-smooth jacobian in minimum-map for friction}
    \nsjacobian_f = 
    \frac{\partial \ncpfunc_f}{\partial \vel} = 
    \begin{cases} 
    \jacobian_f \qquad &\text{if} \; \lam_n > \zeros \\
    \zeros \qquad &\text{if} \; \lam_n \leq \zeros 
    \end{cases}
\end{equation}

\begin{equation}
\begin{aligned}
    \fcompliance 
    & = \frac{\partial \ncpfunc_f}{\partial \lam_f} \\
    & = \begin{cases} 
    \identity \quad &\text{if} \; \lam_n \leq \zeros\\
    \zeros \quad &\text{if} \; \lam_n > \zeros \; \& \; |\dot \pene_f| \leq \nsprecond (\mu \lam_n - |\lam_f|)\\
    \frac{|\dot \pene_f| - \nsprecond (\mu \lam_n - |\lam_f|)}{\mu \lam_n} 
    \quad &\text{if} \; \lam_n > \zeros \; \& \;  |\dot \pene_f| > \nsprecond (\mu \lam_n - |\lam_f|)
    \end{cases}
\end{aligned}
\end{equation}

\subsection{Fischer-Burmeister formulation}

For unilateral constraints:
\begin{equation}
    \ncpfunc_n = \pene_n + \nsprecond \lam_n - \sqrt{\pene_n^2 + \nsprecond^2 \lam_n^2}
\end{equation}

\begin{equation}
\label{eq: non-smooth jacobian in FB for contact}
    \nsjacobian_n = 
    \frac{\partial \ncpfunc_n}{\partial \pos} = 
    \bigg( \vecones - \frac{\pene_n}{\sqrt{\pene_n^2 + \nsprecond^2 \lam_n^2}} \bigg) \jacobian_n
\end{equation}

\begin{equation}
    \ncompliance = 
    \frac{\partial \ncpfunc_n}{\partial \lam_n} = 
    \bigg( \vecones - \frac{\nsprecond \lam_n}{\sqrt{\pene_n^2 + \nsprecond^2 \lam_n^2}} \bigg) \nsprecond
\end{equation}
For friction constraints:

\begin{equation}
    \ncpfunc_f = 
    \nsjacobian_f \vel + \fcompliance \lam_f 
\end{equation}

\begin{equation}
\label{eq: non-smooth jacobian in FB for friction}
    \nsjacobian_f = 
    \frac{\partial \ncpfunc_f}{\partial \vel} = 
    \begin{cases} 
    \jacobian_f \qquad &\text{if} \; \lam_n > \veczeros \\
    \veczeros \qquad &\text{if} \; \lam_n \leq  \zeros 
    \end{cases}
\end{equation}

\begin{equation}
\begin{aligned}
    \fcompliance &= 
    \frac{\partial \ncpfunc_f}{\partial \lam_f} \\
    &= \begin{cases} 
    \identity \quad &\text{if} \; \lam_n \leq \veczeros\\
    \frac{\sqrt{\dot \pene_f^2 + \nsprecond^2 (\mu \lam_n - |\lam_f|)^2} - \nsprecond(\mu \lam_n - |\lam_f|)}{|\dot \pene_f| + \mu \nsprecond \lam_n - \sqrt{\dot \pene_f^2 + \nsprecond^2 (\mu \lam_n - |\lam_f|)^2}} \nsprecond \quad &\text{if} \; \lam_n > \veczeros
    \end{cases}
\end{aligned}
\end{equation}

\end{document}